\documentclass{aastex}
\usepackage{graphicx}
\usepackage{amsfonts}

\newcommand{\rem}[1]{ } 
\DeclareMathAlphabet{\mathpzc}{OT1}{pzc}{m}{it}

\begin{document}

\title{Angular Dependence of Jitter Radiation Spectra from Small-Scale Magnetic Turbulence}

\author{Sarah J. Reynolds, Sriharsha Pothapragada}
\affil{Department of Physics and Astronomy, University of Kansas, Lawrence, KS 66045}

\author{Mikhail V. Medvedev\altaffilmark{1}}

\affil{Niels Bohr International Academy, Niels Bohr Institute, University of Copenhagen, 2100 Copenhagen K, Denmark}

\altaffiltext{1}{Also:  Department of Physics and Astronomy, University of Kansas, Lawrence, KS 66045 and Institute for Nuclear Fusion, RRC ``Kurchatov
Institute'', Moscow 123182, Russia}

\begin{abstract}
Jitter radiation is produced by relativistic electrons moving in turbulent small-scale magnetic fields such as those produced by streaming Weibel-type instabilities at collisionless shocks in weakly magnetized media. Here we present a comprehensive study of the dependence of the jitter radiation spectra on the properties of, in general, anisotropic magnetic turbulence. We have obtained that the radiation spectra do reflect, to some extent, properties of the magnetic field spatial distribution, yet the radiation field is anisotropic and sensitive to the viewing direction with respect to the field anisotropy direction. We explore the parameter space of the magnetic field distribution and its effect on the radiation spectrum. Some important results include: the presence of the harder-than-synchrotron segment below the peak frequency at some viewing angles, the presence of the high-frequency power-law tail even for a monoenergetic distribution of electrons, the dependence of the peak frequency on the field correlation length rather than the field strength, the strong correlation of the spectral parameters with the viewing angle. In general, we have found that even relatively minor changes in the magnetic field properties can produce very significant effects upon the jitter radiation spectra. We consider these results to be important for accurate interpretation of prompt gamma-ray burst spectra and possibly other sources. 
\end{abstract}


\section{Introduction}
\label{s:intro}

Radiation from astrophysical or laboratory plasmas depends upon the strength and topology of the magnetic field within the plasma, as well as on the acceleration of plasma particles.  For relativistic plasmas, radiative output of an individual particle depends significantly on the particle's direction of motion, as relativistic beaming modifies the angular distribution of the radiation emitted to lie within a cone of opening angle 1/$\gamma$ along the particle's velocity.  Particles in uniform magnetic fields produce synchrotron radiation, which is characterized by the sweep of the radiation cone as relativistic particles orbit or spiral along the magnetic field lines.  In turbulent magnetic fields, particles still radiate in a synchrotron regime as long as the scale of the magnetic field variation is longer than the particles' average Larmor radii.  However, if the magnetic field varies on a scale smaller than a Larmor radius, the particles ``jitter'' through a series of small transverse accelerations without a substantial change of direction.  An observer will thus be in a particle's radiative cone over some length of its accelerated path, until the particle direction is sufficiently deviated that the line of sight is outside the radiative cone.  For an isotropic particle distribution, the rates of particles entering and exiting paths within 1/$\gamma$ of the line of sight are equal.  The resulting radiation will be that of particles ``jittering'' with small randomized accelerations, and will be reflective of the magnetic field spectra along a line of sight path through the turbulent magnetic field region \citep{M00}. 
	
Whereas in synchrotron radiation spectra the frequencies depend upon relativistic beaming and the sweep of the radiation cone, in jitter radiation spectra the frequencies instead depend directly upon the turbulent variations of the magnetic field and relativistic beaming serves only to limit the observed radiation to that emitted by trajectories within 1/$\gamma$ of the line of sight \citep{M00}.  The resulting radiative spectrum differs from that of synchrotron radiation and, in particular, its low-frequency spectral index is not limited by $s=1/3$ (i.e., $dW/d\omega\propto\omega^{1/3}$ below the synchrotron peak), referred to in the literature as the synchrotron "line of death" \citep{Preece+98}.
	
Gamma-ray bursts (GRBs) are believed to be a natural astrophysical site for the emergence of a strong small-scale turbulent magnetic field.  Relativistic shocks generate strong magnetic fields that are random on very small (sub-Larmor) scales via the Wiebel-like (particle streaming) instability \citep{ML99}.  In such an instability, a small magnetic field perturbation in the plane transverse to the motions of counter-streaming particles results in the development and growth of filamentary current structures up to a saturation point at which they may persist for some time longer than the dynamical time-scale of the system, creating an extended region of small-scale turbulent magnetic field.  This has been studied extensively in numerical PIC simulations (e.g., \citet{Silva+03,Nish+03,fred04,M+05,Spit05,CSA07,Spit07}) of both baryonic and pair plasmas.  (See \citet{MS09} for a discussion of the applicability and implications of such simulations for GRB physics.)  Recent simulations of magnetic reconnection in pair plasmas have also shown the generation of these strong small-scale magnetic fields via the Weibel instability acting on the streams of accelerated particles in the reconnection exhaust funnels (\citet{Swis+08,ZH08}).  We speculate that magnetic reconnection events in a GRB may produce electron-positron plasmas in situ, even in initially lepton-poor plasmas.  For the resulting regions of strong small-scale magnetic field, synchrotron theory is inapplicable and jitter radiation theory must be considered. Thus, regardless of the particular model of a GRB (e.g., baryonic, leptonic, magnetic, etc.), the jitter radiation mechanism coupled with the relativistic kinematics of the ejected material represents a viable phenomenological model. It has recently been shown that such a model reproduces a number of spectral features of GRB light-curves remarkably well \citep{MPR09}.

Following the approach of \citet{M06}, we define a geometry where the local filamentation axis (the local axis along which counter-streaming particle motion occurs) lies in the $z$-direction so that the field perturbation from the shock or reconnection event is amplified by the Weibel instability in the $xy$-plane (this might be at or upstream from a shock front lying in the $xy$-plane and propagating in $z$, or upstream from reconnection exhaust funnels with filaments pointing in $z$ direction and the Weibel fields being perpendicular to it, as suggested by PIC simulations). The resulting amplified magnetic field is randomly oriented in the $xy$-plane and independently generated at each position in $z$ (shown theoretically in \citet{ML99} and confirmed in PIC simulations such as \citet{Nish+03}, \citet{Silva+03}, and \citet{fred04}).  The decoupled behavior of the magnetic fields along the filamentation axis ($z$) and in the plane ($xy$) transverse to it means that the resulting Fourier spectra are independent of one another and the overall field distribution is highly anisotropic.  Qualitatively, the spectrum of the Weibel-generated magnetic field in the direction transverse to the filamentation axis has been shown to rise and then drop at a scale of order the plasma skin depth \citep{fred04}.  The magnetic field thus has a general spectral form that may be parameterized as
\begin{equation}\label{eq:perpfspec}
f_{xy}(k_{\perp})=\frac{k_{\perp}^{2\alpha_{\perp}}}{\left( \kappa_{\perp}^2+k_{\perp}^2\right)^{\alpha_{\perp}+\beta_{\perp}}},
\end{equation}
where $k_\perp = (k_x^2 + k_y^2)^{1/2}$, and $\alpha_{\perp} > 0$ and $\beta_{\perp} > 0$.  In this form $k_{\perp}$ refers to the magnetic field wavenumber in the plane transverse to the filamentation axis and $\kappa_{\perp}$, $\alpha_{\perp}$, and $\beta_{\perp}$ are free parameters controlling the spectral break and the soft and hard spectral indices, respectively.  The field along the filamentation axis is in general unknown but we expect it to be of a similar form, with independent free parameters $\kappa_{\parallel}$, $\alpha_{\parallel}$, and $\beta_{\parallel}$:  
\begin{equation}\label{eq:pllfspec}
f_{z}(k_{\parallel})=\frac{k_{\parallel}^{2\alpha_{\parallel}}}{\left( \kappa_{\parallel}^2+k_{\parallel}^2\right)^{\alpha_{\parallel}+\beta_{\parallel}}},
\end{equation}
where again $\alpha_{\parallel} > 0$ and $\beta_{\parallel} > 0$.  A plot of $f_{z}(k_{\parallel})$ is shown for a particular choice of parameters in Figure \ref{fig:basicspectra} and demonstrates the basic behavior of this function.  

The variables $k$ and $\kappa$ are presumed to be unitless, their units $k_0$ having been separated into a normalizable coefficient $k_0^{-2\beta}$.  We have modified this form from the \citet{M06} paper, in which the power in the denominator was simply $\beta$ and it was required that $\beta > \alpha$.  In these spectral forms used here, the asymptotes of the functions given in equations \ref{eq:perpfspec}-\ref{eq:pllfspec} are:
\begin{equation}\label{eq:fldasympt}
f(k)=\left\{
 \begin{array}
  {l@{\qquad}l}
	k^{2\alpha} ,& \mbox{if  } k<<\kappa,\\	
	k^{-2\beta} ,& \mbox{if  } k>>\kappa.
	\end{array}
	\right. 
\end{equation}

The theory of jitter radiation with the above magnetic field spectra has been utilized by \citet{M06} to derive the basic equations describing the jitter spectrum and demonstrate its dependence on the angle $\theta$ between the viewing angle and the local filamentation axis.  Taking into account our modification of the form of the magnetic field spectra, the analytical work \citep{M06} indicates that the jitter radition $F_{\nu}$ spectrum should have the following general properties:

\begin{enumerate}
\item two breaks, with locations depending on $\kappa_\perp$, $\kappa_\parallel$, and $\theta$,
\item a high-energy spectral index $\beta^{\prime}$, where $\beta^{\prime}$ approaches $\beta_{\parallel}$ as $\theta$ goes to 0 and $\beta^{\prime}$ approaches $\beta_{\perp}$ as $\theta$ goes to $\pi$/2,
\item a low-energy spectral index $\alpha^{\prime}$, where $\alpha^{\prime}$ approaches 1 as $\theta$ goes to 0 and $\alpha^{\prime}$ approaches 0 as $\theta$ goes to $\pi$/2.
\end{enumerate}

The resulting radiation spectra from numerical calculations for a limited selection of parameter choices and viewing angles were presented in \citet{M06}.  Here we present the results of more extensive numerical calculations illustrating the full range of jitter spectral variation due to viewing angle for a chosen set of field parameters, followed by a more thorough exploration of the magnetic field parameter space and its implications for the jitter radiation spectrum.  Section II of our paper presents the calculated acceleration spectra for a range of viewing angles $\theta$.  In Section III we develop the connection between the acceleration and radiation spectra through an analytical treatment of a simple linearized model of the acceleration spectrum.  In Section IV we present the radiation spectra calculated for various viewing angles $\theta$ and analyze the spectral variation with viewing angle by means of a five-parameter spectral fit.  In Section VI we show the effect of variations in the magnetic field parameters upon the radiation spectra.  Section VII presents discussion and conclusions.

\section{Acceleration Spectra}
\label{s:accspec}

The equations for calculating the spectrum of a particle's acceleration due to magnetic field turbulence generated by a relativistic Weibel-type instability were developed in \citet{M06}, but only calculated in full for a single representative oblique viewing angle of $\theta = 10^o$ in between the head-on ($\theta = 0$) and edge-on ($\theta = 90^o$) cases.  Here we calculate the acceleration spectra for a more complete range of intermediate viewing angles in order to explore the spectral progression with $\theta$.

As derived in \citet{M06}, the volume-averaged temporal Fourier component of a particle's acceleration due to the Lorentz force for the static magnetic field generated by the relativistic Weibel instability is
\begin{eqnarray}\label{eq:accform}
	\left\langle \left| \mathbf{w}_{\omega^{\prime}} \right| ^2 \right\rangle & = & \left(2\pi  V\right)^{-1}\int\left|\mathbf{w_k}\right|^2\delta\left(\omega\prime+\mathbf{k\cdot v}\right)d\mathbf{k}\\
	& = & \frac{C}{2\pi}\left(1+\cos ^2\theta\right) \int f_z(k_\parallel)f_{xy}(k_\perp)\delta\left(\omega\prime+\mathbf{k\cdot v}\right)d^2k_{\perp}dk_{\parallel}
\end{eqnarray}
where $k$ is the magnetic wavevector, $\bf v$ is the particle's velocity vector, and $\mathbf{w}_{\omega^{\prime}}=\int\mathbf{w}e^{i\omega^{\prime}t}$.  For the case of a shock viewed at an oblique angle $\theta$ from the normal of the shock plane (defined as the $z$-axis), ${\bf k\cdot v}=k_{x} v\sin\theta + k_{z} v\cos\theta$, where we have defined the $x$-axis so that the velocity vector $\bf v$ lies in the $xz$-plane.  We can then use the delta function to substitute for either $k_{x}$ or $k_{z}$.  This becomes:
\begin{equation}\label{eq:accxyform}
	\left\langle \left| \mathbf{w}_{\omega^{\prime}} \right| ^2\right\rangle =  \frac{C}{2\pi |v\cos\theta|}(1+\cos^2\theta)\int f_z\left(\frac{\omega^{\prime}}{v\cos\theta}+k_{x}\tan\theta\right)f_{xy}((k_x^2+k_y^2)^{1/2})dk_xdk_y
	\end{equation}	
or
\begin{equation}\label{eq:accz}
	\left\langle \left| \mathbf{w}_{\omega^{\prime}} \right| ^2\right\rangle =  \frac{C}{2\pi |v\sin\theta|}(1+\cos^2\theta)\int f_z(k_z)f_{xy}\bigg(\bigg(\bigg(\frac{\omega^{\prime}}{v\sin\theta}+k_{z}\cot\theta\bigg)^2+k_y^2\bigg)^{1/2}\bigg)dk_ydk_z
	\end{equation}
where $C$ is an arbitrary normalization constant, proportional to $\left\langle B^2\right\rangle$.  The two forms are equivalent; however, as we approach $\theta$ = 0 or $\theta$ = $\pi/2$, the calculation is more convenient if one avoids denominators approaching zero by choosing the appropriate form.  It should be noted that neither form is valid for the endpoints $\theta$ = 0 or $\theta$ = $\pi$/2, which must be treated separately as in \citet{M06}.  

For a single radiating particle, we plug equations (\ref{eq:perpfspec}) and (\ref{eq:pllfspec}) for $f_{xy}$ and $f_z$ into equations (\ref{eq:accxyform}) and (\ref{eq:accz}) to obtain:
\begin{equation}\label{eq:accxyfin}
  \left\langle \left| \mathbf{w}_{\omega^{\prime}} \right| ^2\right\rangle = \eta_1(\theta)\frac{C}{2\pi v} \int_{-\infty}^{\infty} \int_{-\infty}^{\infty}\frac{(\frac{\omega^{\prime}}{v\sin\theta}+k_x)^{2\alpha_{\parallel}}}{(\kappa_{\parallel}^2\cot^2\theta+(\frac{\omega^{\prime}}{v\sin\theta}+k_x)^2)^{\alpha_{\parallel}+\beta_{\parallel}}}\frac{(k_x^2+k_y^2)^{\alpha_\perp}}{(\kappa_{\perp}^2+k_x^2+k_y^2)^{\beta_{\perp}}}dk_xdk_y
\end{equation}
where
\begin{equation}
	\eta_1(\theta) = \frac{(\tan\theta)^{-2\beta_{\parallel}}(1+\cos^2\theta)}{|\cos\theta|}
\end{equation}
or alternatively
\begin{equation}\label{eq:acczfin}
	\left\langle \left| \mathbf{w}_{\omega^{\prime}} \right| ^2\right\rangle =  \eta_2(\theta)\frac{C}{2\pi v}\int_{-\infty}^{\infty}\int_{-\infty}^{\infty} \frac{k_{z}^{2\alpha_{\parallel}}}{( \kappa_{\parallel}^2+k_z^2)^{\alpha_{\parallel}+\beta_{\parallel}}}\frac{((\frac{\omega^{\prime}}{v\cos\theta}+k_z)^2+k_y^2\tan^2\theta)^{\alpha_{\perp}}}{\left( \kappa_{\perp}^2\tan^2\theta+(\frac{\omega^{\prime}}{v\cos\theta}+k_{z})^2+k_y^2\tan^2\theta\right)^{\alpha_{\perp}+\beta_{\perp}}}dk_ydk_z
  \end{equation}
where 
\begin{equation}
\eta_2(\theta) = \frac{(\cot\theta)^{-2\beta_{\perp}}(1+\cos^2\theta)}{|\sin\theta|}
\end{equation}
This may then be numerically integrated to produce the acceleration spectrum as a function of $\omega^{\prime}$.

We normalize the wave-vector to a dimensional constant $k_0$, hence $k$ is dimensionless, as discussed above.  The spectral form remains the same with $\kappa$ = $\kappa{^\prime}k_0$ and a normalizable factor $k_0^{2\beta}$ in front.  The frequencies $\omega$ are normalized to $\omega_0$ such that $v = \omega_0/k_0 = 1$.  The parameters $\alpha_i$, $\beta_i$, and $\kappa_i$ in our magnetic field spectra are in general unknown free parameters that depend upon the spatial distribution of the magnetic field in the turbulent region, and the parameters $\kappa_i$, corresponding to the dimension correlation lengths of the magnetic field in the turbulent region, will vary even within an astrophysical system and affect the resulting peak energy of the radiation spectrum.  For our initial calculations of spectral variations with $\theta$ we choose reasonable values of the parameters as follows:
\begin{eqnarray}
	\kappa_\perp = \kappa_\parallel = 10, \\
	\alpha_\perp = \alpha_{\parallel} = 2, \\
	\beta_\perp = \beta_\parallel = 1.5\,;
\end{eqnarray}
These parameters are varied (jointly and individually) in Section \ref{s:sfvarparam} to explore the resulting effect on the calculated radiation spectrum.  

We present the results for varying $\theta$ in Figure \ref{fig:AccSpecAll}.  Figure \ref{fig:AccSpecPeakDetail} shows further detail of the critical region around the peak of the acceleration spectra. The graphs show linearly-connected $\log_e-\log_e$ data, with data point intervals of 0.1.  We have arbitrarily normalized the spectra so that the low-$k$ part of the $\theta = 2$ spectrum asymptotes at unity. The acceleration spectra are flat for low $\omega^{\prime}$, then at a certain $\omega^{\prime}$ they turn rapidly into a sloped region, then there is a second transition to a steep decline for high $\omega^{\prime}$.  For $\theta$ near 0 the spectra have a clear peak, but as $\theta$ increases the peak recedes and eventually disappears altogether.  As $\theta$ approaches $\pi$/2 a peak again becomes evident but the location of the peak has shifted by about 0.4 from its position for low $\theta$.  The transition point for flattening at low $\omega^{\prime}$ appears to move off rapidly to lower $\omega^{\prime}$ as $\theta$ approaches 0.   

Figure \ref{fig:AccSpecAmps} plots the amplitude of the acceleration spectra at our lower calculation boundary, at the approximate location of the peak for small $\theta$, and at the approximate location of the peak for $\theta$ close to $\pi$/2.  The crossing of the lines on this graph correspond to spectral transitions as the peak disappears and then eventually reappears at a new location for higher $\theta$.  
The graph of the slope of the $\log-\log$ plot in Figure \ref{fig:AccSpecSlope} also illustrates the disappearance and reemergence of the peak, but further shows that even for the unpeaked spectra there is a flattening of the slope of the mid-range $\theta$ spectra in the region in between the positions of the peaks that appear at higher and lower $\theta$.  For unpeaked spectra, there remains a  transition region of some extent between the low-frequency and high-frequency power laws; consequently, the unpeaked spectra may still be better fit by division into three power law regions as opposed to two.  

An analysis of equations (\ref{eq:accxyfin}) and (\ref{eq:acczfin}) indicates that $\omega^{\prime}$ functions as a shift of the center of the magnetic spectral form in which we have made the delta-function substitution.  Since this $\omega^{\prime}$ factor is always positive, the function's offset always occurs in the direction of negative wavenumber components $k_x$ or $k_z$.  An example of this behavior for the product of two functions such as equations (\ref{eq:perpfspec}) and (\ref{eq:pllfspec}) over a range of offsets is shown in Figure \ref{fig:offsetvar}.  The resulting integral is highly sensitive to the shape of the two functions and the offset between them, which control the resonance like behavior that produces the resulting peak and transition points in the acceleration spectrum.  The angle $\theta$ plays a role in determining both the width and the offset of one function relative to the other: the $\kappa_{\parallel}^2\cot^2\theta$ or $\kappa_{\perp}^2\tan^2\theta$ terms in the denominator influence the width of the function under consideration, while the offset is given by $\frac{\omega^\prime}{v\cos\theta}$ or ${\frac{\omega^\prime}{v\sin\theta}}$.  These are linked such that as $\theta$ increases, both the width and the offset of the function containing $\theta$ increases. The spectral indices $\alpha_\perp$, $\alpha_{\parallel}$, $\beta_\perp$, and $\beta_\parallel$ can also influence the width of the functions and hence also affect the location of the transition points in the acceleration spectra.  The overall effect of such variations on the resulting jitter radiation spectrum will be explored more in Section 5.  

Summarizing the results of this section, we find that the acceleration spectra are generally characterizable to a good approximation by division into three regions: a flat low-$k$ region and two power law regions, as shown in Figure \ref{fig:wcartoon}. These may altogether be described via one amplitude, two non-zero spectral indices being functions of $\alpha$s and $\beta$s of the field spectra, and two breaks, which depend primarily on the magnetic spectral peaks $\kappa_\perp$, $\kappa_{\parallel}$, and the viewing angle $\theta$.  In the next section we develop a simple linearized model of the acceleration spectrum using these five parameters and show that they can be used to analytically predict the behavior of the final jitter radiation spectrum.
	
\section{From Acceleration to Radiation Spectra}
\label{s:acctorad}

The angle-averaged emissivity of a relativistic particle undergoing a series of small transverse accelerations not substantially affecting its overall velocity is as follows (\citealp{LL,M00}, and others):
\begin{equation}\label{eq:powereq}
	\frac{dW}{d\omega}=\frac{e^2\omega}{2\pi c^2}\int_{\omega/2\gamma^2}^{\infty}\frac{\left|\mathbf{w}_{\omega^{\prime}}\right|^2}{\omega^{\prime2}}\left(1-\frac{\omega}{\omega^{\prime}\gamma^2}+\frac{\omega^2}{2\omega^{\prime2}\gamma^4}\right)d\omega^{\prime}
	\end{equation}
Analysis of the high-$\omega$ and low-$\omega$ asymptotic behavior of the radiation spectra for shocks viewed head-on and edge-on, as carried out in \citet{M06}, yields: 
\begin{eqnarray}\label{eq:radasympt}
\mbox{for } \theta=0,\qquad \frac{dW}{d\omega}\propto \left\{
\begin{array}
  {l@{\qquad}l}
	\omega^1 & \mbox{if  } \omega\ll\kappa_\| v \gamma^2\\	
	\omega^{-2\beta_{\parallel}} & \mbox{if  } \omega\gg\kappa_\| v \gamma^2\\
	\end{array}
	\right. \nonumber \\ \nonumber \\
\mbox{for } \theta=\pi/2,\qquad \frac{dW}{d\omega}\propto \left\{
\begin{array}
  {l@{\qquad}l}
	\omega^0 & \mbox{if  } \omega\ll\kappa_\perp v \gamma^2\\	
	\omega^{-2\beta_{\perp}+1} & \mbox{if  } \omega\gg\kappa_\perp v \gamma^2\\
	\end{array}
	\right. 
\end{eqnarray}
where $\theta$ is the angle between the line of sight and the normal to the shock front \citep{M06} and $\alpha_\perp,\ \alpha_\|>1/2$.  We thus expect the radiation spectra at oblique viewing angles to vary between these two forms, dominated by the parallel or the transverse spectra as we vary between the two extremes. 

The $\omega$ dependence contributed by the integral primarily originates in the lower limit $\omega/(2\gamma^2)$ and where it falls on the acceleration spectrum.  In section \ref{s:accspec}, we found that the acceleration spectrum could be simply characterized as three regions of approximately power-law behavior: a flat initial amplitude at low $\log\omega^{\prime}$ (Region I), a region of positive or negative slope (Region II), and a region with a more steeply negative slope (Region III).  Using this to make a simple approximation for our acceleration spectrum in three regions, we can calculate an approximate analytical solution to equation \ref{eq:powereq}.  

We define our simple approximation to the acceleration spectrum using five free parameters, all of which may be fit to the spectrum on a $\log_e-\log_e$ plot: $\mathpzc{A}=\log a_0$ is the amplitude of the low-$\omega^{\prime}$ limit; $\mathpzc{T}_1=\log \omega_1^{\prime}$ is the transition point between the first and second regions; $\mathpzc{S}_1$ is the spectral index in the second region; $\mathpzc{T}_2=\log \omega_2^{\prime}$ is the transition point between the second and third regions; and $-\mathpzc{S}_2$ is the spectral index in the third region ($\mathpzc{S}_2 > 0$).  The acceleration spectrum then has the form (shown in Figure \ref{fig:wcartoon}):
\begin{eqnarray}\label{eq:accap1}
\label{eq:accap1.1}
\lefteqn{\mbox{for } \omega^{\prime} < e^{\mathpzc{T}_1}:} \nonumber \\
& & \left\langle \left| \mathbf{w}_{\omega^{\prime}} \right| ^2\right\rangle = e^\mathpzc{A} = a_0, \\
\label{eq:accap1.2}
\lefteqn{\mbox{for } e^{\mathpzc{T}_1} < \omega^{\prime} < e^{\mathpzc{T}_2}:} \nonumber \\
& & \left\langle \left| \mathbf{w}_{\omega^{\prime}} \right| ^2\right\rangle = e^{\mathpzc{A}-\mathpzc{S}_1\mathpzc{T}_1}\omega^{\prime \mathpzc{S}_1} = a_0\left(\frac{\omega^{\prime}}{\omega^{\prime}_1}\right)^{\mathpzc{S}_1},\\
\label{eq:accap1.3}
\lefteqn{\mbox{for } \omega^{\prime} > e^{\mathpzc{T}_2}: } \nonumber \\
& & \left\langle \left| \mathbf{w}_{\omega^{\prime}} \right| ^2\right\rangle = e^{\mathpzc{A}+\mathpzc{S}_1\mathpzc{T}_2-\mathpzc{S}_1\mathpzc{T}_1+\mathpzc{S}_2\mathpzc{T}_2}\omega^{\prime -\mathpzc{S}_2} = a_0 \left(\frac{\omega^{\prime}_2}{\omega^{\prime}_1}\right)^{\mathpzc{S}_1}\left(\frac{\omega^{\prime}}{\omega^{\prime}_2}\right)^{-\mathpzc{S}_2}
\end{eqnarray}
In Figure \ref{fig:modaccspec} we show example models for the acceleration spectra at $\theta$ = 10 degrees and $\theta$ = 60 degrees and their comparison to the original acceleration spectra (as in Figure \ref{fig:AccSpecAll}) for a particular choice of fitting rules.  

To first order (neglecting the second and third terms in Eq. (\ref{eq:powereq})), the resulting radiation spectrum is as follows:
\begin{eqnarray}
\label{eq:radap1.1}
\mbox{for } \omega/2\gamma^2 < e^{\mathpzc{T}_1}: & & \nonumber \\
 \left(\frac{dW}{d\omega}\right)_I &=& \frac{e^2}{2 \pi c^2} e^{\mathpzc{A}}\left[(2\gamma^2)-\omega\frac{\mathpzc{S}_1 e^{-\mathpzc{T}_1}}{\mathpzc{S}_1-1}+\omega \frac{(\mathpzc{S}_1+\mathpzc{S}_2)e^{-\mathpzc{T}_1\mathpzc{S}_1+\mathpzc{T}_2(\mathpzc{S}_1-1)}}{(\mathpzc{S}_1-1)(\mathpzc{S}_2+1)}\right] \nonumber \\
&=& \frac{a_0 e^2}{2\pi c^2}\left[2\gamma^2-\frac{\mathpzc{S}_1}{\mathpzc{S}_1-1}\left(\frac{\omega}{\omega_1^{\prime}}\right)+\frac{\mathpzc{S}_1+\mathpzc{S}_2}{(\mathpzc{S}_1-1)(\mathpzc{S}_2+1)}\left(\frac{\omega_2^{\prime}}{\omega_1^{\prime}}\right)^{\mathpzc{S}_1}\left(\frac{\omega}{\omega_2^{\prime}}\right)\right], \\
\label{eq:radap1.2}
\mbox{for } e^{\mathpzc{T}_1} < \omega/2\gamma^2 < e^{\mathpzc{T}_2}: & & \nonumber \\
\left(\frac{dW}{d\omega}\right)_{II} &=& \frac{e^2}{2 \pi c^2}e^{\mathpzc{A}-\mathpzc{T}_1\mathpzc{S}_1}\left[\frac{-\omega^{\mathpzc{S}_1}}{(\mathpzc{S}_1-1)(2\gamma^2)^{\mathpzc{S}_1-1}}+\omega\left(\frac{1}{\mathpzc{S}_1-1}+\frac{1}{\mathpzc{S}_2+1}\right)e^{\mathpzc{T}_2(\mathpzc{S}_1-1)}\right] \nonumber \\
&=& \frac{a_0 e^2}{2 \pi c^2}  \left[-\frac{(2\gamma^2)^{1-\mathpzc{S}_1}}{(\mathpzc{S}_1-1)}\left(\frac{\omega}{\omega_1^{\prime}}\right)^{\mathpzc{S}_1}+
\frac{\mathpzc{S}_1+\mathpzc{S}_2}{(\mathpzc{S}_1-1)(\mathpzc{S}_2+1)}
\left(\frac{\omega_2^{\prime}}{\omega_1^{\prime}}\right)^{\mathpzc{S}_1}\frac{\omega}{\omega_2^{\prime}}\right],  \\
\label{eq:radap1.3}
\mbox{for } \omega/2\gamma^2 > e^{\mathpzc{T}_2}: & & \nonumber \\
\left(\frac{dW}{d\omega}\right)_{III} &=& \frac{e^2}{2\pi c^2}e^{\mathpzc{A}-\mathpzc{T}_1\mathpzc{S}_1+\mathpzc{T}_2(\mathpzc{S}_1+\mathpzc{S}_2)}\frac{(2\gamma^2)^{1+\mathpzc{S}_2}}{\left|\mathpzc{S}_2+1\right|}\omega^{-\mathpzc{S}_2} \nonumber \\
&=& \frac{a_0 e^2}{2 \pi c^2} \frac{(2\gamma^2)^{1+\mathpzc{S}_2}}{\left|\mathpzc{S}_2+1\right|}\left(\frac{\omega_2^{\prime}}{\omega^{\prime}_1}\right)^{\mathpzc{S}_1}\left(\frac{\omega}{\omega_2^{\prime}}\right)^{-\mathpzc{S}_2} 
\end{eqnarray}

The radiative power spectrum (equations (\ref{eq:radap1.1}) - (\ref{eq:radap1.3})) obtained analytically by our fit-based approximation for the acceleration spectrum agrees with that obtained via full numerical integration in the following section within about 10\%.  We have chosen our fitting method for the acceleration spectrum to most closely capture the spectral indices; a different choice of fit may allow for a better determination of peak positions.  

Calculations of such spectra for parameters fitted to our acceleration spectra at $\theta$ = 10 degrees and $\theta$ = 60 degrees are shown in Figures \ref{fig:modradspec10} and \ref{fig:modradspec60}.  The region boundaries in these figures indicate that the radiation spectrum cannot be described by a simple linear approximation in the three regions that were defined by the breaks in our acceleration spectrum.  The key features (position of spectral peak and spectral breaks) of the radiation spectrum originate in the additional terms in Equations (\ref{eq:radap1.1}) and (\ref{eq:radap1.2}).  Consequently, the transition points in our radiation spectrum do not directly correspond to the transition points in the acceleration spectrum.  

The asymptotic behavior of the spectrum at high and low energies can be easily obtained from Eqs. (\ref{eq:radap1.1}) and (\ref{eq:radap1.3}):
\begin{eqnarray}
\label{eq:radap1..1}
\mbox{for } \omega/2\gamma^2\ < \omega_1^{\prime}: & & \nonumber \\
 \left(\frac{dW}{d\omega}\right)_I &\propto& \omega^0, \\
\label{eq:radap1..3}
\mbox{for } \omega/2\gamma^2 > \omega_2^{\prime}: & & \nonumber \\
\left(\frac{dW}{d\omega}\right)_{III}  &\propto& \omega^{-\mathpzc{S}_2} 
\end{eqnarray}
Thus the high and low-energy asymptotic behavior of the radiation spectrum will be identical to the high and low-energy behavior of the acceleration spectrum. 

The behavior of the spectrum in the intermediary Region II can be solved for as well.  We take the derivative of Equation (\ref{eq:radap1.2}) to solve for the position $P = \log(\omega_p/2\gamma^2)$ of the spectral peak, in cases where it exists:  
\begin{equation}
\label{eq:radpeak}
 P = \frac{1}{1-\mathpzc{S}_1}\log\left[\frac{\mathpzc{S}_1(\mathpzc{S}_2+1)}{\mathpzc{S}_1+\mathpzc{S}_2}\right]+\mathpzc{T}_2 \\
\end{equation}
We note that the peak position becomes undefined in the case $\mathpzc{S}_1<0, \left|\mathpzc{S}_1\right|>\mathpzc{S_2}$, for which Equation (\ref{eq:radap1.2}) is everywhere decreasing.  The case $\mathpzc{S}_1<0,\left|\mathpzc{S}_1\right|<\mathpzc{S_2}$, in which the acceleration spectrum would decline more steeply in Region II than in Region III is impermissible, so we find that the radiation spectrum will be unpeaked whenever the mid-range spectral index $\mathpzc{S}_1$ of the acceleration spectrum is negative (i.e., for $\mathpzc{S}_1<0$).  

An exploration of the behavior of peak point $P$ relative to the region boundaries shows that for certain values of $\theta$ the peak of the Region II function exists, but has crossed the boundary into Region I and consequently does not appear in the resulting radiation spectrum.  Thus, while a calculation of $P$ from the results of fitting the acceleration spectrum appears to indicate the re-emergence of a peak in the radiation spectrum as $\theta$ approaches $\pi/2$, this peak falls beyond the Region II lower boundary and is not observed.  We note that the first term in Equation (\ref{eq:radpeak}) is negative for both $0 < \mathpzc{S}_1 < 1$ and $\mathpzc{S}_1 > 1$, so the peak is always located below the transition point between Regions II and III.  

An analysis of the behavior of Equation(\ref{eq:radap1.2}) below the Region II peak indicates the following behavior:  
\begin{equation}
\label{eq:radap1..2}
\mbox{for } \omega_1^{\prime} < \omega/2\gamma^2\ < \omega_2^{\prime}: \\
\left(\frac{dW}{d\omega}\right)_{II} \propto
\cases{
\omega^1 & if $\mathpzc{S}_1>1$,  \cr 
\omega^{\mathpzc{S}_1} & if $\mathpzc{S}_1<1$, } \\
\end{equation}
 
Thus, from Equations (\ref{eq:radap1..1})-(\ref{eq:radap1..2}), it is evident that the spectral indices $\mathpzc{S}_1$ and $\mathpzc{S}_2$ of the acceleration spectra will generally correspond to spectral indices $s_1$ and $s_2$ in two power law regions of the radiation spectra.  In the case of the high-frequency spectral index $s_2$ this correspondence is exact; however, the relation between mid-range (i.e. intermediate-frequency) spectral indices $s_1$ and $\mathpzc{S}_1$ is modified an upper limit of unity on $s_1$ and also breaks down when the first term in equation \ref{eq:radpeak} is undefined or larger in magnitude than the distance between the acceleration's spectral transition points $\mathpzc{T}_2 - \mathpzc{T}_1$.  

The asymptotic form in Equation (\ref{eq:radap1..1}) suggests that we may neglect the second and third terms in Equation (\ref{eq:radap1.1}) and solve for the transition point $T = \log(\omega_t/2\gamma^2)$ at which the dominating term in Region II becomes significant.  We find that:
\begin{eqnarray}
\label{eq:radtranspt}
\mbox{for } \mathpzc{S}_1 < 1: & & \nonumber \\
 & \left(\frac{\omega_t}{2\gamma^2}\right)^{\mathpzc{S}_1} & = (1-\mathpzc{S}_1) \omega_1^{\prime \mathpzc{S}_1} \nonumber \\   
 & T & = \frac{1}{\mathpzc{S}_1}\log(1-\mathpzc{S}_1)+\mathpzc{T}_1, \\
\mbox{for } \mathpzc{S}_1 > 1: & & \nonumber \\
 & \left(\frac{\omega_t}{2\gamma^2}\right)^{\mathpzc{S}_1} & = \frac{(\mathpzc{S}_1-1)(\mathpzc{S}_2+1)}{\mathpzc{S}_1+\mathpzc{S}_2}\left(\frac{\omega_1^{\prime \mathpzc{S}_1}}{\omega_2^{\prime (\mathpzc{S}_1-1)}}\right) \nonumber \\
 & T & = \log\left[\frac{(\mathpzc{S}_1-1)(\mathpzc{S}_2+1)}{\mathpzc{S}_1+\mathpzc{S}_2}\right] + \mathpzc{S}_1\mathpzc{T}_1+(\mathpzc{S}_1-1)\mathpzc{T}_2 
\end{eqnarray}

We have demonstrated that a simple approximation for our acceleration spectrum allows us to analytically obtain some of the key features of the resulting jitter radiation spectrum, notably that it will have a similar three-region form with $dW/d{\omega} \propto \omega^0$ for small $\omega$ and $dW/d{\omega} \propto \omega^{-s_2}$ for large $\omega$ and a possibly-peaked transition region.  Unlike the acceleration spectrum, the intermediary region in the jitter radiation spectrum will have a maximum slope of 1 and may be unpeaked at angles $\theta$ at which the acceleration spectrum was peaked.  We emphasize that our spectral calculations are all for a single emitting electron, {\em not} a power-law distribution of electrons. Yet, the power-law photon spectrum emerges at high energies (above the second jitter spectral break $\omega_2$) in this jitter mechanism, in contrast to the synchrotron exponential spectral decay above the synchrotron frequency.

\section{Analysis of Radiation Spectra}
\label{s:radspec}

Now we turn to full numerical calculations of the jitter radiation spectrum, generated by successive numerical integrations of equations (\ref{eq:powereq}) and (\ref{eq:accxyfin}) or (\ref{eq:acczfin})  The results for varying $\theta$ are presented in Figure \ref{fig:RadSpecAll}, with data point intervals of 0.2 on the $\log \omega$ scale.  Once again we have normalized the spectra such that the low-energy asymptotic value of the $\theta = 2$ spectrum is unity.  The detailed view of the peak region in Figure \ref{fig:RadSpecPeakDetail} shows results in this region for intervals of 0.05 in $\log(\omega)$.  The spectral shapes and trends are, as expected, much like the acceleration spectra but broadened and flattened overall.  No peak reemerges in the spectra as $\theta$ approaches $\pi$/2.  Figure \ref{fig:vFvRadSpecAll} shows the $\nu F_{\nu}$ spectrum such as is commonly presented for GRBs and used in GRB spectral analysis.   

These jitter radiation results show a significant evolution in the spectrum emitted at different viewing angles relative to the main filamentation axis of the magnetic field spectrum.  Note that this viewing angle effect in our calculations is entirely due to particles with velocities directed along the line of sight providing the dominant contribution to the radiation emitted to any particular viewing angle.  We are neglecting the angular distribution of the radiation emitted by each particle and using the angle-averaged emissivity for the spectrum emitted in the forward direction by particles with a particular orientation angle relative to the magnetic field filamentation.  

Like the acceleration spectrum, the radiation spectrum can be generally described in terms of three regions (two spectral breaks), and an amplitude or slope in each.  To conveniently summarize the spectral features and their evolution, we have developed a five-parameter fit which describes the spectral behavior in these three regions, which we designate as Regions R-I, R-II, and R-III to avoid confusion with the Regions I, II, and III as defined earlier for the acceleration spectrum. (Recall that the transition points for the acceleration spectra do not correspond to the apparent transition points in the radiation spectra.)  As before, we have chosen our technique to optimize our results for the spectral indices, rather than the spectral transition points.  

Since for jitter radiation the soft spectral index varies continuously and approaches 0 for low $\omega$, the results of a simple two-region fit to these spectra would depend significantly upon where the lower bound of the data window falls relative to the peak.  Within a fixed data window, a two-region spectral fit would tend to produce an artificial reduction in the soft spectral index for spectra with higher-frequency spectral peaks or breaks.  Unfortunately there is no simple way to characterize the behavior of the middle range of the spectrum because of the transition from peaked to unpeaked spectra as $\theta$ varies.  Even in unpeaked spectra, the extent and curve of the transition region between the flat low $\omega^{\prime}$ part of the spectrum and the strongly negatively-sloped high $\omega^{\prime}$ part of the spectrum varies substantially.  Consequently, we have chosen to model the unpeaked spectra still as a three-region spectra rather than solely by its upper and lower asymptotes. 
We characterize the spectra by defining three lines (each requiring a slope and a reference point) and finding the transition points at which they intersect.  

\emph{Region R-I (flat, amplitude $A$)}: 
The low-frequency region R-I is flat, with a slope close to zero.  To describe this region, we take our initial calculated amplitude to be $A$, the low-frequency amplitude.  For very small $\theta$ our lower calculation boundary may be insufficient to capture the initial flat part of the spectrum, since our first spectral transition point approaches -$\infty$ as $\theta$ goes to 0.  

\emph{Region R-II (positive or negative slope $s_1$)}: 
The intermediate-frequency region R-II may have positive slope resulting in a peak or a slight negative slope (of notably less magnitude than the slope in region R-III).  Since not all the spectra are peaked, we have chosen to avoid using the peak value for our fit.  Instead we define the ``drop point" as the region where the second derivative reaches its minimum value.  As the place of largest negative change in slope, this coincides well with the ``knee" or second break of the function, and is always at slightly higher frequencies than the peak itself.  We then find the slope and a reference point in this region by either:

\begin{itemize}
	\item Method a: for peaked spectra, we take the maximum value of the numerical derivative and its associated data point.

	\item Method b: for unpeaked spectra, we take the average value of the numerical derivative in the region between the drop point and the 	``deviation point" where the spectrum first drops below $A-0.01$ (this corresponds to a deviation of about 1\% from its original value).  Our reference point is the data point halfway between or next highest to halfway between the deviation and drop points.  
\end{itemize}

\emph{Region R-III (negative slope, defined as -$s_2$)}:
The high-energy frequency region R-III has a large negative slope compared to the rest of the function.  This slope is still changing over the region close to the second spectral break that we are considering, so we determine a representative slope by calculating the slope of a line between the drop point, which is the minimum of the numerical second derivative, and the higher frequency point that is the absolute minimum of the numerical second derivative (the data point closest to where the second derivative crosses zero in this region).  These points are well-defined for all our radiation spectra as long as the calculation boundary extends a couple orders of magnitude in $e$ above the drop point.  Either point may be used as a reference point in this region.  

The first spectral transition point $\tau_1$ is obtained by solving for the intersection of the lines defined in Regions R-I and R-II, and the second spectral transition point $\tau_2$ is obtained by solving for the intersection of the lines in Regions R-II and R-III.  

We have chosen to work with the $F_{\nu}$ spectrum because of the convenience of its distinctive flat (spectral index of 0) initial amplitude at very low frequencies, but it is easy enough to translate $F_{\nu}$ spectral features into spectral features of the $\nu F_{\nu}$ spectrum or the photon spectrum $N(E)$, as the spectral indices will simply be increased or decreased by 1 and the transition points between the power law regions will roughly coincide, with a slight shift based on normalization.  In terms of the Band function fit commonly used for GRB spectra \citep{band}, the relation between the high-energy spectral indices is $\beta_{Band} = s_2-1$.  The relation between the low-energy spectral indices is complicated by the fact that the Band function is a two-region fit and not sensitive to multiple spectral indices below the spectral peak; consequently $\alpha_{Band}$ will range between -1 and $s_1-1$ depending on where the data fitting window falls relative to our first spectral break $\tau_1$.  The $\nu F_{\nu}$ peak energy, which is $E_p$ in the Band function, will correspond to a slope of -1 in the $F_{\nu}$ spectrum, and will lie roughly in the vicinity of the second spectral break $\tau_2$ in our fit. 

Figures \ref{fig:fullfitamp} - \ref{fig:fullfits2} show spectral fit results obtained using our above technique on the radiation spectra.  We have also applied the same technique to the acceleration spectra presented in section \ref{s:accspec} and plotted them for comparison.  In addition to data resolution effects, some discontinuity in fitting the peaked vs. unpeaked form of the spectrum is unavoidable and is reflected in our results. 

Figures \ref{fig:fullfitamp} and \ref{fig:fullfits2} indicate that the amplitude $A$ and the high-frequency spectral index $s_2$ are close in both their values and their evolution with $\theta$ for the two types of spectra.  The mid-range spectral index $s_1$ varies similarly in Fig. \ref{fig:fullfits1} for both spectra, but appears to approach different asymptotic values as it approaches $\theta = 0$ and $\theta = \pi/2$, clearly showing the expected $s_1<1$ limiting behavior.  The second spectral break $\tau_2$ shows that the radiation transition point tends to be about 1/2 a power of $e$ lower than the second spectral behavior for the acceleration case, but shows similar evolution with $\theta$ in both cases.  

Figure \ref{fig:fullfitdroppk} shows the angular dependence of the spectral peak in both our acceleration and radiation $F_{\nu}$ spectra, the $\nu F_{\nu}$ spectral peak (peak data point in Figure \ref{fig:vFvRadSpecAll}, and the drop point, which we have defined as the minimum in the numerical second derivative.  We clearly see the usefulness of the drop point in tracking the spectral behavior across the full range of $\theta$, and that it closely tracks the behavior of the $\nu F_{\nu}$ spectral peak ($E_p$).  Both Figures \ref{fig:fullfitdroppk} and \ref{fig:fullfitpkheight} clearly show the re-emergence of peaked acceleration spectra as $\theta$ approaches $\pi/2$ and the lack of peaked radiation spectra for similar values of $\theta$.

\section{Spectral Features and Exploration of the Spectral Parameter Space}
\label{s:sfvarparam}

We have explored the influence of changes in the magnetic field spectral parameters on the acceleration experienced by the particle and hence its resulting radiative profile.  Sections \ref{s:accspec} and \ref{s:radspec} presented the acceleration and radiation spectra calculated from magnetic field spectra of the form given in Equations (\ref{eq:perpfspec}) and (\ref{eq:pllfspec}) with our original choice of parameters $\alpha$ = $\alpha_{\perp}$ = $\alpha_{\parallel}$ = 2.0, $\beta$ = $\beta_{\perp}$ = $\beta_{\parallel}$ = 1.5, $\kappa$ = $\kappa_{\perp}$ = $\kappa_{\parallel}$ = 10.  In this section, we present the results of varying these parameters.  We vary the joint parameters $\alpha$, $\beta$, and $\kappa$ in the parallel and perpendicular magnetic field spectra.  We also vary the parameters $\alpha_{\parallel}$, $\alpha_{\perp}$, $\beta_{\parallel}$, and $\beta_{\perp}$ individually.  Finally, we vary the ratio $K = \kappa_{\perp}$/$\kappa_{\parallel}$.  For each variation of the initial parameters, we have calculated the radiation spectrum for three representative angles at $\theta$ = $10^o$, $45^o$, and $80^o$.  The results are presented according to their impact on the characteristics of the radiation spectrum as developed in section \ref{s:radspec}, namely the initial amplitude $A$, the spectral breaks $\tau_1$ and $\tau_2$, and the spectral indices $s_1$ and $s_2$.  We also present the results for the peak ''strength'', the height of the spectral peak above the initial amplitude.  

The spectra are divided by $\langle B^2\rangle \propto \int{f(\kappa_{\parallel})f_(\kappa_{\perp})d^3k} \propto k_0^{-2(\beta_{\parallel}+\beta_{\perp})}$ to appropriately normalize the amplitudes relative to one another, but in all cases we have arbitrarily normalized the final spectra such that the low-energy asymptotic value of the $\theta = 10$ spectrum with our original choice of parameters is unity (zero on the logarithmic scale).  

As in Section \ref{s:radspec}, the initial, low-frequency amplitude A is the first calculated value of the angle-averaged radiative power emitted per frequency $dW/d\omega$.  This value is generally a good approximation for the asymptotic value of the function as it approaches lower $\omega$, though it may deviate somewhat from this value for $\theta$ approaching 0, as the spectra becomes sloped rather than flattened at our lower calculation boundary in $\omega$.  

Among the resulting figures \ref{fig:alphaamp}-\ref{fig:ampcomp}, variations in the magnetic field parameters $\kappa_i$ produce the largest effect on the low-frequency amplitude, causing changes of about 4 orders of magnitude in $e$ when varied individually via the ratio $K = \kappa_{\perp}/\kappa_{\parallel}$, and up to 7 orders of magnitude in $e$ when varied together as $\kappa = \kappa_{\parallel} = \kappa_{\perp}$.  Variation with changes in the magnetic field spectral indices $\alpha_i$ and $\beta_i$ are small in comparison, on the scale of about 1-2 orders of magnitude.  The amplitude increases with increasing $K$ for $\theta = 10^o$ and generally decreases with increasing $K$ for $\theta=80^o$; thus, it increases when $\kappa_{\perp}$ dominates at small $\theta$ and when $\kappa_{\parallel}$ dominates at large $\theta$.  

The mid-range spectral index in peaked spectra is the maximum slope below the peak, as determined by taking the numerical first derivative of our calculated values.  For unpeaked spectra, we find the spectral index as the average slope between the point at which the graph deviates by more than 0.01 from the initial amplitude A and the drop point at which the numerical second derivative reaches a minimum.  

Figures \ref{fig:alphas1}-\ref{fig:kappas1} present the effect of magnetic field parameter variations on $s_1$.  As can be seen in Figure \ref{fig:s1comp}, the mid-range spectral index is strongly affected by variations in both the parameters $\beta_i$, especially for $\theta = 10^o$, and in the relative strength of the $\kappa_i$ values.  The peaked $\theta = 10^o$ spectra is notably more sensitive to magnetic field variations than the unpeaked $45^o$ and $80^o$ spectra. The ratio $K = \kappa_{\perp}/\kappa_{\parallel}$ has the largest influence at all three representative viewing angles, with $s_1$ increasing as $\kappa_{\parallel}>\kappa_{\perp}$).  We note that even at $\theta$ approaching $\pi/2$, we obtain a positive slope (and hence a peaked spectrum) for $K = 1/10$. 

For both peaked and unpeaked spectra, the high-frequency spectral index -$s_2$ is determined by taking the slope between the drop point and the absolute minimum of the second derivative above the peak (i.e., the closest data point to where the second derivative crosses zero).  Figures \ref{fig:alphas2} - \ref{fig:kappas2} show the effects of variations in the magnetic field parameters on -$s_2$.  We find that as expected analytically, this spectral index is primarily influenced by the magnetic field parameters $\beta_i$.  In particular, $s_2$ is most strongly influenced by $\beta_{\perp}$, the high-wavenumber spectral index of the magnetic field spectrum transverse to the current filamentation.  As seen in figure \ref{fig:betas2}, $\beta_{\parallel}$ affects $s_2$ only at small angles $\theta$, and its influence even then is less than that of varying $\beta_{\perp}$.  The apparently strong influence of $\kappa$ is largely an artificial effect as the $\kappa$ parameter's strong shifting of the function (as indicated in our analysis of the spectral breaks below) towards higher frequencies interferes with the calculation of $s_2$ by shifting the absolute minumum of the second derivative outside our calculation boundaries.  This causes an artificial reduction in the steepness of the slope for $\kappa = 100$, as evident also in Figure \ref{fig:kappas2}.  

We calculate the first spectral break (i.e. transition point) as the intersection between the line $\log \left\langle \left| \mathbf{w}_{\omega^{\prime}} \right| ^2\right\rangle$ = A and the line of slope $s_1$ through the point of maximum positive slope for peaked spectra or through the data point in the middle of the range over which we averaged to obtain slope $s_1$ for unpeaked spectra.  (If the middle of the range does not fall on a data point, we take the next larger data point.)  We find, as shown in figures \ref{fig:alphatp1}-\ref{fig:tp1comp}, that the first spectral break is strongly influenced by changes in $\alpha$ and $\kappa$ or the $\kappa$-ratio $K$.  The break position shifts to higher frequency by about an order of magnitude in $e$ as we increase $\alpha$ from 1 to 10 jointly in the parallel and perpendicular magnetic field spectra.  In varying the $\alpha_i$ separately we see that $\alpha_{\parallel}$ has a larger influence at $\theta = 10^o$ and $\alpha_{\perp}$ has a larger influence at $\theta = 80^o$.  Increasing the $\kappa$ jointly by powers of 10 results in shifting the first spectral break to higher frequencies by roughly 4 orders of magnitude in e.  Varying the $\kappa$ parameters relative to one another results in a similarly strong shift, towards higher frequencies for $\kappa_{\perp} > \kappa_{\parallel}$.   

The second spectral break $\tau_2$ is calculated as the intersection between a line of slope $s_1$ through the point of maximum positive slope (for peaked spectra) or through the mid-point of the averaging region (for unpeaked spectra), and the line of slope $-s_2$ through the ''drop point" at which the second derivative reaches a minimum (i.e., the largest negative change in the slope).  Our results (in figures \ref{fig:alphatp2}-\ref{fig:tp2comp}) indicate that the second transition point is most strongly influenced by the $\kappa_i$ varied jointly or via the ratio $K$.  The low-wavenumber magnetic field spectral index also demonstrates a fairly strong influence, with larger $\alpha$ shifting $\tau_2$ to higher frequencies.  A comparison of the influence of $\alpha$ on the two spectral break points (as seen in Figures \ref{fig:alphatp1} and \ref{fig:alphatp2}) indicates a very similar shift in both break points; thus increasing $\alpha$ shifts the entire spectrum towards higher frequencies.  

Figures \ref{fig:alphapkheight}-\ref{fig:kappapkheight} show the variation in the peak strength (which we have defined as the height of the spectral peak above the low-frequency amplitude $A$ with changes in the magnetic field parameters.  For $\theta=10^o$, the only of our three representative angles that has peaked spectrum for $K=1$, peak strength increases with increasing $\alpha$ and $\beta$.  Individually, increasing $\beta_{\perp}$ has the largest effect in increasing the peak strength, while increasing $\alpha_{\perp}$ lowers it.  Similarly, increasing $\alpha_{\parallel}$ increases the peak strength while increasing $\beta_{\parallel}$ lowers it.  The largest effect overall is produced by variation of the ratio $K$ between the perpendicular and parallel field parameters $\kappa_i$.  For $\kappa_{\parallel} > \kappa_{\perp}$ (i.e. $K<1$), the peak strength appears to persist to higher angles $\theta$, while for $\kappa_{\perp}>\kappa_{\parallel}$ the peak can be small or non-existent even at $\theta = 10^o$.  Thus the ratio between $\kappa_{\perp}$ and $\kappa_{\parallel}$, the respective peaks of the magnetic field perpendicular and parallel spectra, strongly influences the progression of the spectral evolution between its $\theta$ = 0 and $\theta$ = $\pi/2$ limiting values, as expected from our earlier analysis in Section \ref{s:accspec}.  

We have seen that relatively minor changes in the magnetic field spectra can produce very significant effects upon the jitter radiation spectra, particularly in the appearance of the spectral peak or break region.  Furthermore, while we have included in this section only spectra from a few representative viewing angles $\theta$, the angular dependence demonstrated suggests that the connection of such features to the transverse or parallel magnetic field spectra can be tested by observing their variation with viewing angle.

\section{Conclusions}

	We have calculated the angle-averaged power spectra of jitter radiation emitted by {\em a single relativistic electron} undergoing small Lorentz-force accelerations transverse to its overall velocity. Note that the obtained spectra are equivalent to the ensemble-averaged spectra per one electron from a collection of monoenergetic relativistic electrons. The radiation spectra are calculated using a smoothly connected broken power-law model of a magnetic field mimicking the structure of magnetic fields generated by the Weibel instability.  The shapes of the resulting jitter radiation spectra are shown to depend on the magnetic field spatial spectrum and to vary with the angle $\theta$ of the electron velocity (being also the line of sight) with respect to the direction of the field anisotropy ($z$-axis). The effect of varying parameters in the magnetic field spectra has been explored and indicates that the jitter radiation spectral features, such as the strength of the spectral peak or the extent of a sloped transition region, are quite sensitive to the parameters controlling the magnetic field spectra. 
	
	Despite the high sensitivity of the jitter radiation spectra to the magnetic field spatial spectrum or, in general, the field correlation tensor, $K_{ij}({\bf k})=B_{\bf k}^i B_{\bf k}^{*j}$ \citep{M06}, one can draw some fairly robust conclusions. When the parallel and perpendicular magnetic field spectra are similar, one has just four essential parameters, their low-$k$ and high-$k$ spectral slopes, $\alpha>1/2$ and $\beta>0$, the peak representing a typical correlation length $\kappa$, and the viewing angle of the line of sight with respect to the magnetic filament direction, $\theta$. The power (i.e., $F_\nu$) spectrum produced by monoenergetic electrons moving towards the observer with the Lorentz factor $\gamma$, in general, has three power-law segments: a flat low-energy part, an intermediate-energy region which rises or slightly falls with a slope of less than unity (the exact value depending on $\theta$), and the more steeply falling off part with the slope being between -2$\beta$ and -2$\beta+1$, again, depending on $\theta$.  
	
  The shape of the spectrum changes significantly with the angle $\theta$ between the radiating particle's velocity and the axis of the current filamentation generated by the counterstreaming Weibel instability.  As $\theta \rightarrow 0$, the low-frequency spectral break $\tau_1$ approaches $-\infty$ and the maximum spectral slope (mid-range spectral index $s_1$ approaches the value of 1 (the trend of our results agreeing well with the $\theta = 0$ case in \citet{M06}).  As $\theta$ increases, the spectral peak weakens as $s_1$ decreases and $\tau_1$ shifts towards the peak region.  The disappearance of the spectral peak at some particular $\theta$ appears to be a result of both these spectral changes, and there is an extended transition region between the low-energy and high-energy power law trends.  Consequently, we find that both the peaked and unpeaked spectra are well described by a three-region fit.  Two-region fits are likely to miss out on the variation in the spectral slope below the peak at small $\theta$; consequently the resulting low-energy spectral index will be extremely sensitive to where the peak falls relative to the lower bound of a measured spectral window.  This will be true even if the low-energy spectral index is taken at a common energy, as in the ``effective'' low-energy spectral index $\alpha_{eff}$ commonly taken as the tangential slope of the logarithmic spectrum at 25 keV.  

In comparing the radiation spectra for the full range of $\theta$, we have found that the ``drop point'', which we determined as the minimum of the numerical second derivative of the logarithmic data (i.e. the largest negative change in the spectral slope) serves as a good common reference point for both peaked and unpeaked spectra; in addition, the ``drop point'' in the radiation power spectrum evolves with the angle $\theta$ much like the $\nu F_{\nu}$ spectral peak energy $E_p$.  
  
In section \ref{s:acctorad}, we developed in detail the relation between the radiation spectrum and the underlying Fourier spectrum of the particle's acceleration.  In particular, we find that the radiation spectrum has much the same shape as the acceleration spectrum but that the apparent transition points in the two spectra do not simply coincide for most angles of $\theta$.  Furthermore, although the acceleration spectrum sees the re-emergence of a spectral peak for $\theta \rightarrow \pi/2$, the radiation spectrum does not.  We have also demonstrated that a simple fit to the acceleration spectrum allows for the generation of a model radiation spectrum which approximates the realistic one with 10\% accuracy. 

We have found that variations in the magnetic field spectral parameters influence the final radiation spectrum by controlling the width and peak-positions of the functions within the integrand and the extent to which this directly modifies the effect of the offset, which is proportional to $\omega^{\prime}$.  If we consider the general progression of the radiation spectra from being strongly peaked at small $\theta$ to unpeaked at $\theta$ near $\pi$/2, the trends shown here indicate that the speed of the progression of the spectral shape between the two extremes is dependent on the relative strengths of the parameters in the magnetic field spectra transverse and parallel to the shock front.  

We have confirmed that the jitter radiation high-energy spectral index is determined primarily by the high-$k$ magnetic field spectral index $\beta$, which otherwise has little influence on the spectrum.  The low-$k$ magnetic field spectral index $\alpha$ is shown to have a significant influence on the low-energy and mid-range portions of the radiation spectrum when varied jointly in both magnetic field spectrum, although this influence is substantially reduced when only one $\alpha_i$ is varied.  

The parameters $\kappa_\perp$ and $\kappa_\parallel$ represent the dimensionless correlation lengths of the magnetic field distribution in the direction along the Weibel current filaments (and the direction of shock propagation in the case of a GRB) and in the perpendicular plane (parallel to the shock plane for a GRB).  We find that increasing $\kappa_{\perp}$ and $\kappa_{\parallel}$ jointly shifts the entire spectrum to higher energies with relatively little effect on the spectral shape.  Thus, as expected, the location of the spectral peak and break energies (and the corresponding peak energy $E_p$ of the $\nu F_{\nu}$ spectrum) are determined primarily by the correlation length of the magnetic field turbulence.  The progression of the spectral shape between the head-on and edge-on cases is sensitive to the variation of the $\kappa$ parameter in one function relative to the other, such that for a particular viewing angle $\theta$ either peaked or unpeaked spectra can be attained via modification of the $\kappa$ ratio $K$.  In the extreme that $\kappa_\parallel$ is 2 orders of magnitude larger than $\kappa_\perp$ we recover a peaked spectra for the angles as high as $\theta\sim80$ degrees.   It is also notable that the spectral peak and transition points undergo relatively little horizontal shift as $K$ varies when $\theta$ = 10 degrees, but shift quite dramatically (3-4 orders of magnitude) during this variation for $\theta$ = 80 degrees.  
	
We shall summarize the most notable properties of our jitter radiation results with particular significance to interpreting radiation spectra from astrophysical sources.    
	First, the jitter radiation spectra are significantly harder than synchrotron spectra in the region just below the spectral peak.  This may be a significant mechanism in astrophysical sources such as gamma-ray bursts where a substantial population is seen to violate the synchrotron limit.
	Second, both the maximum slope in the region below the peak and the extent of this sloped region (between the low-frequency spectral break and the peak) are strongly influenced by $\theta$, with the largest slope and largest extent of the sloped region at small $\theta$.  The angle $\theta_{np}$ at which the peak disappears is determined primarily by the ratio between the magnetic field correlation lengths perpendicular to and along the filamentation axis.     
	Third, the position of the spectral peak represents the characteristic correlation length of the magnetic field, which in case of the Weibel turbulence depends on the density as $n^{1/2}$. This is at odds with the synchrotron radiation in which the spectral peak measures the magnetic field strength. This result is important for accurate interpretation of the observed spectra as well.
	Fourth, the high-energy part of the spectrum is represented by a power-law, even though the electrons are monoenergetic.  Thus, no power-law distribution of Fermi-accelerated electrons is required to produce the observed power-law spectra in prompt GRBs. This has important implication for the interpretation of the observed data, provided the electrons are radiating in the jitter regime. 
	Fifth, the angular dependence exhibited, combined with relativistic kinematics of a curved shock front can explain certain puzzling features of the GRB prompt spectral variability \citep{MPR09}. The sensitivity of the jitter spectra to the magnetic field anisotropy makes it a possible tool for diagnostics in sites of small-scale magnetic field turbulence and also a basis for analysis of astrophysical sources where the magnetic field orientations relative to the direction of observation are changing over time.   
	
	Although we often appeal to GRBs as sites where jitter radiation can likely be produced, we cannot exclude other astrophysical objects, e.g. jets in active galactic nuclei, early supernova shocks and other violent sources, provided that small-scale magnetic fields may be produced and maintained in them. At last but not least, one can use jitter radiation as diagnostic tool in laser-plasma interaction experiments, e.g., Hercules \citep{GRB+Hercules08,HerculesKU07}, aimed at studies of Weibel turbulence and conditions in GRBs within the Laboratory Astrophysics and High-Energy-Density Physics programs.

\acknowledgements 

This work has been supported by NASA, NSF, DOE via grants NNX07AJ50G, NNX08AL39G, AST-0708213, DE-FG02-04ER54790, DE-FG02-07ER54940.

\renewcommand{\thefigure}{\arabic{figure}(a)}
\begin{figure}
	\plotone{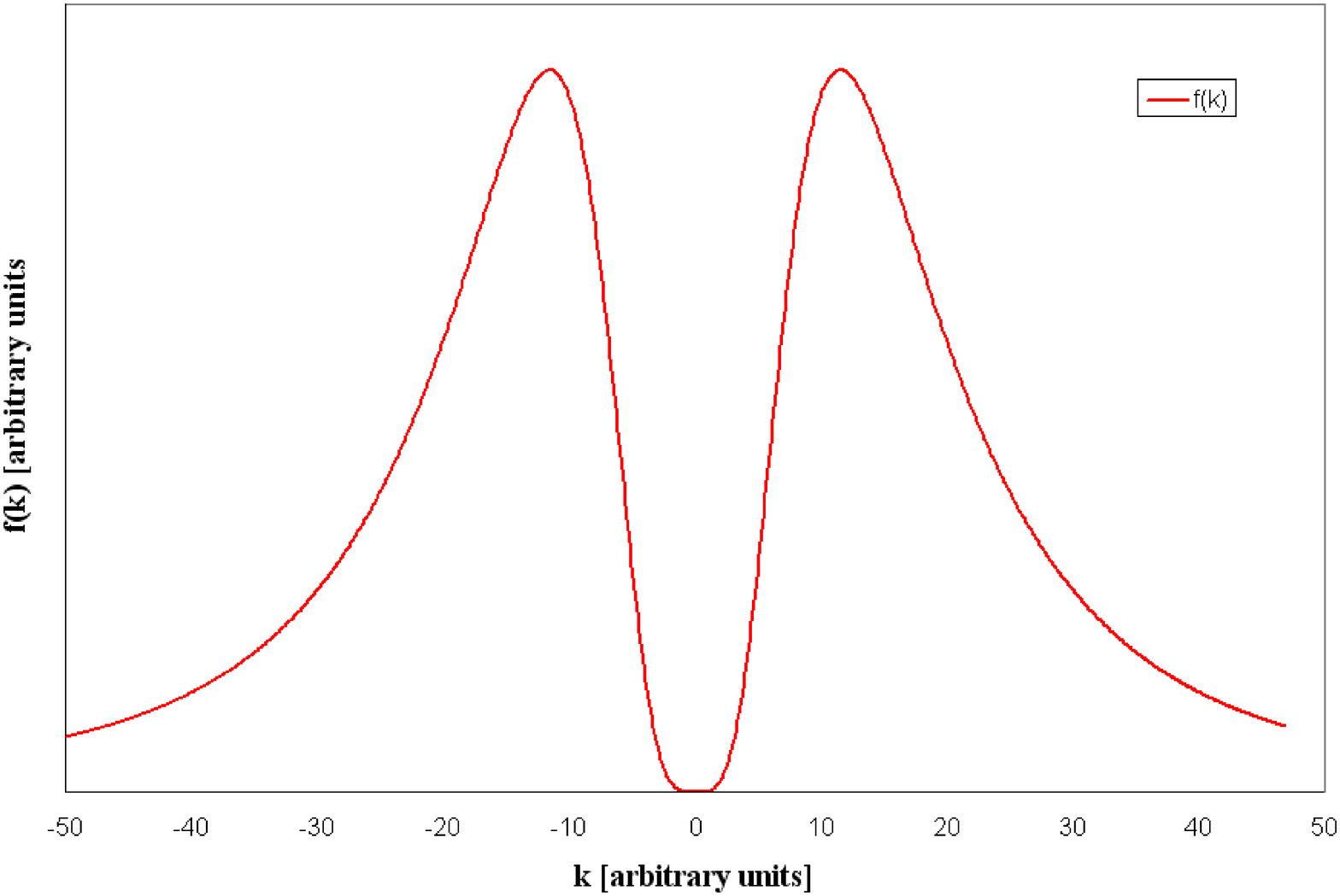}
	\caption{A plot of the general form of the spectrum $f(k)$ used in our paper for the magnetic field spectrum in the transverse and parallel directions.}
	\label{fig:basicspectra}
\end{figure}
%

\renewcommand{\thefigure}{\arabic{figure}(a)}
\begin{figure}
	\plotone{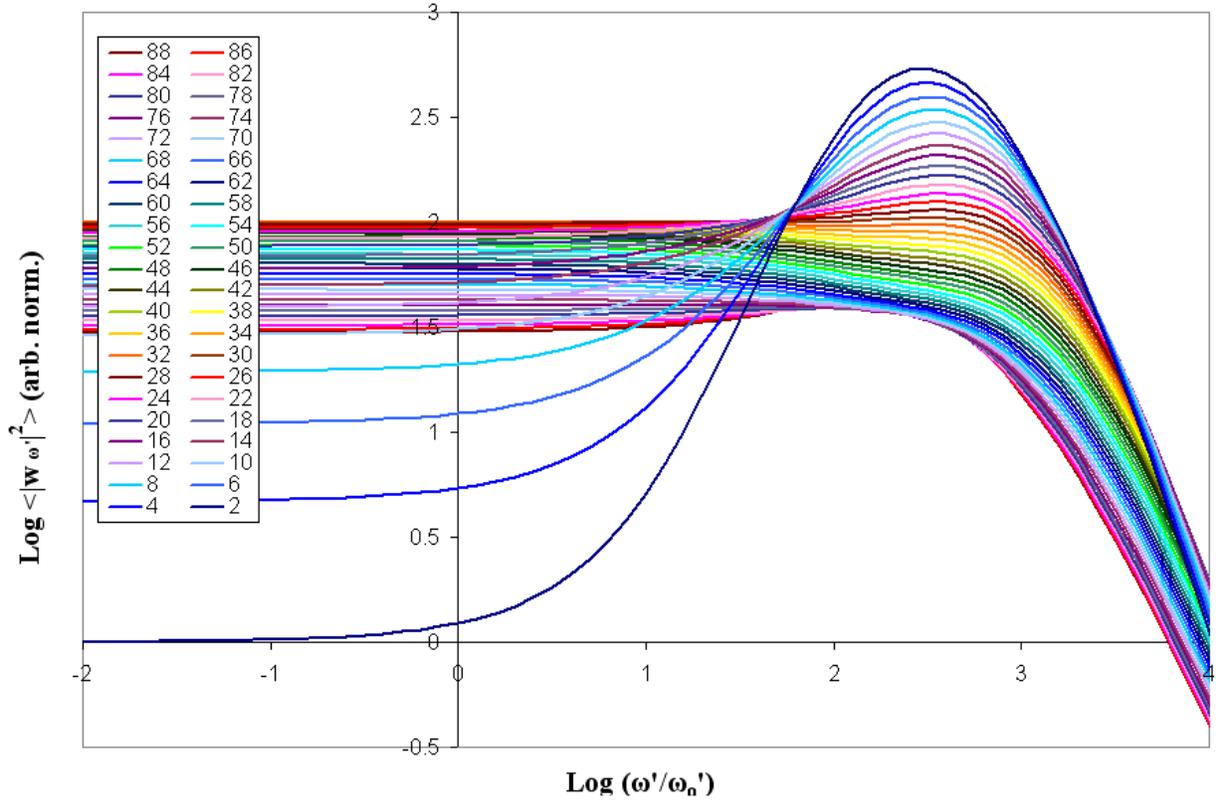}
	\caption{The acceleration spectra $\left\langle \left| \mathbf{w}_{\omega^{\prime}} \right| ^2\right\rangle$ for $\theta$ ranging from 2 through 88 degrees. The spectra is numerically calculated for a step size of 0.1 in $\log(\omega^{\prime})$ and is normalized via division by the 3-dimensional integral over the magnetic field spectra.
	\label{fig:AccSpecAll}}
\end{figure} 
\addtocounter{figure}{-1}
\renewcommand{\thefigure}{\arabic{figure}(b)}
\begin{figure}
	\plotone{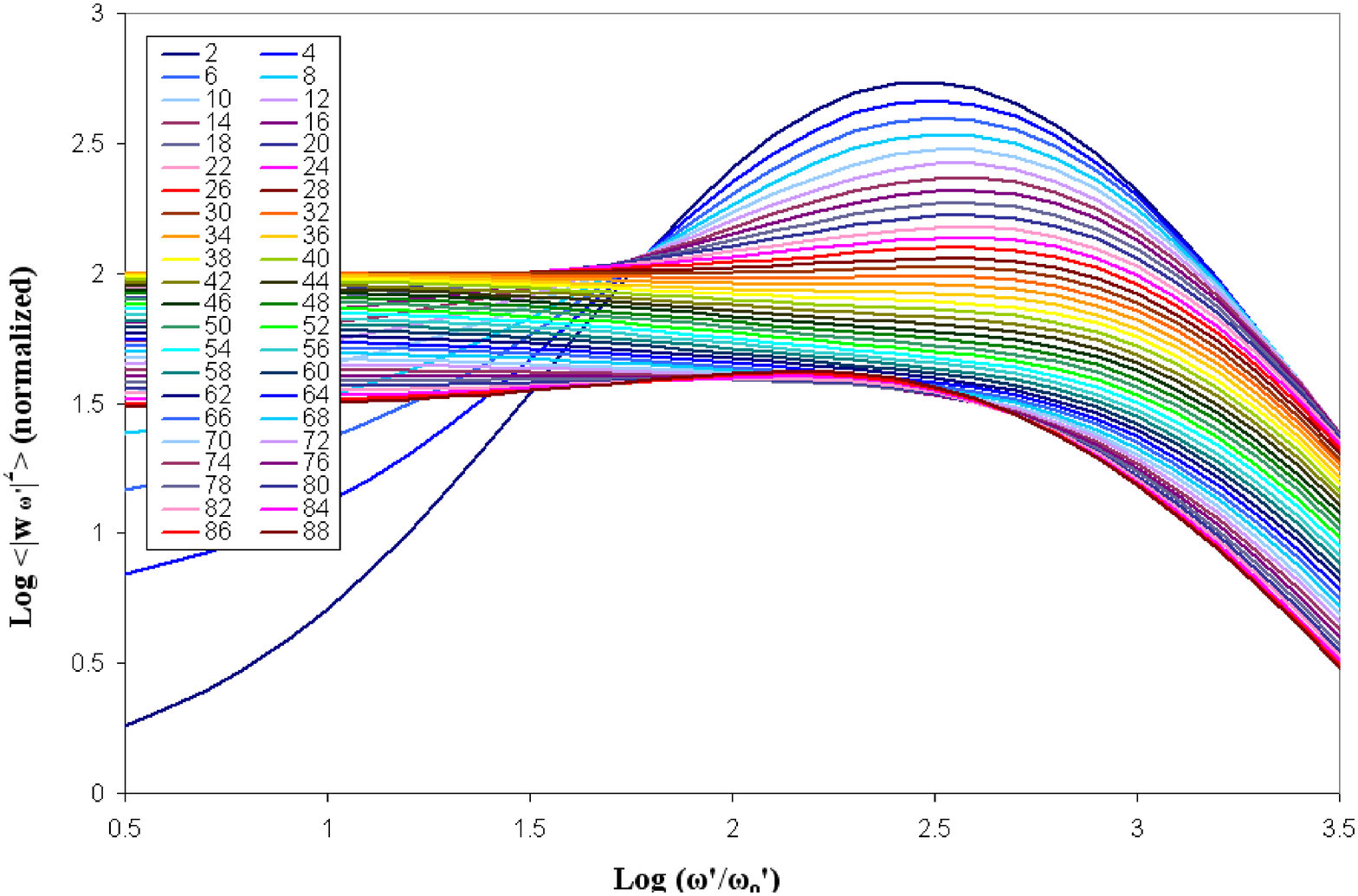}
	\caption{View showing detail of peak region of acceleration spectra $\left\langle \left| \mathbf{w}_{\omega^{\prime}} \right| ^2\right\rangle$ for $\theta$ shown in Figure \ref{fig:AccSpecAll}.  One can see the disappearance of the peak and flattening of the spectra for mid-range $\theta$, followed by its reappearance at slightly lower $\log(\omega^{\prime})$ at $\theta$ of about 76 degrees.} 
	\label{fig:AccSpecPeakDetail}
\end{figure}

\renewcommand{\thefigure}{\arabic{figure}}
\begin{figure}
	\plotone{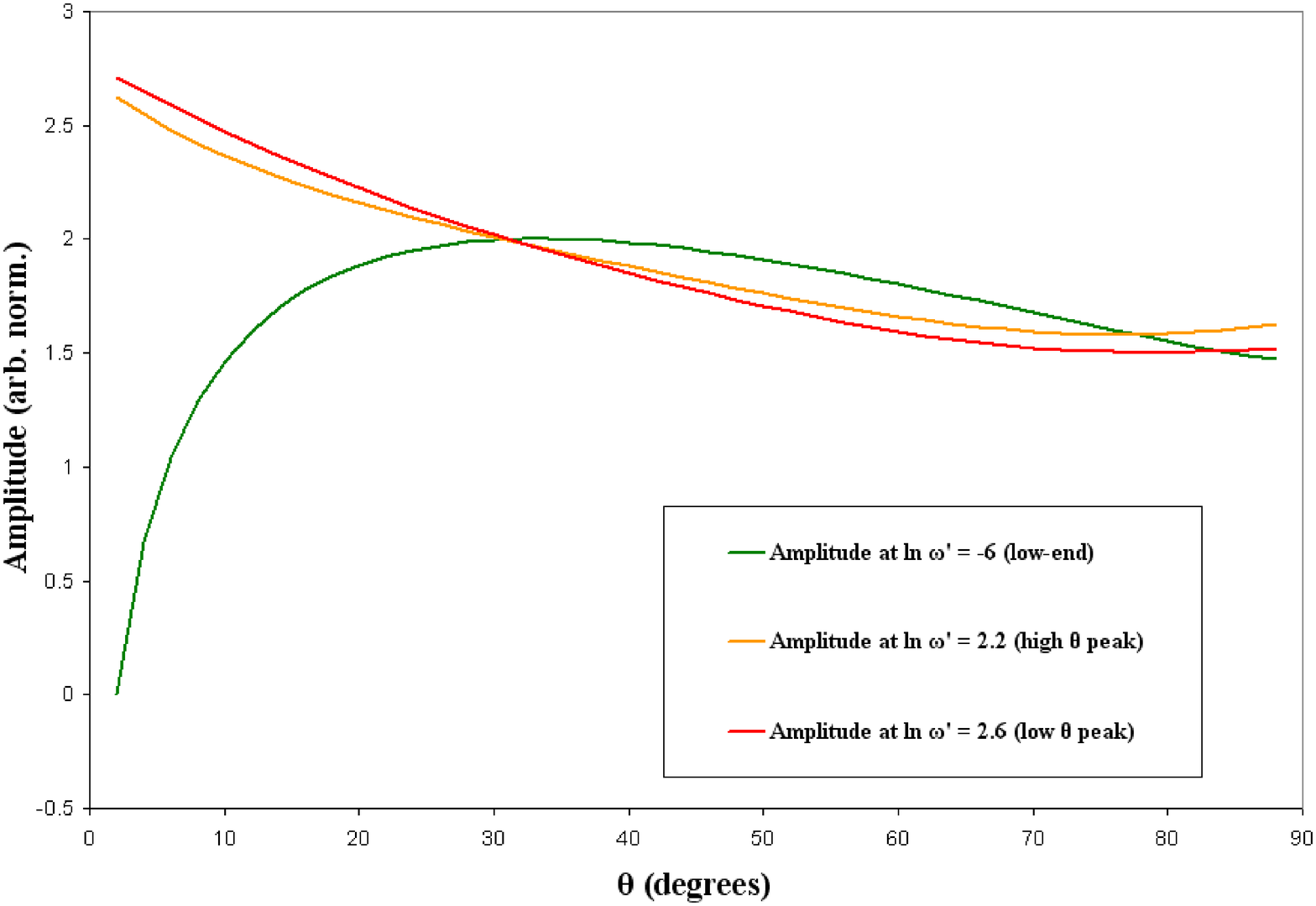}
	\caption{Plot of the spectral amplitudes taken at the low-$\omega^{\prime}$ end of the calculated spectra, and at the approximate locations of the spectral peak for low-$\theta$ and high-$\theta$.  The dominance of one amplitude over the other illustrates the transition of the spectra from peaked to unpeaked as $\theta$ progresses from 0 to $\pi$/2.}  
	\label{fig:AccSpecAmps}
\end{figure}
 
\begin{figure}
	\plotone{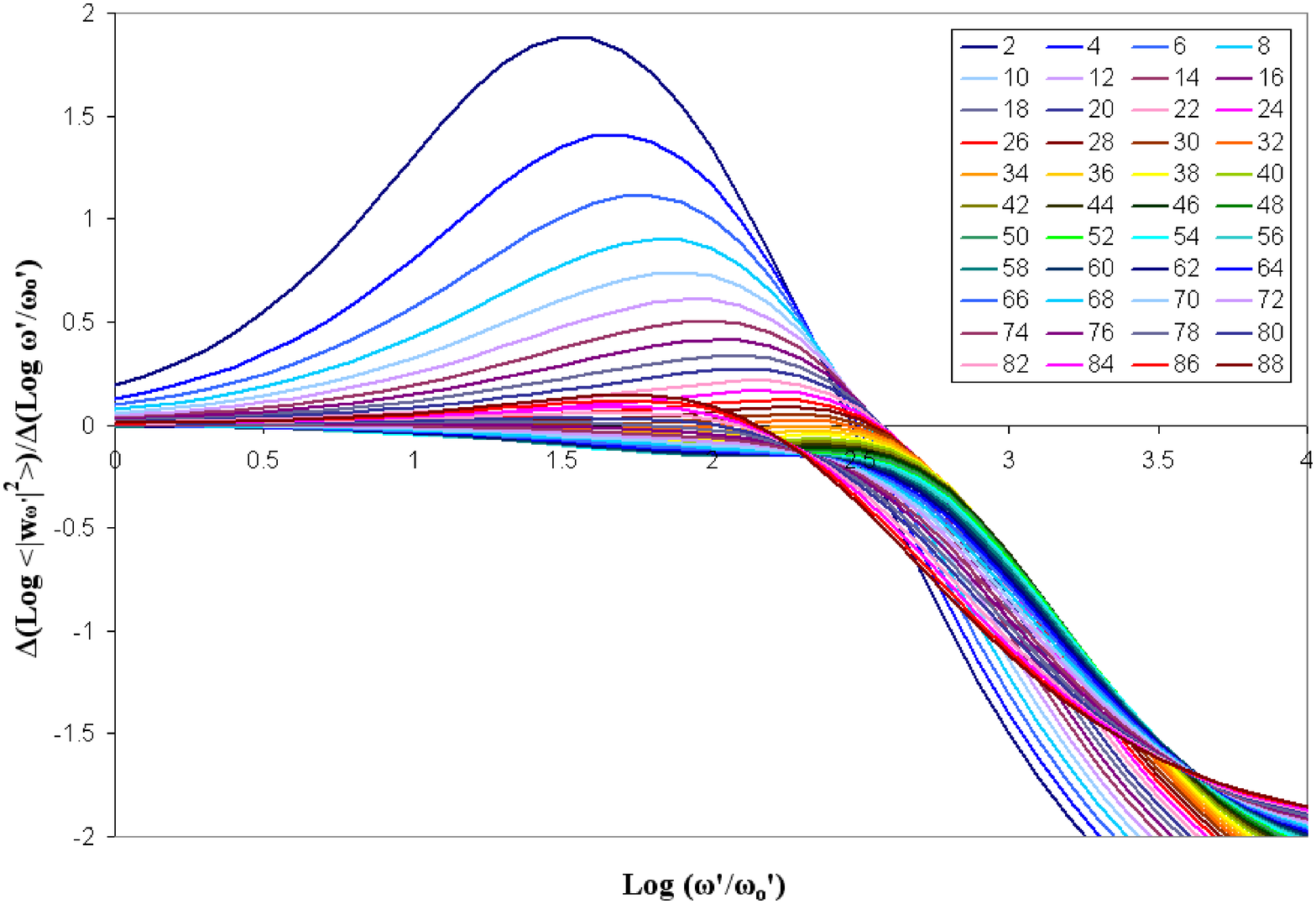}
	\caption{The slope of the spectra in the $\log-\log$ plot in Figure \ref{fig:AccSpecAll}.  Note that even for unpeaked spectra there is a flattening of the spectral slope and in some cases a local maximum (around $\log(\omega^{\prime}/\omega^{\prime}_o)=2.5$).}
	\label{fig:AccSpecSlope}
\end{figure}

\renewcommand{\thefigure}{\arabic{figure}}
\begin{figure}
	\plotone{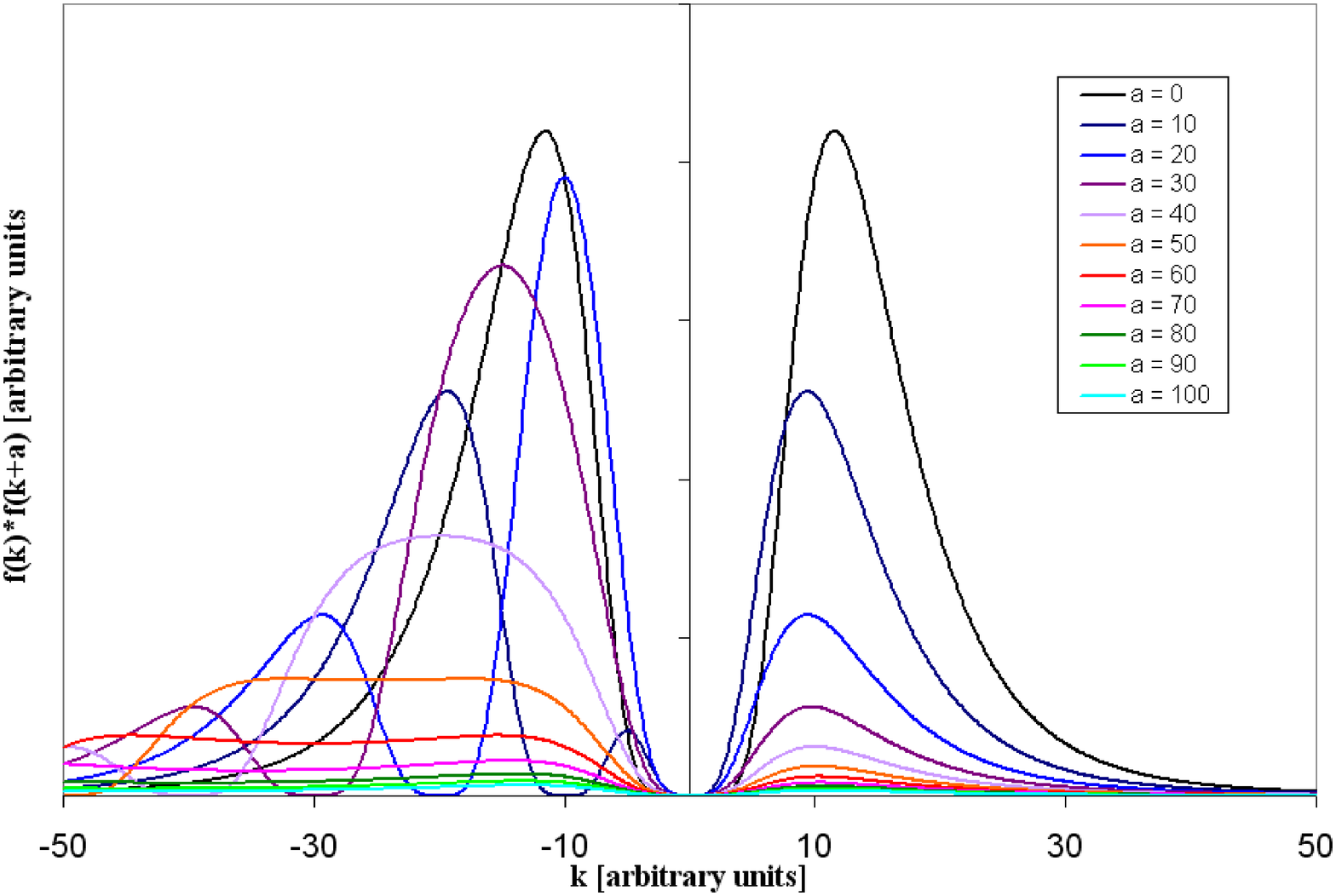}
	\caption{Plots of the product $f(k)f(k+a)$ of two functions of the form of our magnetic field spectra (Equations (\ref{eq:perpfspec}) and (\ref{eq:pllfspec})), illustrating the effect of the offset $a$.  As described in section \ref{s:accspec}, the acceleration spectrum is the integral of a product of such functions, with an offset controlled by the acceleration Fourier frequency $\omega^{\prime}$.}
	\label{fig:offsetvar}
\end{figure}

\clearpage
 
\begin{figure}
	\plotone{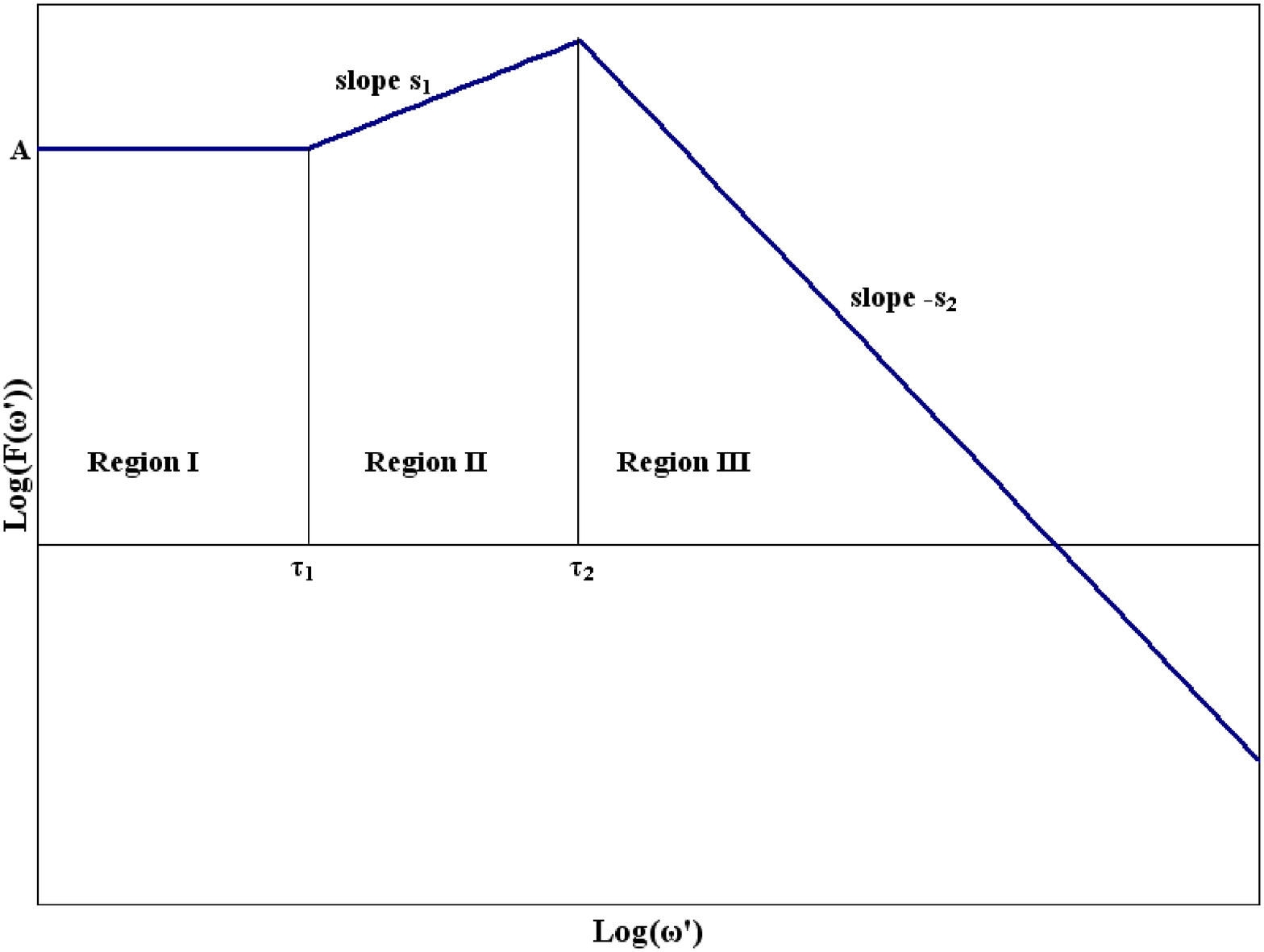}
	\caption{Here we show a model acceleration spectrum, as described in Equations (\ref{eq:accap1.1}) - (\ref{eq:accap1.3}).  The spectrum is flat in Region I ($\log(\omega^{\prime})<\tau_1$), then becomes a power law $\omega^{\prime s_1}$ in Region II ($\tau_1<\log(\omega^{\prime})<\tau_2$), and a power law $\omega^{\prime -s_2}$ in Region III ($\log(\omega^{\prime})>\tau_2$). }  
	\label{fig:wcartoon}
\end{figure}

\begin{figure}
	\plotone{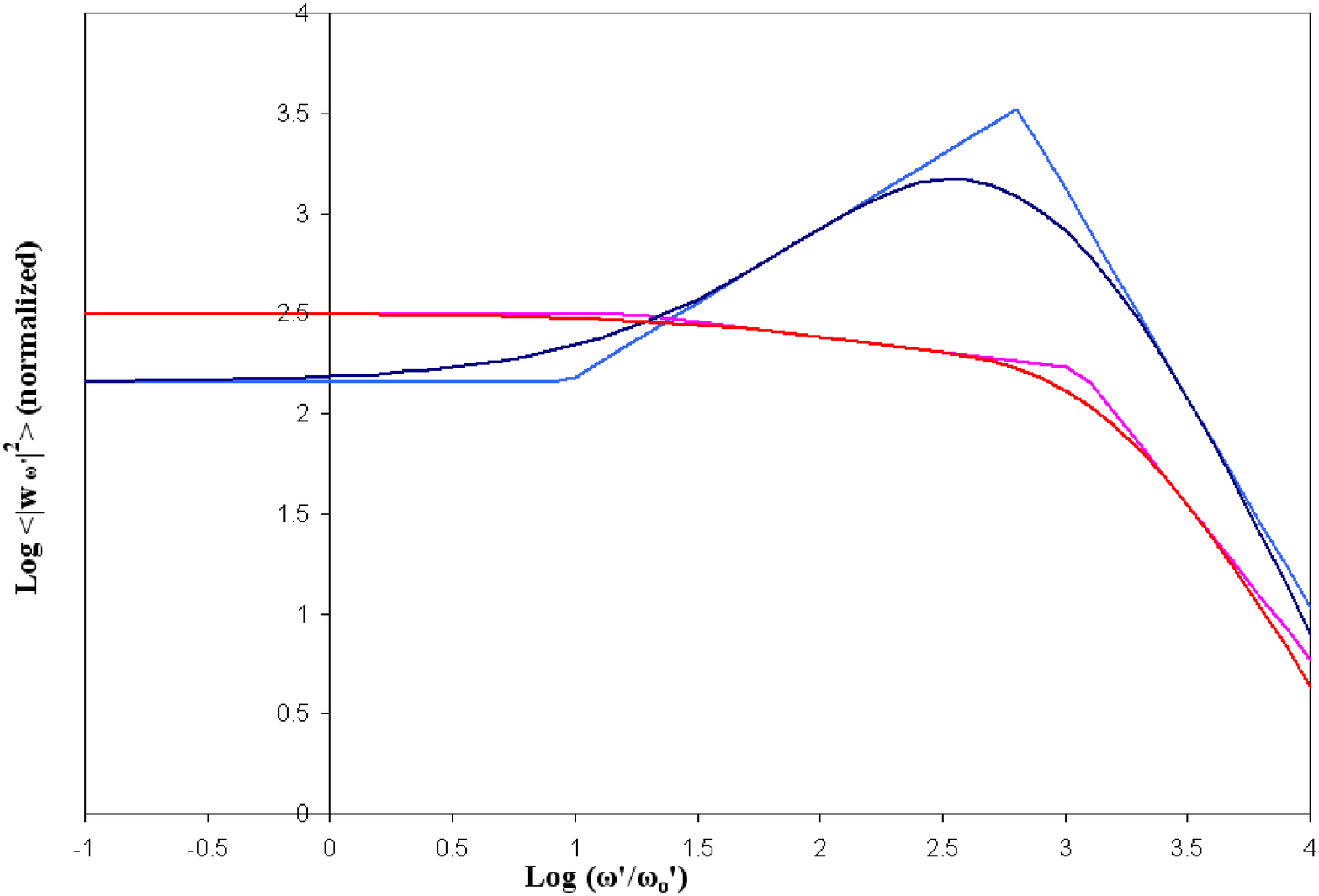}
	\caption{Here we show a comparison of our original calculated spectra (as presented in Figure \ref{fig:AccSpecAll}) for $\theta$=10 (peaked) and $\theta$=60 (unpeaked), and our model acceleration spectra as described in Equations \ref{eq:accap1.1}-\ref{eq:accap1.3}.  In Region I, we extrapolate the initial calculated amplitude.  In Region II, we fit the calculated spectra using the maximum slope in the region and fitting through the point of maximum slope.  In Region III we have done a simple fit using the slope and position at point $\omega$=4.6.  Our choice of fit overestimates the peak of the calculated acceleration spectra.}  
	\label{fig:modaccspec}
\end{figure}

\renewcommand{\thefigure}{\arabic{figure}(a)}
\begin{figure}
	\plotone{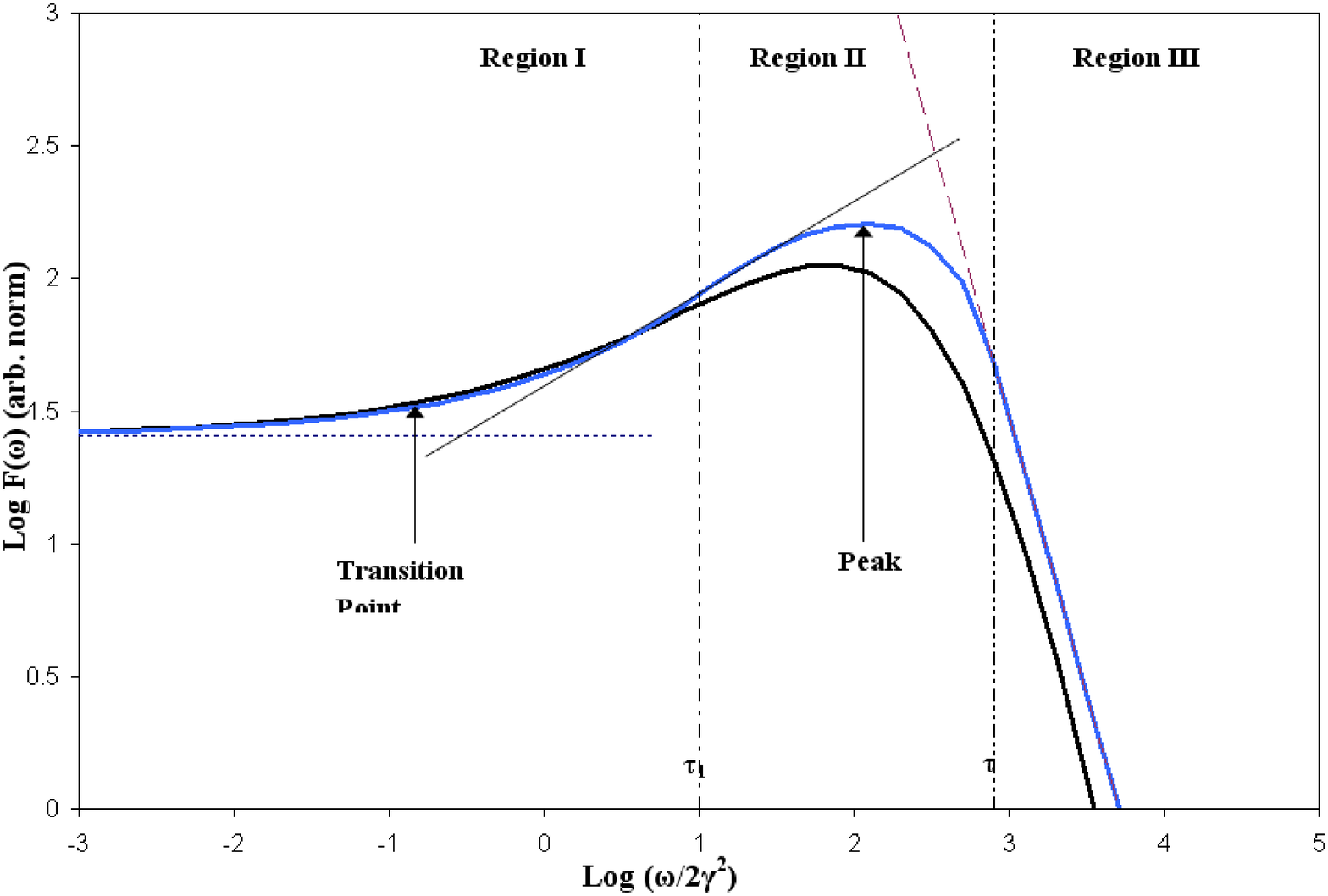}
	\caption{The radiation spectrum for $\theta$=10 degrees as obtained analytically using equations \ref{eq:radap1.1} - \ref{eq:radap1.3} and the fit values for the model acceleration spectra as shown in Figure \ref{fig:wcartoon}.  The result calculated (black) via full double numerical integration (as in Figure \ref{fig:RadSpecAll}) is shown here for comparison. The spectral transition point defined by equation \ref{eq:radtranspt} and the peak defined by equation\ref{eq:radpeak} are marked with arrows.	We note that the boundaries (dotted lines) between Regions I, II, and III as defined by our model acceleration spectra do not clearly correspond to transition points in the resulting radiation spectra.  We have extrapolated slopes for Regions I, II, and III to demonstrate the variation.  The model again slightly overestimates the peak energy, but matches the spectral shape well and agrees with the integrated spectra within about $10\%$.} 
	\label{fig:modradspec10}
\end{figure}

\addtocounter{figure}{-1}
\renewcommand{\thefigure}{\arabic{figure}(b)}
\begin{figure}
	\plotone{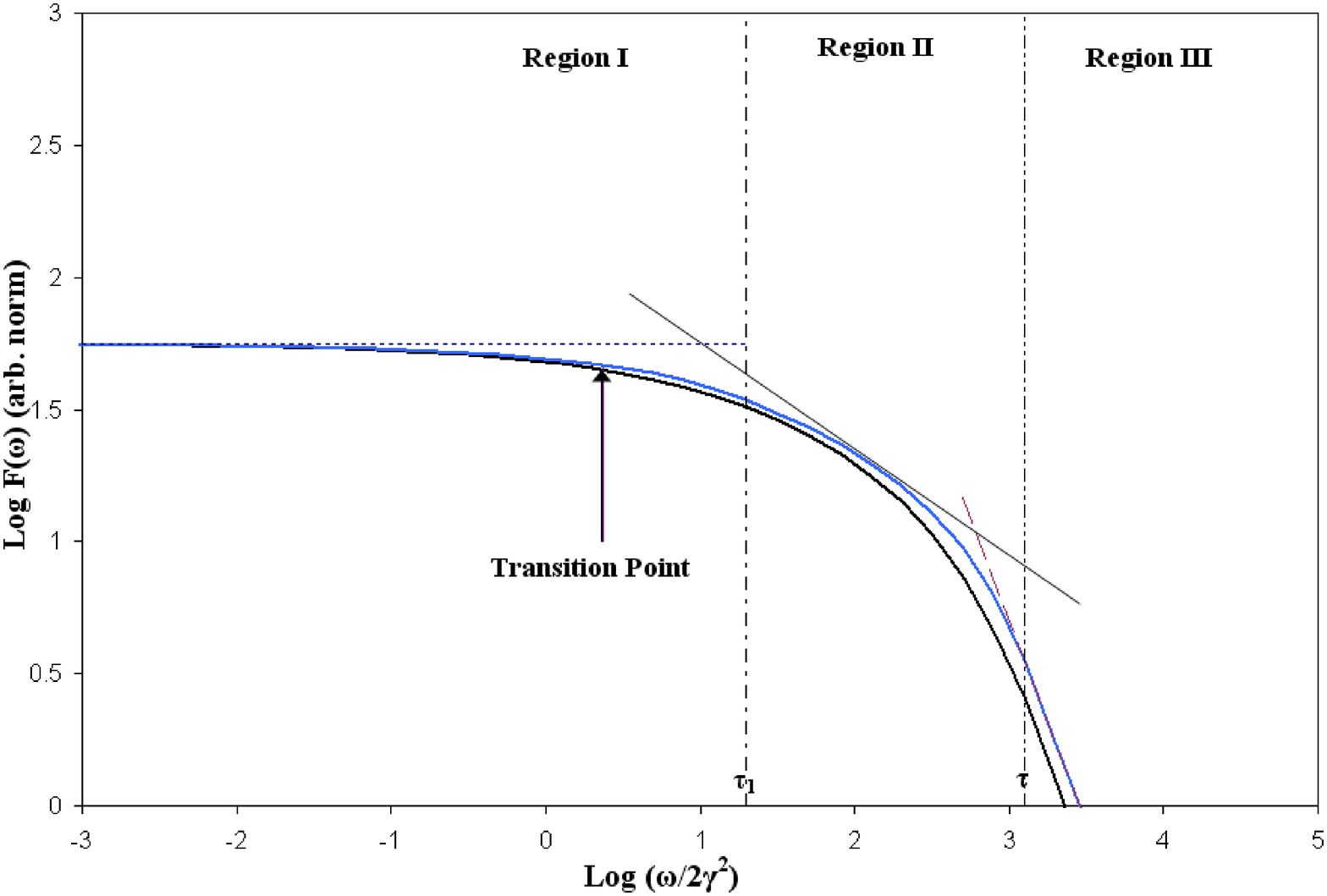}
	\caption{The radiation spectrum for $\theta$=60 degrees as obtained analytically from our simplified model acceleration spectrum (as shown in Figure \ref{fig:wcartoon}) versus the full double numerical integration result (black).  As in the previous figure, we have extrapolated the slope in Regions I, II, and III. We again obtain a good agreement with the overall spectral shape, but now with an unpeaked form appropriate to this range of theta.  The spectral transition point defined by equation \ref{eq:radtranspt} is marked with an arrow.}
	\label{fig:modradspec60}
\end{figure}
%
\renewcommand{\thefigure}{\arabic{figure}(a)}
\begin{figure}
	\plotone{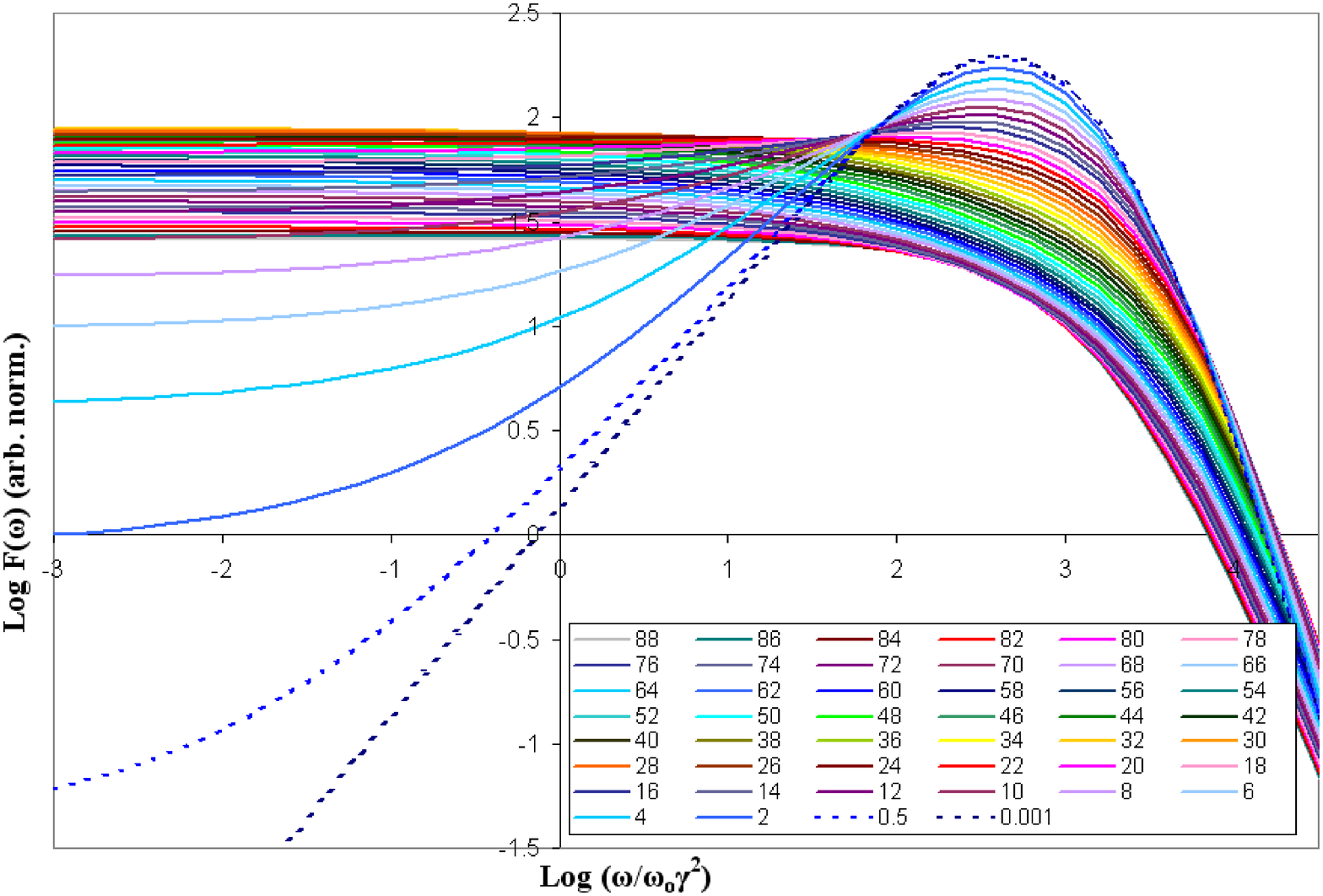}
	\caption{The angle-averaged radiation power spectra ($F_\nu$) of jitter radiation, numerically calculated for a step size of 0.2 in $\log(\omega)$ for every two degrees in $\theta$.  The dotted lines are numerical calculations for values close to $\theta$=0, illustrating the behavior of the spectrum in this limit.  The spectra are arbitrarily normalized so that the first calculated value for the $\theta = 2$ spectrum is 0.  The radiation spectra are flat for low-$\omega$, then slope upwards to a notable peak for low $\theta$.  As $\theta$ progresses to $\pi$/2, the low-$\omega^{\prime}$ amplitude increases and the size of the sloped region decreases until the spectra becomes relatively flat until its sharp decline at high-$\omega$.}
	\label{fig:RadSpecAll}
\end{figure}
\addtocounter{figure}{-1}
\renewcommand{\thefigure}{\arabic{figure}(b)}
\begin{figure}
	\plotone{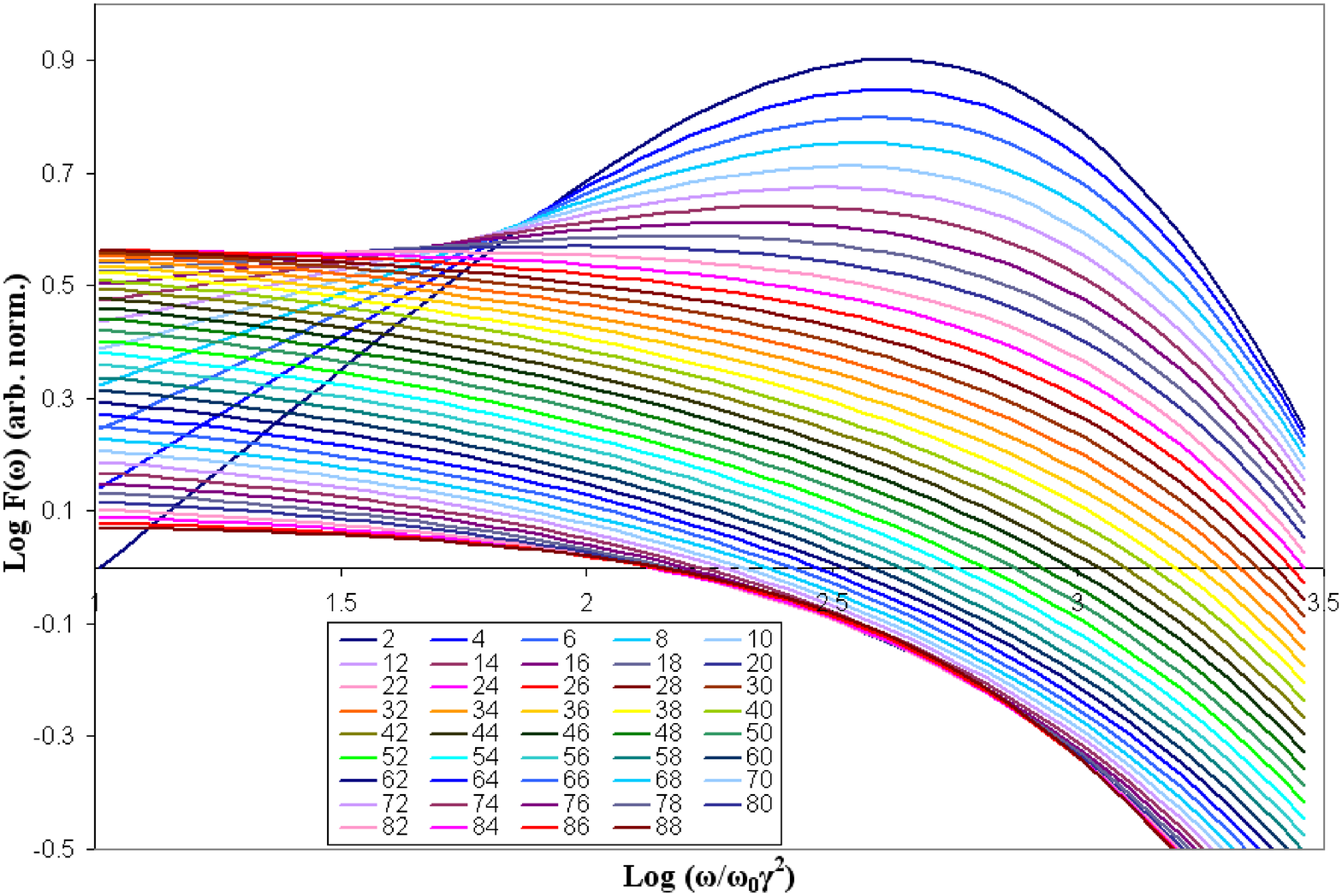}
	\caption{A more detailed view of the radiation spectra in the region of the spectral peak. The spectral values are calculated every 0.05 in $\log(\omega$) for finer resolution of the detail in this region.}
	\label{fig:RadSpecPeakDetail}
\end{figure}
\renewcommand{\thefigure}{\arabic{figure}}
\begin{figure}
	\plotone{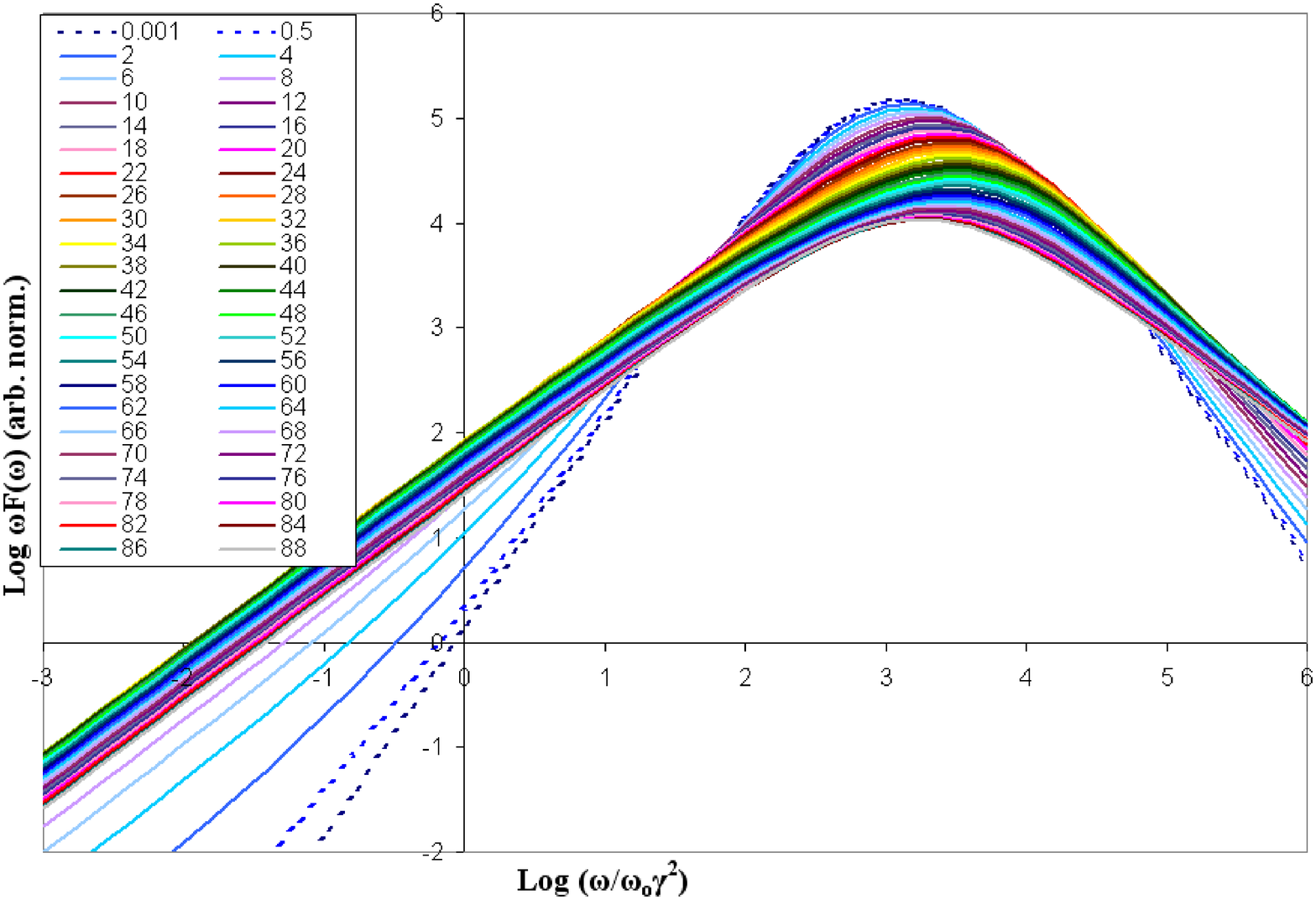}
	\caption{The $\nu F_\nu$ spectra calculated from our jitter radiation power spectral results.  The $\nu F_{\nu}$ spectral peak is the peak energy $E_p$ used in the Band functional fit commonly used for GRB spectra \citep{band, Kaneko}. }
	\label{fig:vFvRadSpecAll}
\end{figure}

\begin{figure}
	\plotone{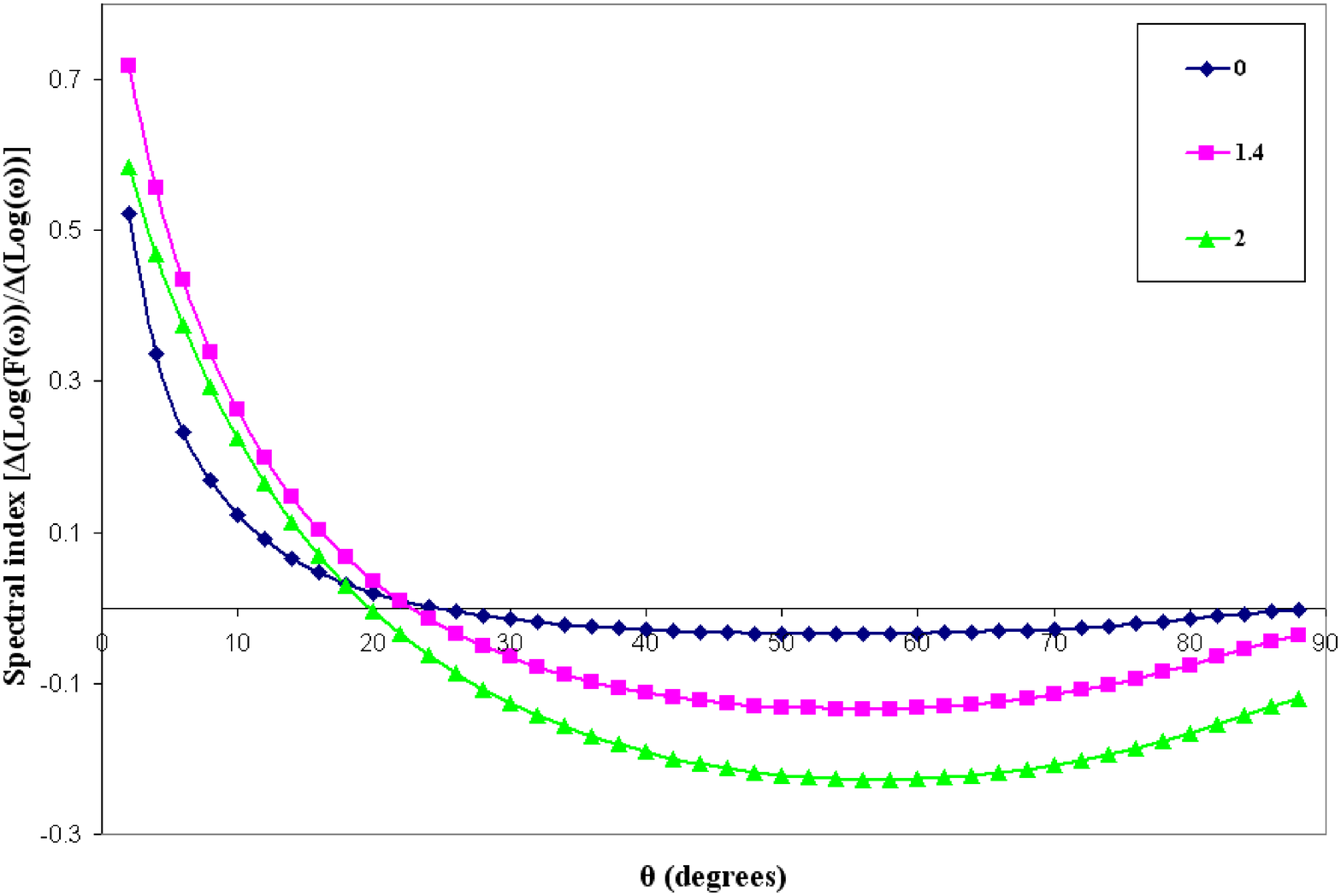}
	\caption{A plot of the slope of the radiation spectra taken at several values of $\log(\omega/\omega_0\gamma^2)$ over the range of $\theta$.} 
	\label{fig:RadSpecIndex}
\end{figure}
\clearpage
\begin{figure}
	\plotone{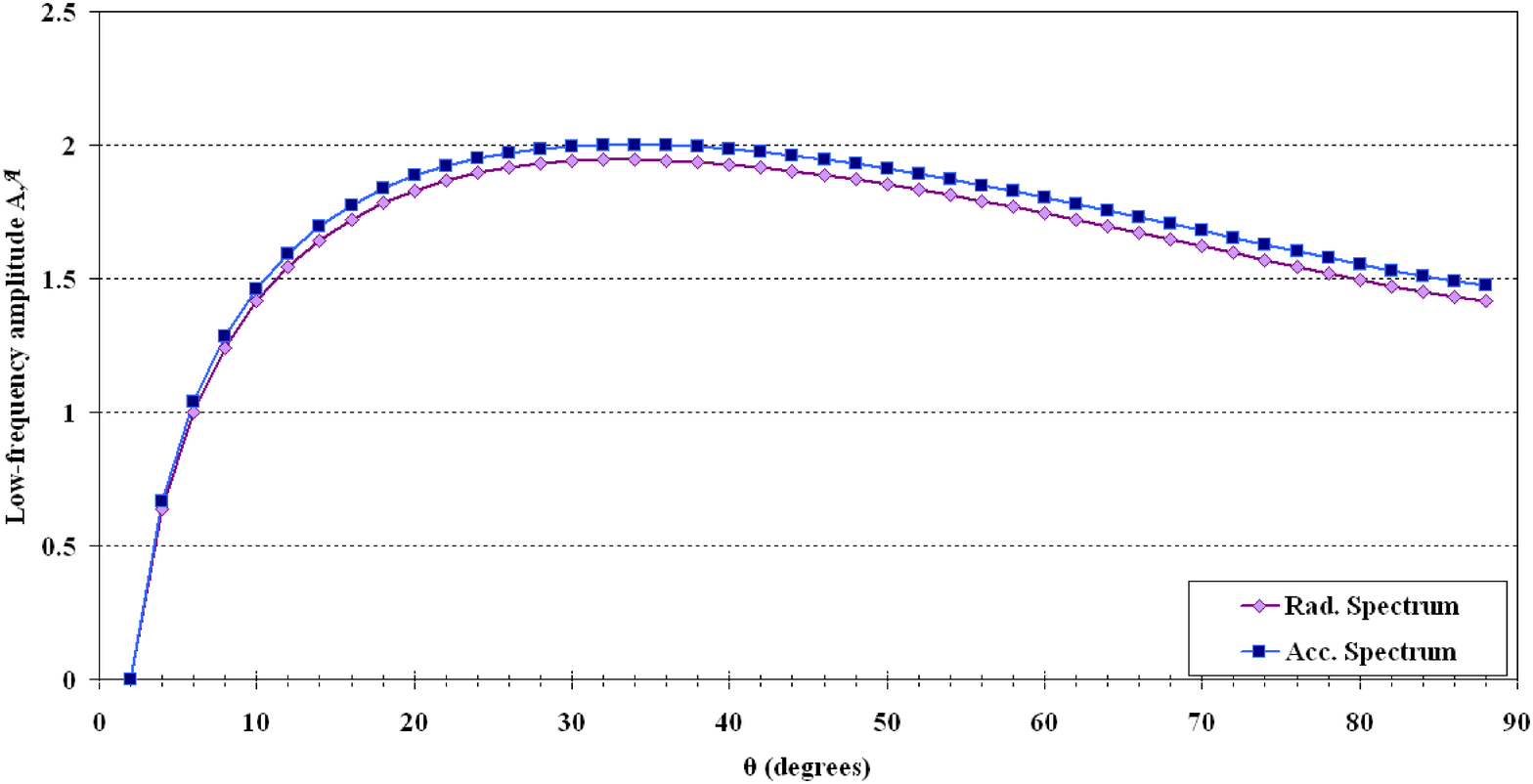}
	\caption{The angular dependence of the low-frequency amplitude $A$ of our calculated jitter radiation spectra and the corresponding variation of the low-frequency amplitude $\mathpzc{A}$ of our calculated acceleration spectra.  The low-energy amplitude in each case is taken to be the first calculated value of the spectrum; for our choice of parameters and calculation window, this initial value is well below the first spectral break for all $\theta > 2^o$.  In both cases, we have normalized our spectra such that the low-energy amplitude at $\theta = 2^o$ is 0.}
	\label{fig:fullfitamp}
\end{figure}
\begin{figure}
	\plotone{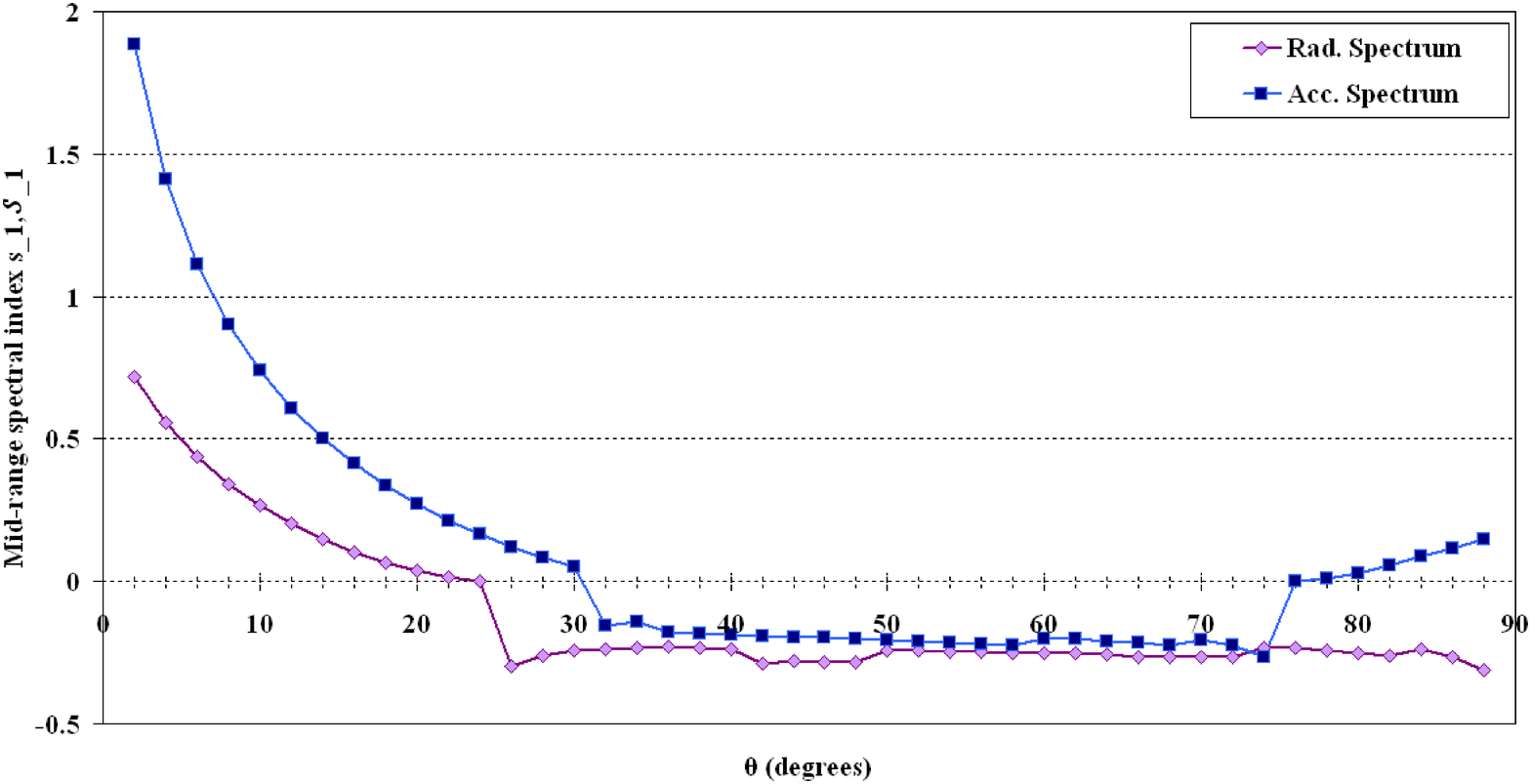}
	\caption{The angular dependence of the mid-range spectral indices $s_1$ of our calculated jitter radiation spectra and $\mathpzc{S}_1$ of our calculated acceleration spectra.  For peaked spectra the mid-range spectral index is the maximum slope below the spectral peak; for unpeaked spectra the mid-range spectral index is the average slope between the point at which the spectrum falls below $A-0.01$ and the point at which the numerical second derivative reaches its minimum value (the ''drop point").}
	\label{fig:fullfits1}
\end{figure}
\begin{figure}
	\plotone{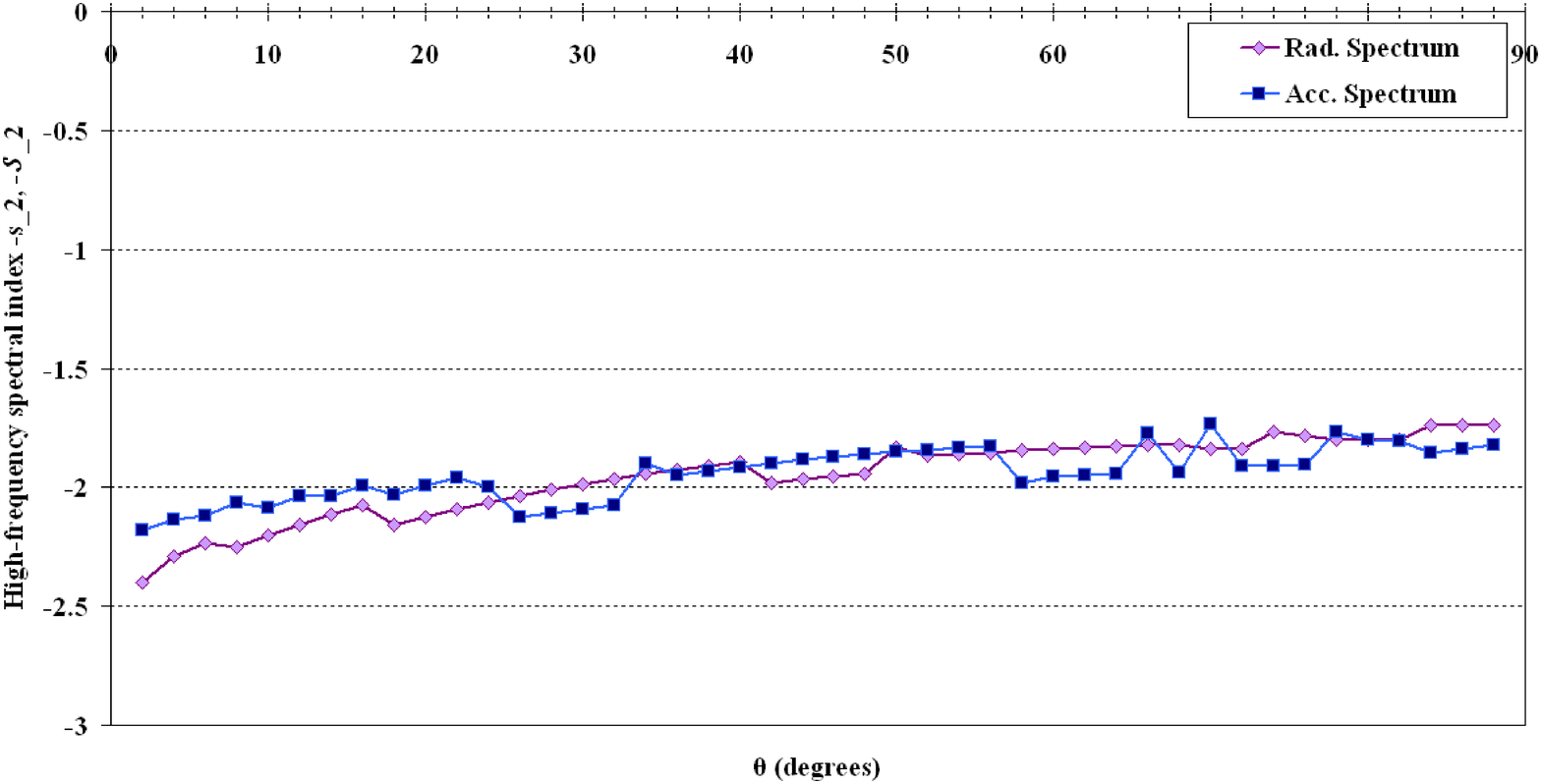}
	\caption{The angular dependence of the high-frequency spectral indices $s_2$ for our calculated jitter radiation spectra and $\mathpzc{S}_2$ of the corresponding acceleration spectra.  The high-frequency spectral index is calculated as the slope between the drop point (the position of the largest negative change in slope) and the higher-frequency position at which the numerical second derivative is closest to 0.}  
	\label{fig:fullfits2}
\end{figure}
\begin{figure}
	\plotone{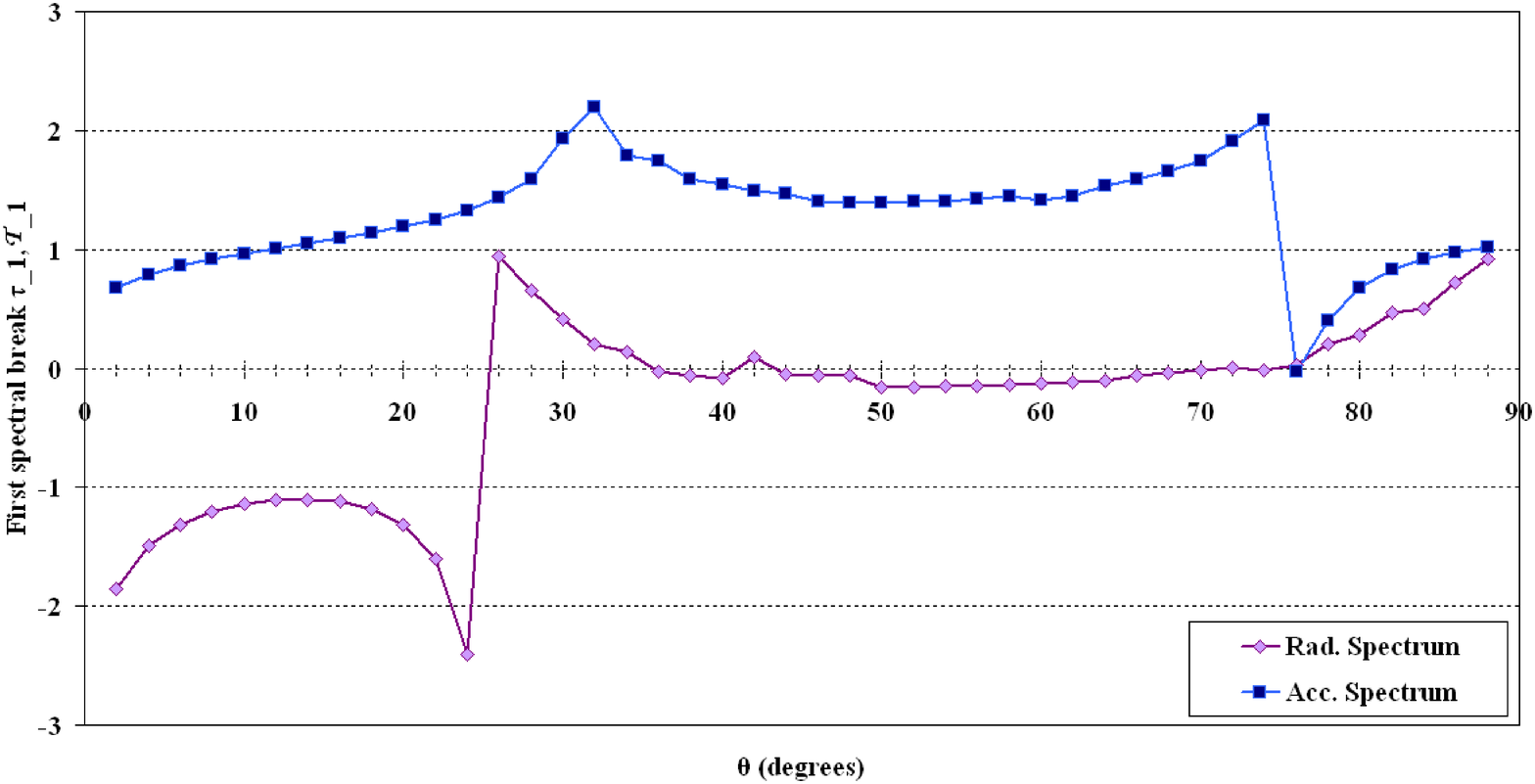}
	\caption{The angular dependence of the first spectral break positions $\log_e(\omega/\omega_o\gamma^2) = \tau_1$ and $\log_e(\omega^{\prime}/\omega^{\prime}_o) = \mathpzc{T}_1$ in our calculated jitter radiation spectra and acceleration spectra, respectively.  In each case these transition points are found as the intersection between the low-frequency fit line of slope 0 and the mid-range fit line of slope $s_1$ (radiation) or $\mathpzc{S}_1$ (acceleration), as found by the fit described in detail in section \ref{s:radspec}.}
	\label{fig:fullfittp1}
\end{figure}
\begin{figure}
	\plotone{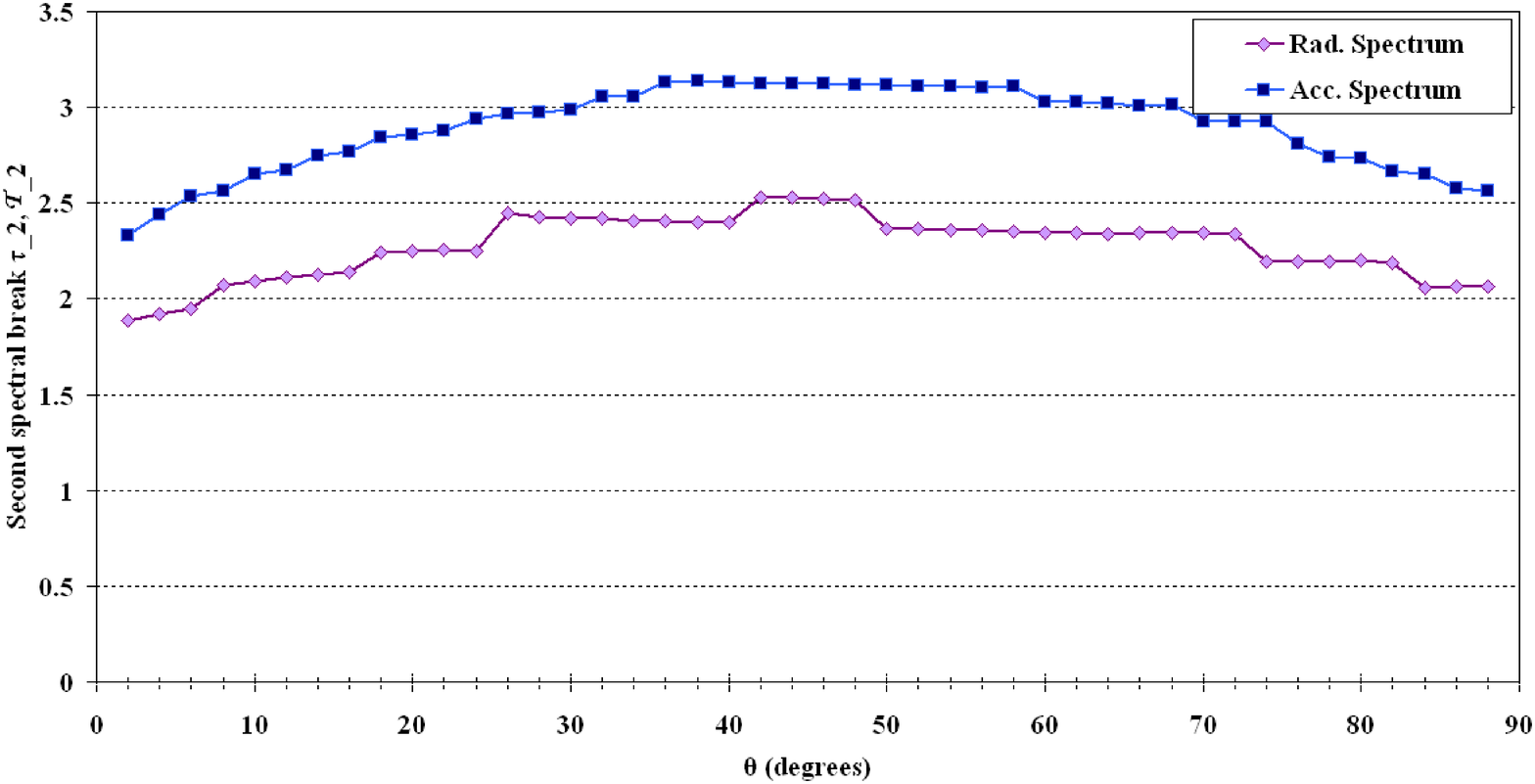}
	\caption{The angular dependence of the second spectral break $\tau_2$ of the jitter radiation spectra and $\mathpzc{T}_2$ of the acceleration spectra.  The transition point is the calculated intersection between the mid-range fit line of slope $s_1$ (radiation) or $\mathpzc{S}_1$ (acceleration) and the high-frequency fit line of slope $-s_2$ (radiation) or $-\mathpzc{S}_2)$ (acceleration).}
	\label{fig:fullfittp2}
\end{figure}
\begin{figure}
	\plotone{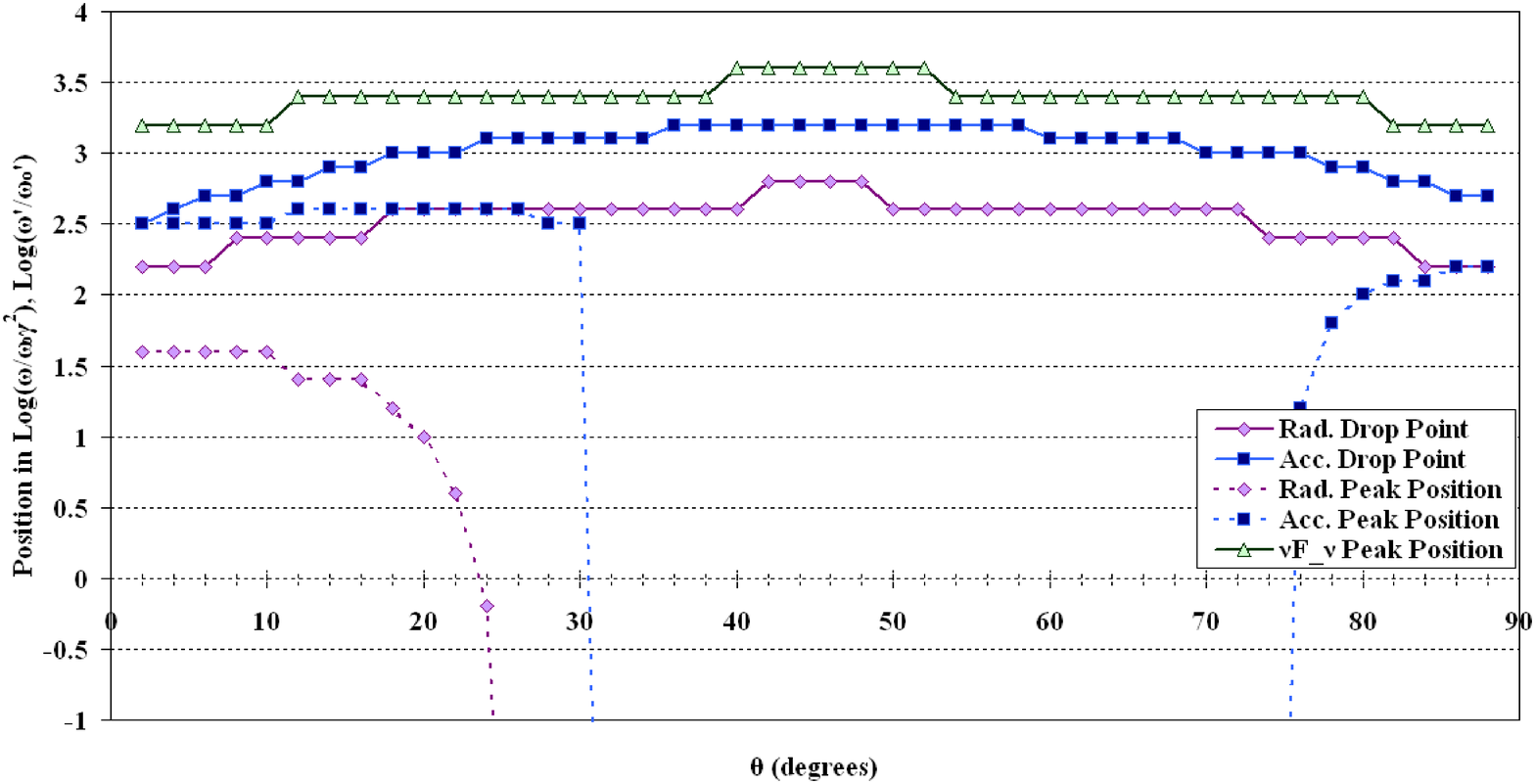}
	\caption{A comparison of the angular dependence of the positions of the radiation and acceleration spectral peaks and the spectral drop points, where the numerical second derivatives of our calculated spectra reach a minimum.  We have also plotted the angular dependence of the peak in the $\nu F_{\nu}$ spectrum.  The drop point in our $F_{\nu}$ radiation spectrum nicely tracks the behavior of the peak in the $\nu F_{\nu}$ spectrum, which is peak energy $E_p$ in the Band function \citep{band} commonly used to fit GRB spectra.}
	\label{fig:fullfitdroppk}
\end{figure}
\begin{figure}
	\plotone{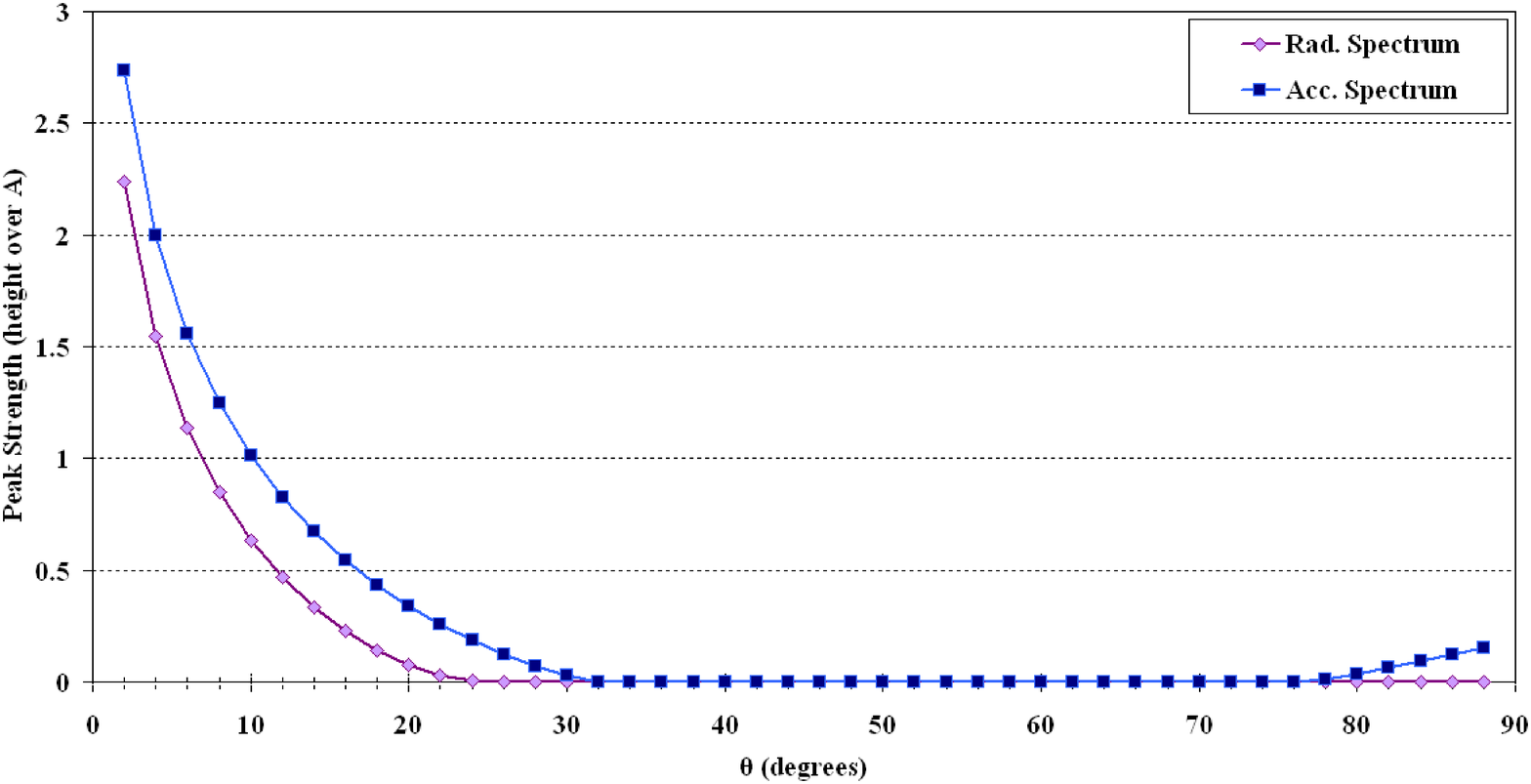}
	\caption{The angular dependence of the strength of the spectral peak, i.e. the height of the peak above the initial low-frequency spectral amplitude A.  We see that the peak disappears in the radiation spectrum at $\theta$ roughly $8^o$ less than in the acceleration spectrum, and does not reappear at $\theta$ close to $90^o$.}
	\label{fig:fullfitpkheight}
\end{figure}
%
\renewcommand{\thefigure}{\arabic{figure}(a)}
\begin{figure}
	\plotone{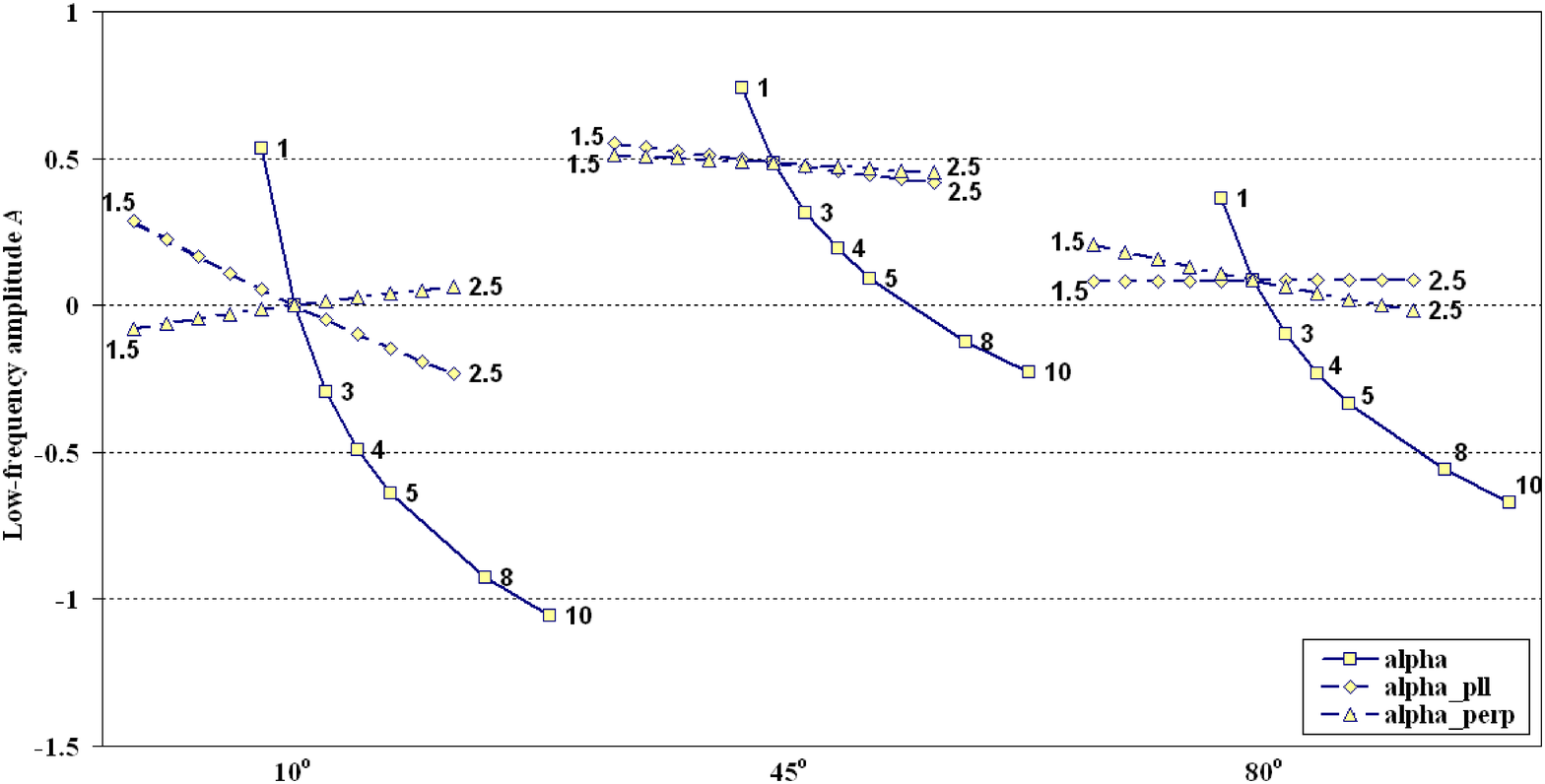}
	\caption{This figure shows the variation of the low-frequency amplitude of the radiation spectrum (vertical axis) with changes in the magnetic field parameters $\alpha_i$, which are varied progressively in the horizontal dimension as described below, for each of the fixed viewing angles $\theta = 10^o$, $45^o$, and $80^o$ (first, second, and third clusters respectively).  The solid line with square data points indicates the behavior when $\alpha$ is jointly varied (over a range from 1 to 10, as indicated) in both the transverse and parallel magnetic field equation: $\alpha = \alpha_{\perp}=\alpha_{\parallel}$.  The dotted lines show the effect of individually varying $\alpha_{\perp}$ (triangular data points) and $\alpha_{\parallel}$ (diamond data points), which chracterize the magnetic field transverse to and along the instability filamentation axis, respectively.  The parameters $\alpha_{\perp}$ and $\alpha_{\parallel}$ are varied from a starting value of 1.5 to an ending value of 2.5, in increments of 0.1.  Our original value of $\alpha = \alpha_{\perp}=\alpha_{\parallel} = 2.0$ is the central data point where the lines intersect.}
		\label{fig:alphaamp}
\end{figure}
\addtocounter{figure}{-1}
\renewcommand{\thefigure}{\arabic{figure}(b)}
\begin{figure}
	\plotone{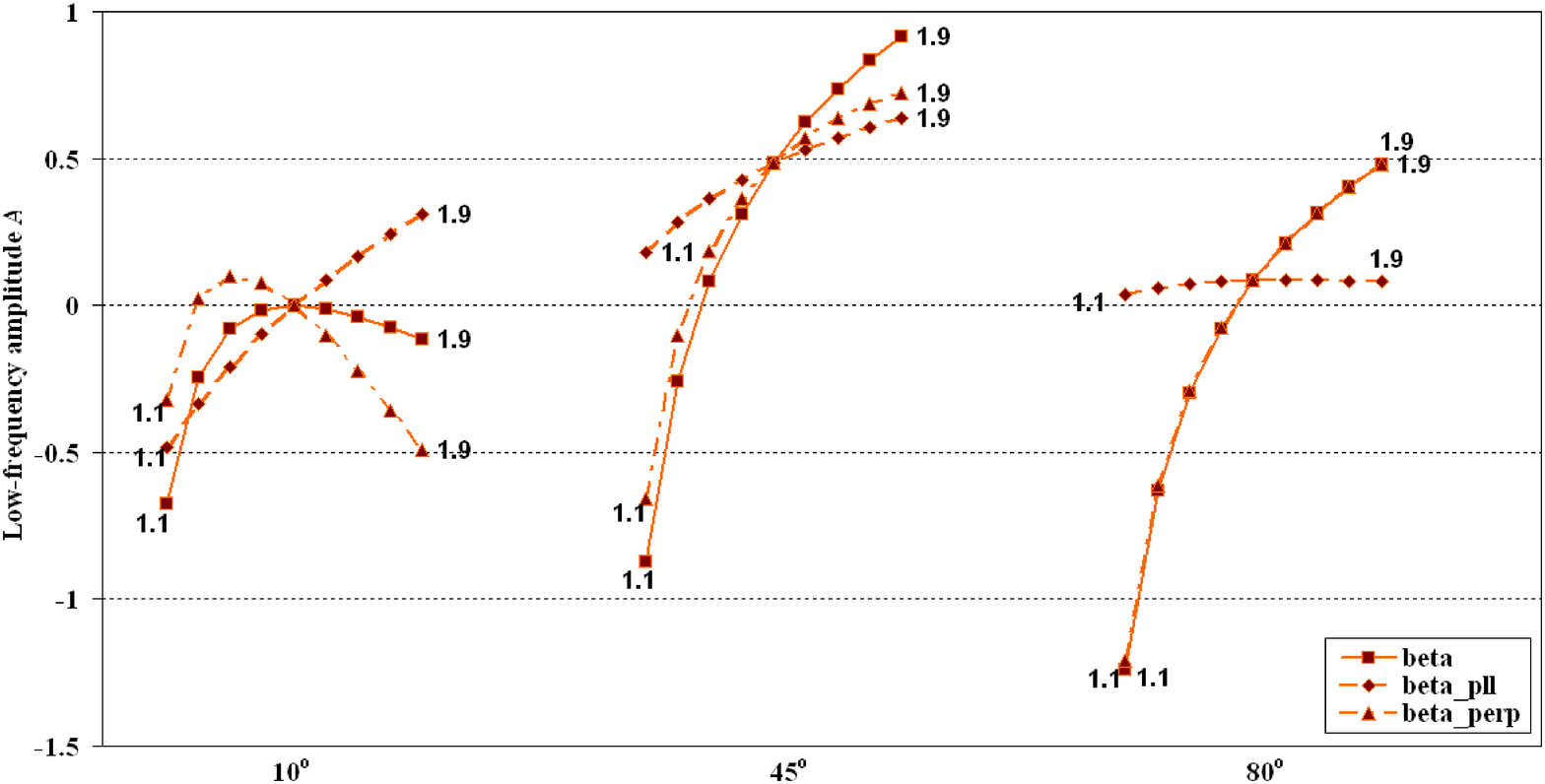}
	\caption{This figure shows the variation of the low-frequency amplitude of the radiation spectrum (vertical axis) with changes in the magnetic field parameters $\beta_i$, which are varied progressively in the horizontal dimension for the fixed viewing angles $\theta = 10^o$, $45^o$, and $80^o$ (first, second, and third clusters respectively).  The solid line with square data points indicates the behavior when $\beta$ is jointly varied from 1.1 to 1.9, in increments of 0.1, in both the transverse and parallel magnetic field equation: $\beta = \beta_{\perp}=\beta_{\parallel}$.  The dotted lines show the effect of varying $\beta_{\perp}$ (triangular data points) and $\beta_{\parallel}$ (diamond data points) individually, also from 1.1 to 1.9, in increments of 0.1.  Our original value of $\beta = \beta_{\perp}=\beta_{\parallel} = 1.5$ is the central data point where the lines intersect.}
	\label{fig:betaamp}
\end{figure}
\addtocounter{figure}{-1}
\renewcommand{\thefigure}{\arabic{figure}(c)}
\begin{figure}
	\plotone{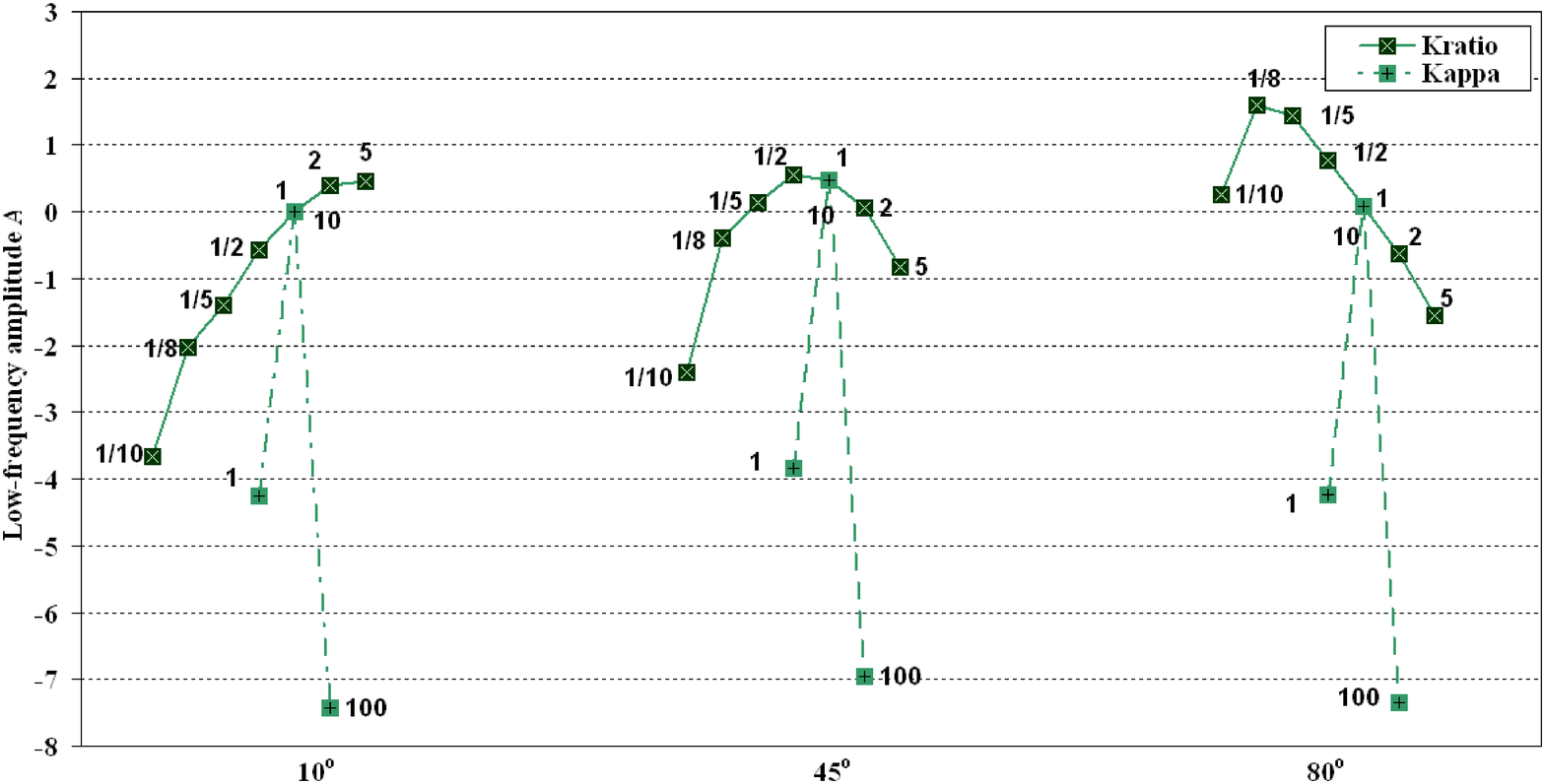}
	\caption{This figure shows the variation of the low-frequency amplitude of the radiation spectrum with changes in the magnetic field parameters $\kappa$, which are varied progressively in the horizontal dimension for the fixed viewing angles $\theta = 10^o$, $45^o$, and $80^o$ (first, second, and third clusters respectively).  We vary $\kappa$ jointly ($\kappa = \kappa_{\perp} = \kappa_{\parallel}$) by powers of 10, from 1 to 100, as indicated.  We also vary $\kappa_{\perp}$ and $\kappa_{\parallel}$ relative to one another by changing the ratio $K = \kappa_{\perp}/\kappa_{\parallel}$ through a range of values as indicated.}   
	\label{fig:kappaamp}
\end{figure}
\renewcommand{\thefigure}{\arabic{figure}}
\begin{figure}
\plotone{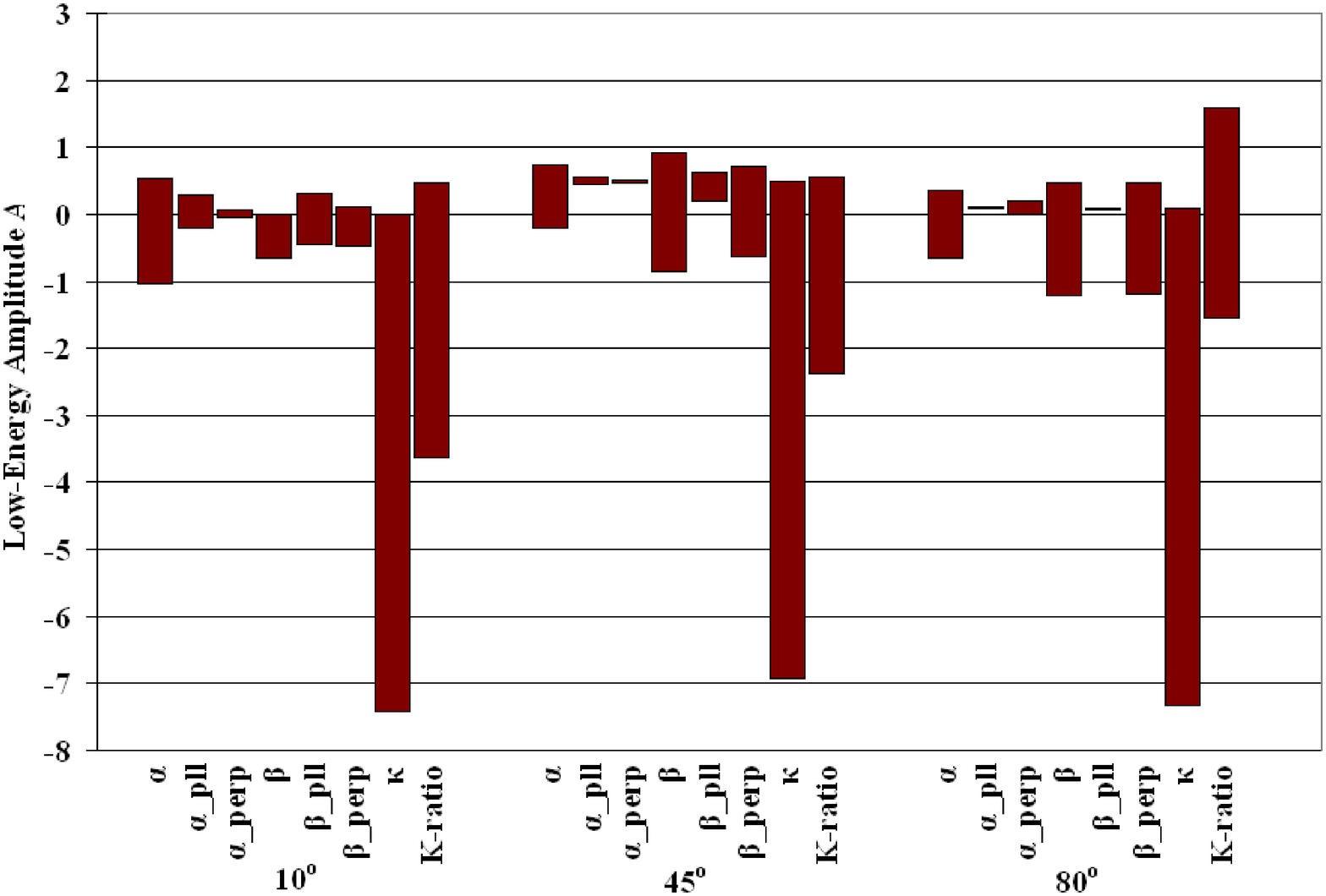}
\caption{This figure compares the influences of the magnetic field spectral parameter variations on the low-frequency amplitude $A$ of the radiation spectrum obtained for representative viewing angles $\theta = 10^o$, $45^o$, and $80^o$.  For each spectral parameter (indicated on the bottom axis) the graph indicates the range between the maximum and minimum values of $A$ obtained by our variations of that parameter. (The parameter variations are as indicated in the previous figures and described in detail in section \ref{s:sfvarparam}.)}
\label{fig:ampcomp}
\end{figure}
%
\renewcommand{\thefigure}{\arabic{figure}(a)}
\begin{figure}
	\plotone{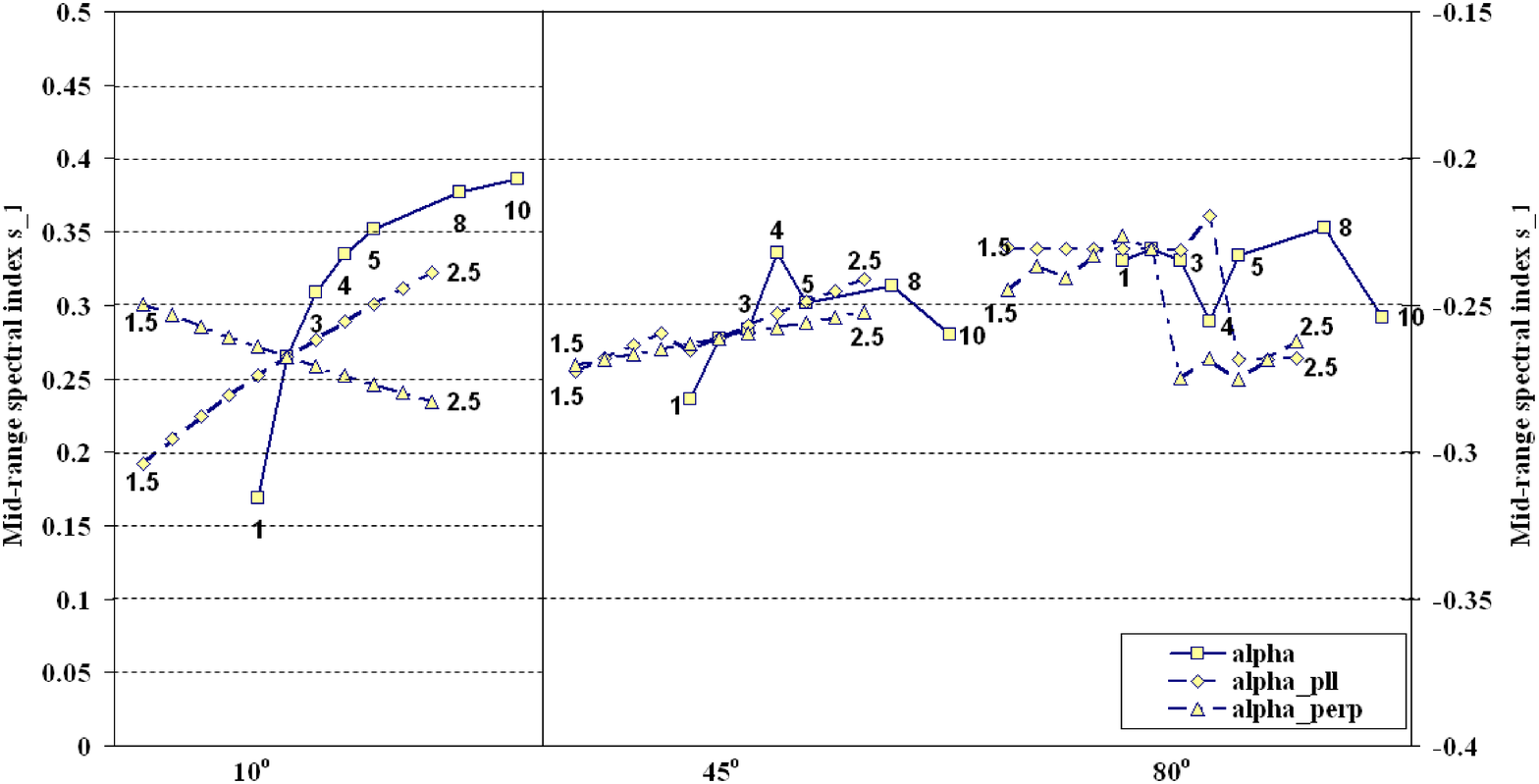}
	\caption{This figure shows the variation of the mid-range spectral index (as determined by the fit described in section \ref{s:sfvarparam}) with changes in the magnetic field parameters $\alpha_i$.  This is the maximum slope below the peak for peaked spectra, and the average slope in intermediary region for unpeaked spectra, and is shown for representative viewing angles $\theta = 10^o$, $45^o$, and $80^o$.  The solid line with square data points indicates the behavior when $\alpha$ is jointly varied (over a range from 1 to 10, as indicated) in both the transverse and parallel magnetic field equation: $\alpha = \alpha_{\perp}=\alpha_{\parallel}$.  The dotted lines show the effect of varying $\alpha_{\perp}$ (triangular data points) and $\alpha_{\parallel}$ (diamond data points) individually, from 1.5 to 2.5, in increments of 0.1. }
	\label{fig:alphas1}
\end{figure}
\addtocounter{figure}{-1}
\renewcommand{\thefigure}{\arabic{figure}(b)}
\begin{figure}
	\plotone{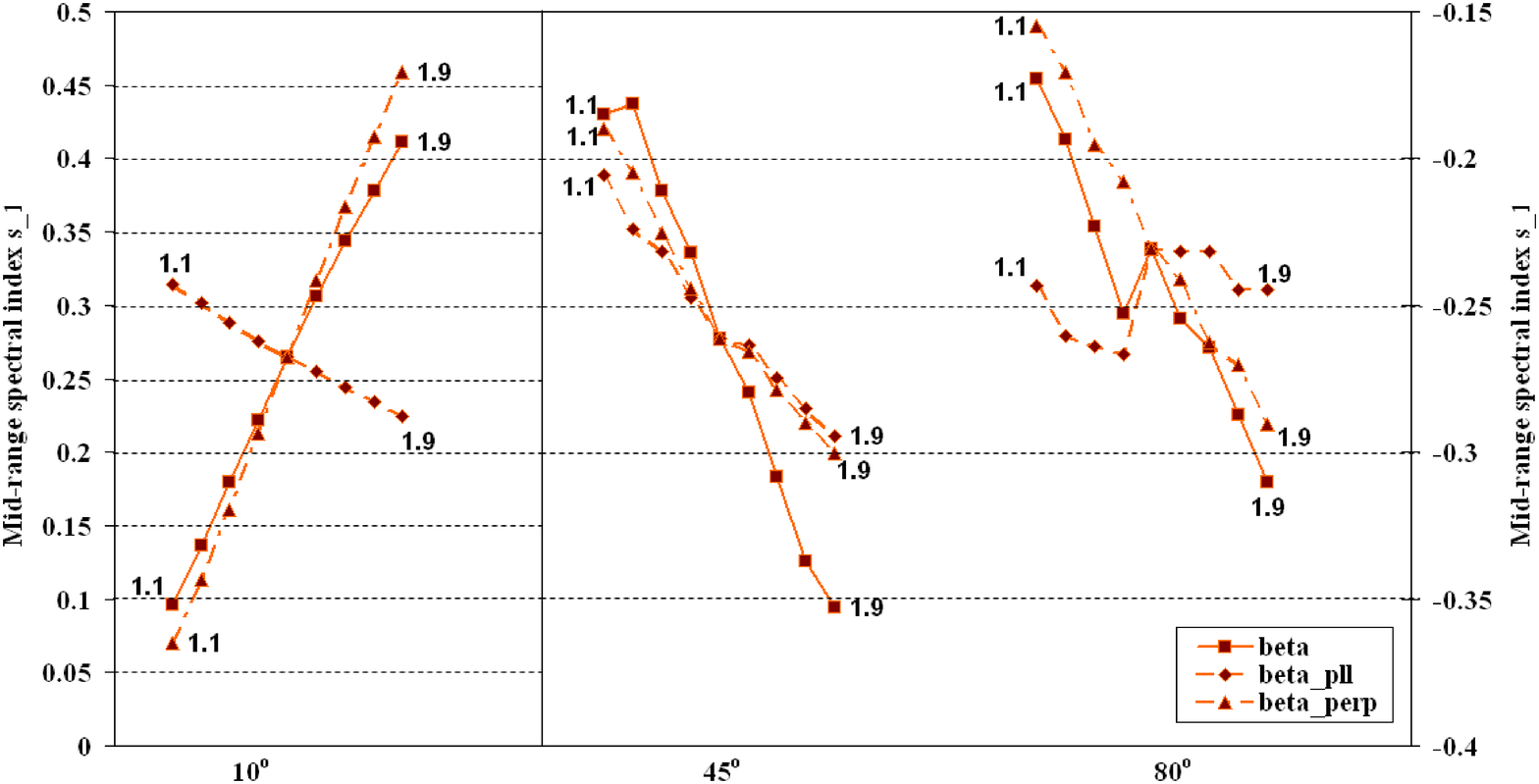}
	\caption{This figure shows the variation of the mid-range spectral index with changes in the magnetic field parameters $\beta_i$.  The solid line with square data points indicates the behavior when $\beta$ is jointly varied from 1.1 to 1.9, in increments of 0.1, in both the transverse and parallel magnetic field equation: $\beta = \beta_{\perp}=\beta_{\parallel}$.  The dotted lines show the effect of varying $\beta_{\perp}$ (triangular data points) and $\beta_{\parallel}$ (diamond data points) individually, also from 1.1 to 1.9, in increments of 0.1.}
	\label{fig:betas1}
\end{figure}
\addtocounter{figure}{-1}
\renewcommand{\thefigure}{\arabic{figure}(c)}
\begin{figure}
	\plotone{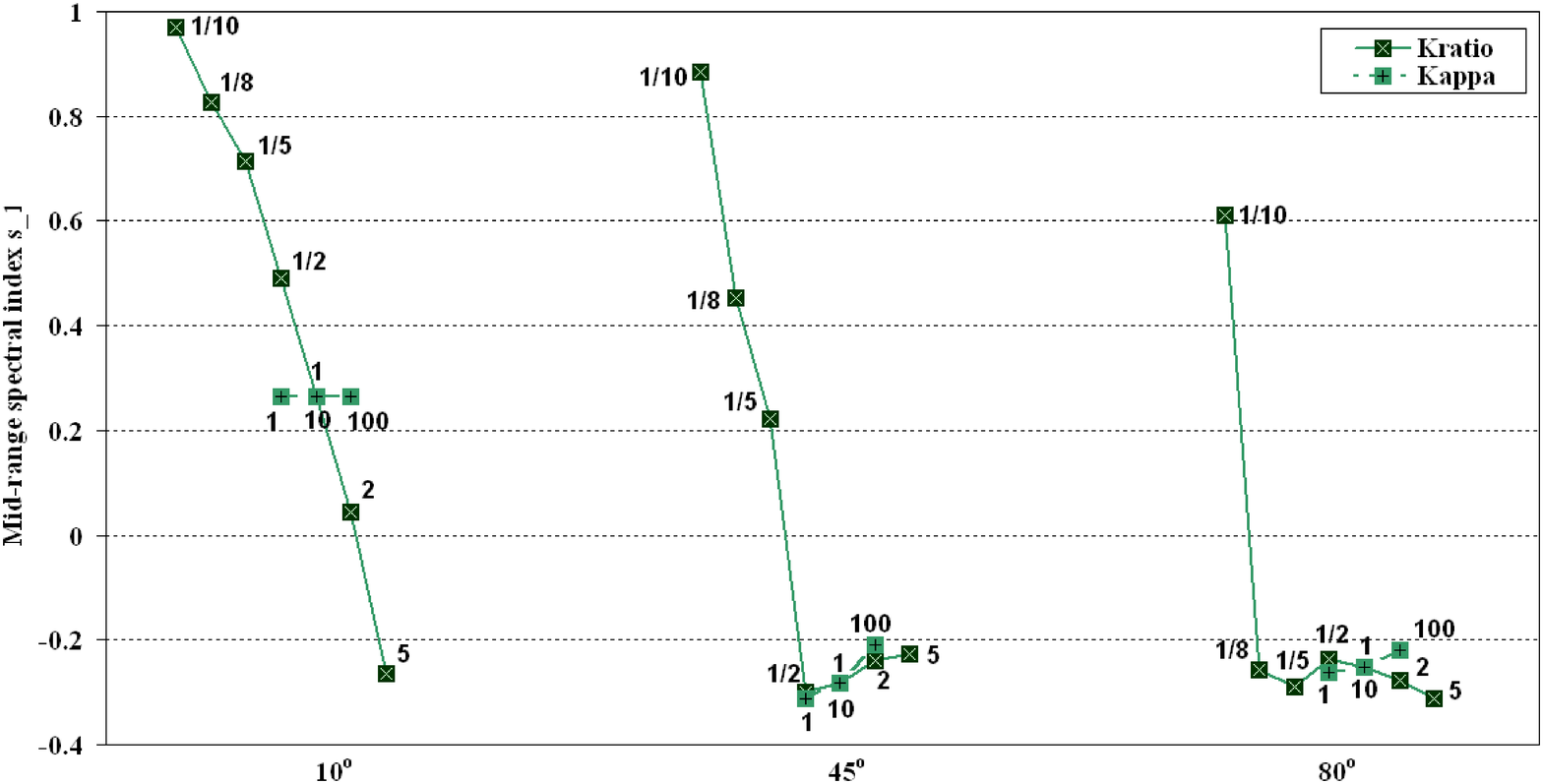}
	\caption{This figure shows the variation of the mid-range spectral index with changes in the magnetic field parameters $\kappa$.  We vary $\kappa$ jointly ($\kappa = \kappa_{\perp} = \kappa_{\parallel}$) by powers of 10, from 1 to 100, as indicated by the dotted line.  We also vary $\kappa_{\perp}$ and $\kappa_{\parallel}$ relative to one another by changing the ratio $K = \kappa_{\perp}/\kappa_{\parallel}$ through a range of values as indicated by the solid line.}
	\label{fig:kappas1}
\end{figure}

\renewcommand{\thefigure}{\arabic{figure}}
\begin{figure}
	\plotone{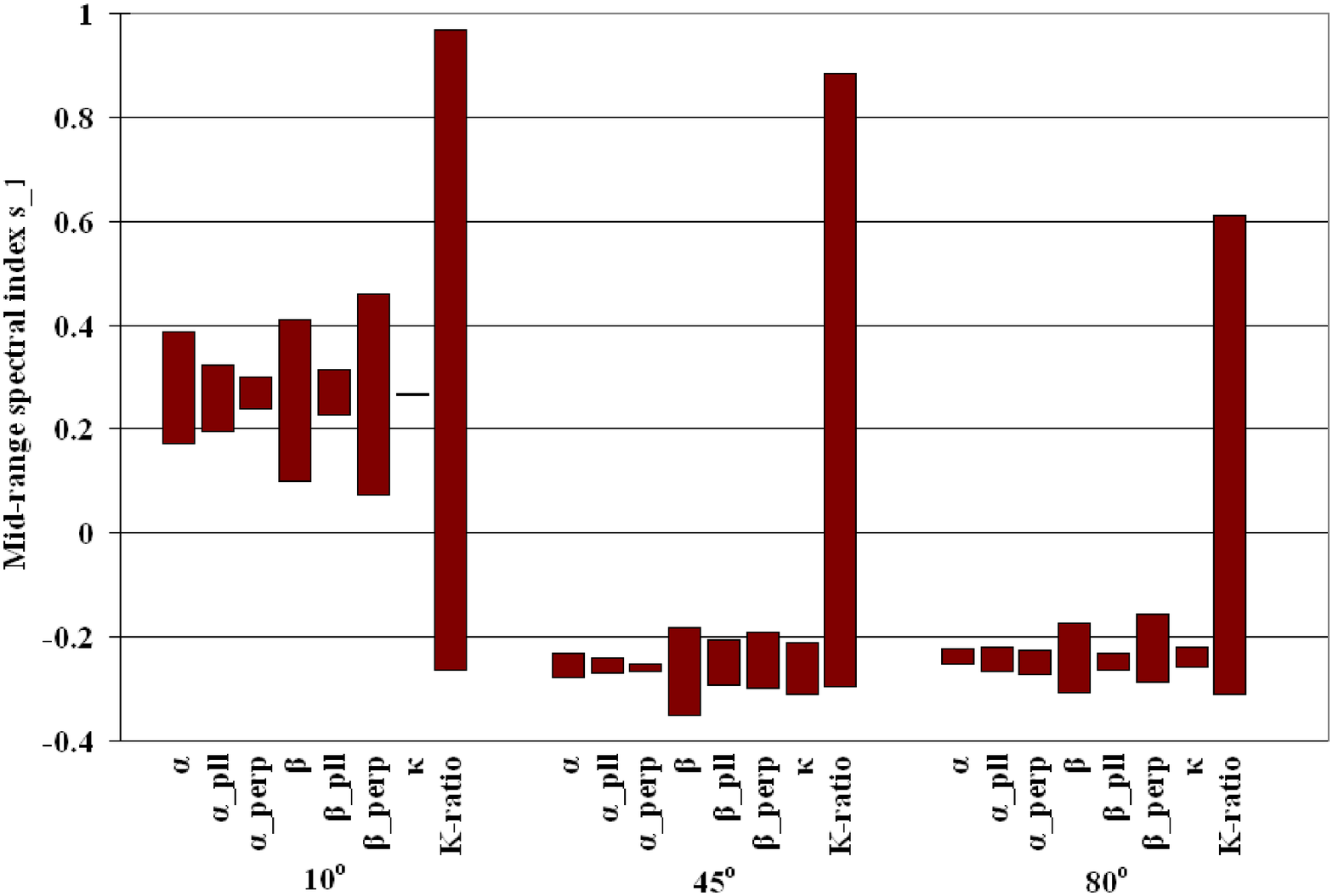}
	\caption{This figure compares the influences of different magnetic field spectral parameters on the mid-range spectral index $s_1$ of the radiation spectrum obtained for representative viewing angles $\theta = 10^o$, $45^o$, and $80^o$.  For each spectral parameter (indicated on the bottom axis) the graph indicates the range between the maximum and minimum values of $s_1$ obtained by our variations of that parameter (The parameter variations are as indicated in the previous figure and described in detail in section \ref{s:sfvarparam}.)}
	\label{fig:s1comp}
\end{figure}

\renewcommand{\thefigure}{\arabic{figure}(a)}
\begin{figure}
	\plotone{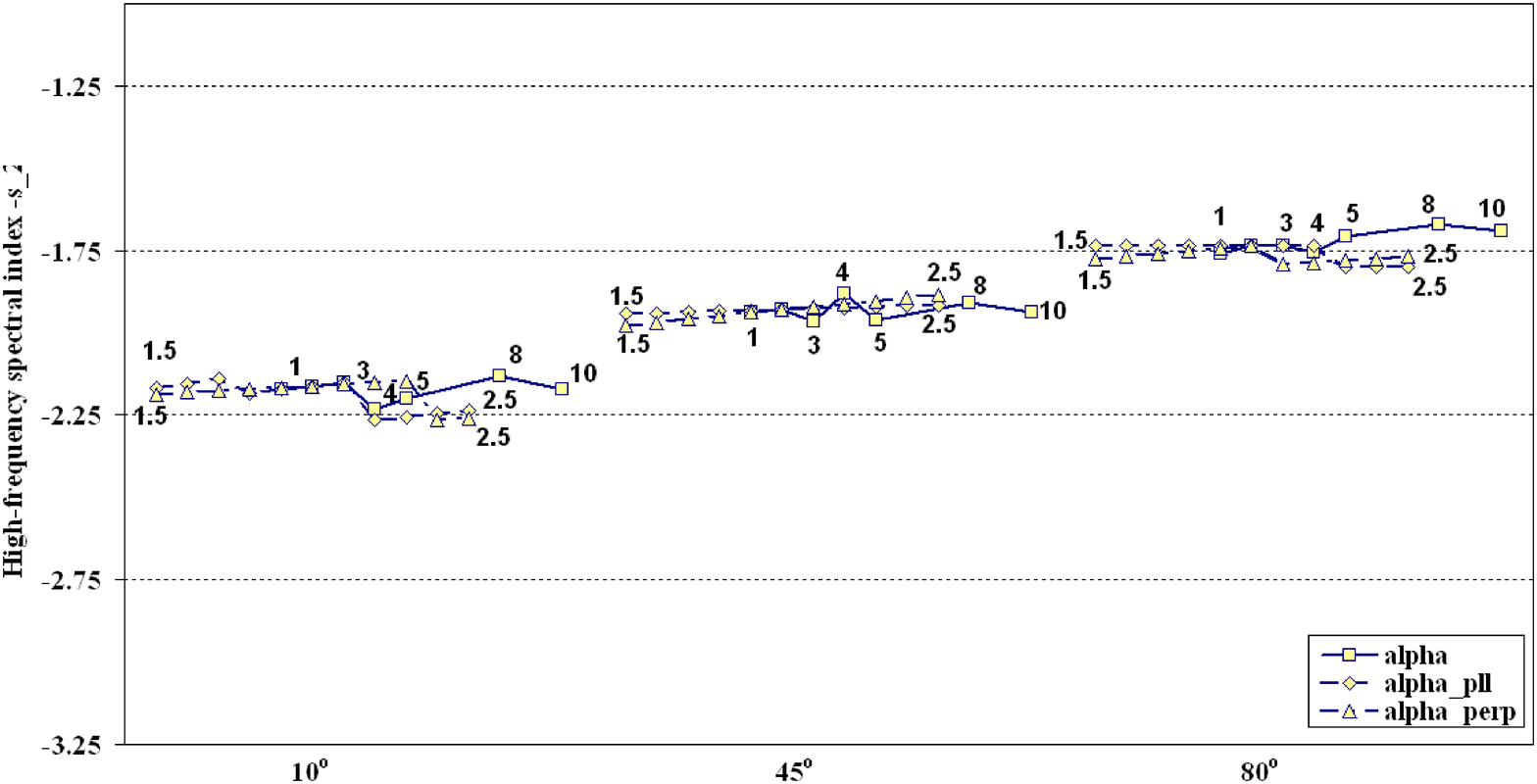}
	\caption{This figure shows the variation of the high-frequency spectral index (as determined by the fit described in section \ref{s:sfvarparam}) with changes in the magnetic field parameters $\alpha_i$.  The solid line with square data points indicates the behavior when $\alpha$ is jointly varied (over a range from 1 to 10, as indicated) in both the transverse and parallel magnetic field equation: $\alpha = \alpha_{\perp}=\alpha_{\parallel}$.  The dotted lines show the effect of varying $\alpha_{\perp}$ (triangular data points) and $\alpha_{\parallel}$ (diamond data points) individually, from 1.5 to 2.5, in increments of 0.1.}
	\label{fig:alphas2}
\end{figure}
\addtocounter{figure}{-1}
\renewcommand{\thefigure}{\arabic{figure}(b)}
\begin{figure}
	\plotone{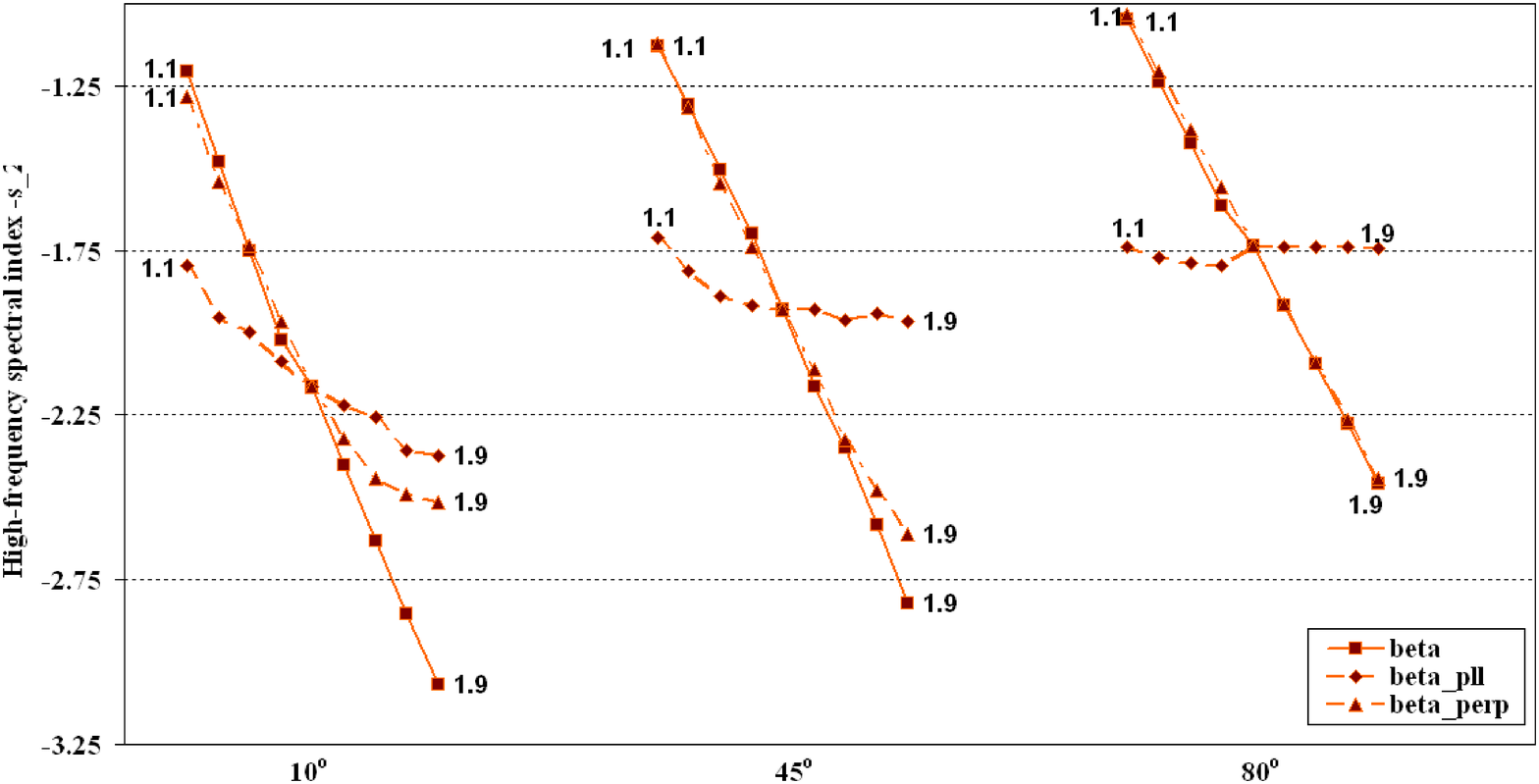}
	\caption{This figure shows the variation of the high-frequency spectral index with changes in the magnetic field parameters $\beta_i$.  The solid line with square data points indicates the behavior when $\beta$ is jointly varied from 1.1 to 1.9, in increments of 0.1, in both the transverse and parallel magnetic field equation: $\beta = \beta_{\perp}=\beta_{\parallel}$.  The dotted lines show the effect of varying $\beta_{\perp}$ (triangular data points) and $\beta_{\parallel}$ (diamond data points) individually, also from 1.1 to 1.9, in increments of 0.1.}
	\label{fig:betas2}
\end{figure}
\addtocounter{figure}{-1}
\renewcommand{\thefigure}{\arabic{figure}(c)}
\begin{figure}
	\plotone{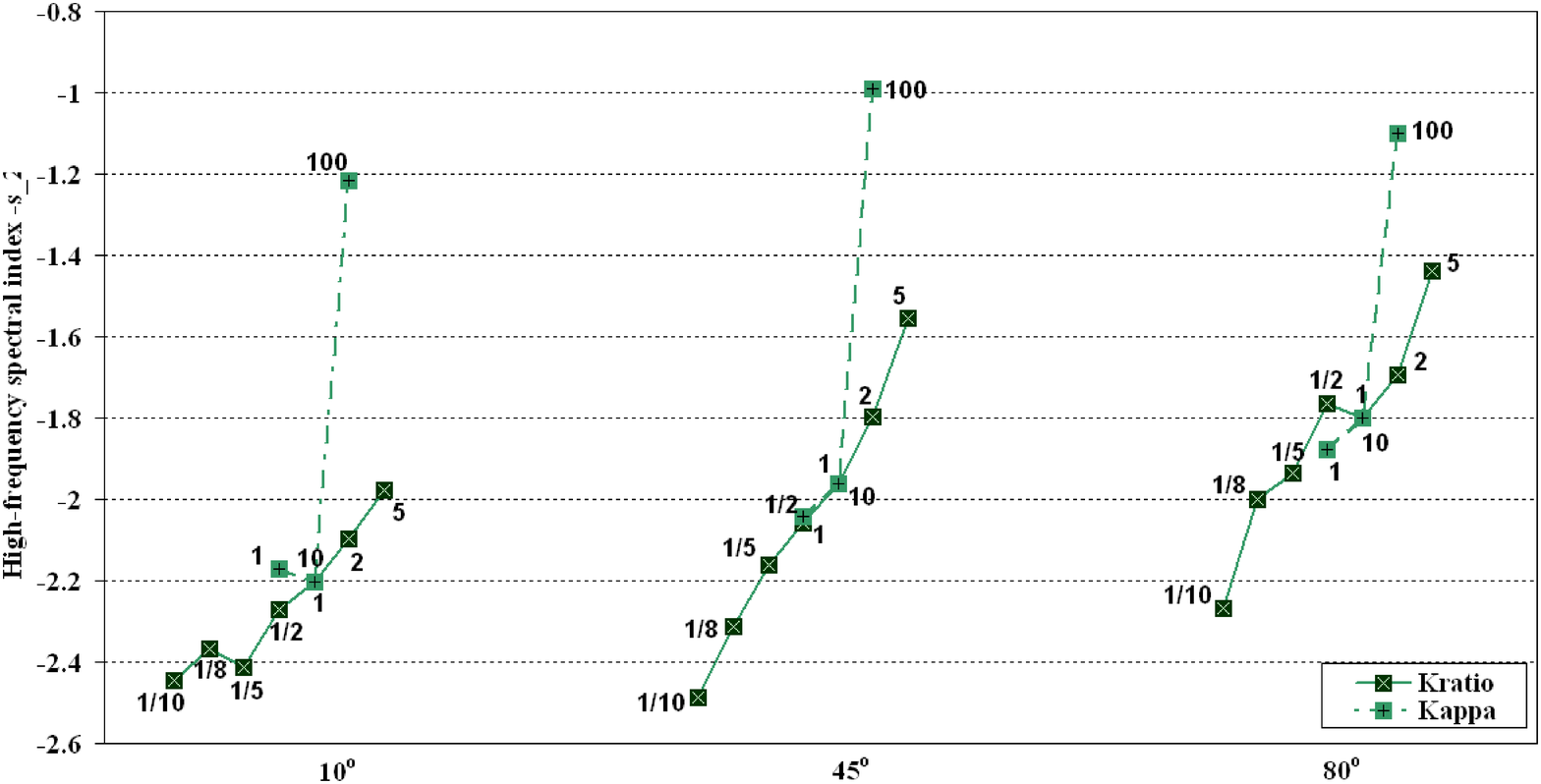}
	\caption{This figure shows the variation of the high-frequency spectral index with changes in the magnetic field parameters $\kappa$.  We vary $\kappa$ jointly ($\kappa = \kappa_{\perp} = \kappa_{\parallel}$) by powers of 10, from 1 to 100, as indicated.  We also vary $\kappa_{\perp}$ and $\kappa_{\parallel}$ relative to one another by changing the ratio $K = \kappa_{\perp}/\kappa_{\parallel}$ through a range of values as indicated.}
	\label{fig:kappas2}
\end{figure}
\clearpage

\renewcommand{\thefigure}{\arabic{figure}}
\begin{figure}
	\plotone{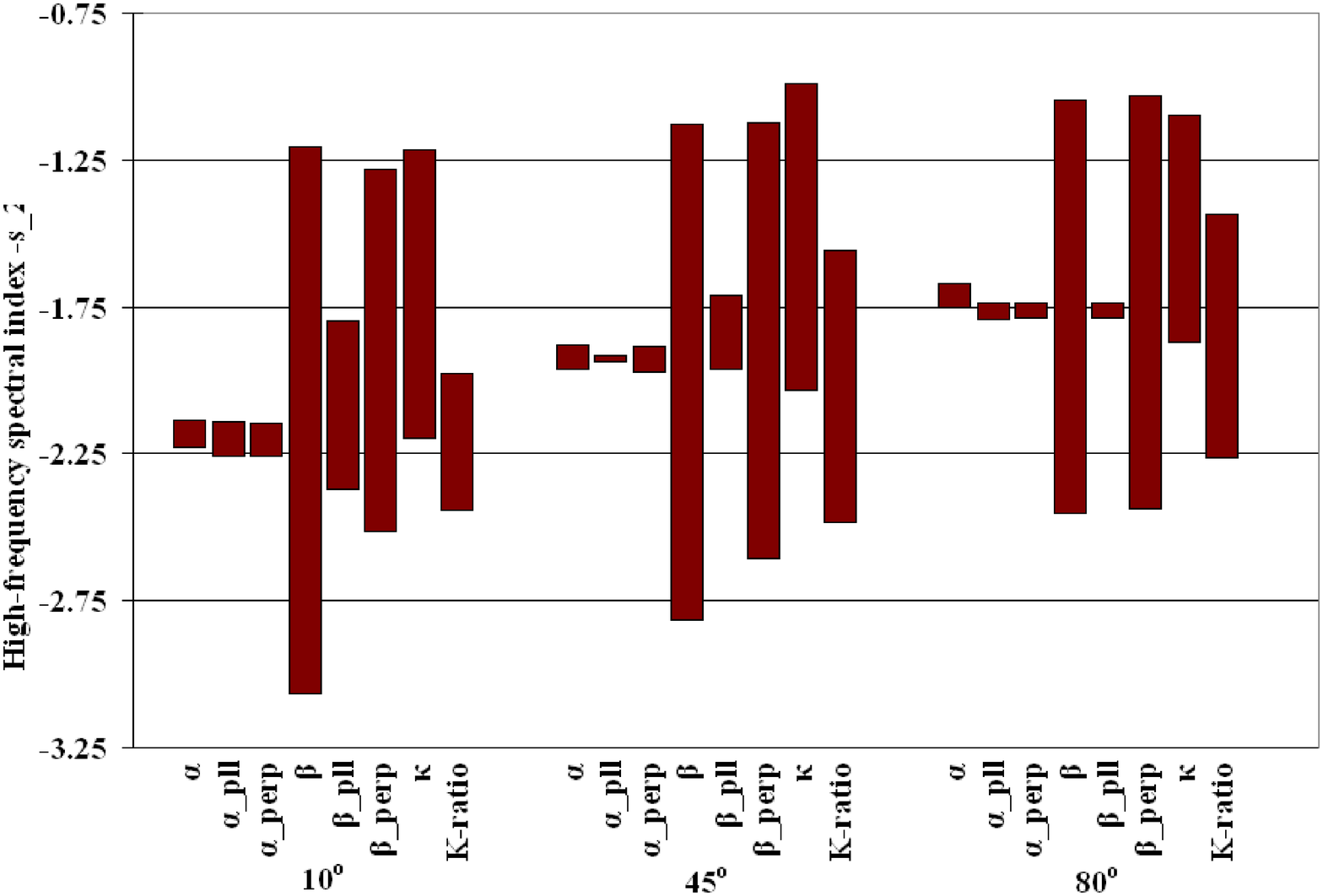}
	\caption{This figure compares the influences of different magnetic field spectral parameters on the high-frequency spectral index $-s_2$ of the radiation spectrum.  For each spectral parameter (indicated on the bottom axis) the graph indicates the range between the maximum and minimum values of $-s_2$ obtained by our variations of that parameter (The parameter variations are as indicated in the previous figures and described in detail in section \ref{s:sfvarparam}.)}
	\label{fig:s2comp}
\end{figure}

\clearpage
\renewcommand{\thefigure}{\arabic{figure}(a)}
\begin{figure}
	\plotone{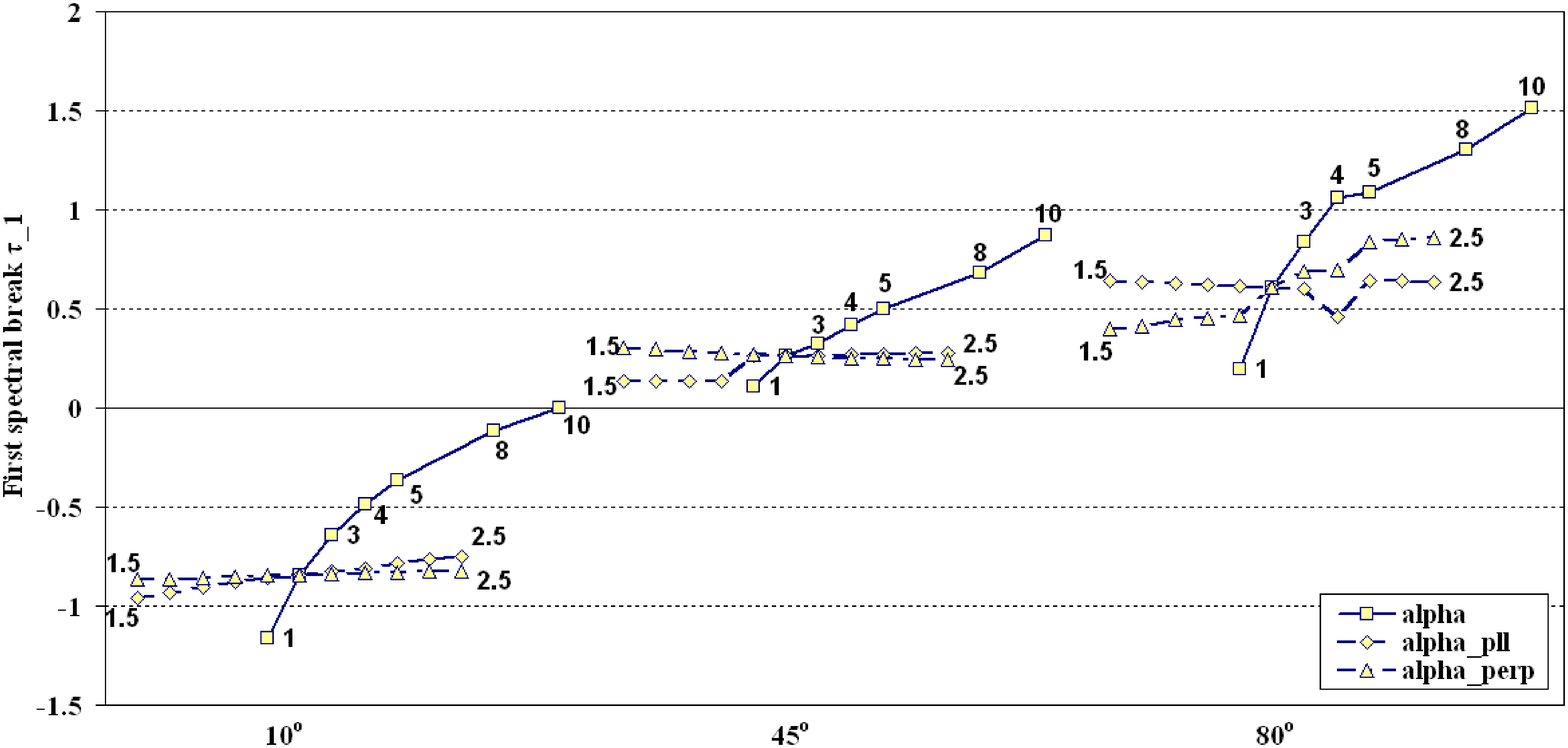}
	\caption{This figure shows the variation of the first spectral transition point (as described in section \ref{s:sfvarparam}) with changes in the magnetic field parameters $\alpha_i$, for $\theta = 10^o$, $45^o$, and $80^o$.  The solid line with square data points indicates the behavior when $\alpha$ is jointly varied (over a range from 1 to 10, as indicated) in both the transverse and parallel magnetic field equation: $\alpha = \alpha_{\perp}=\alpha_{\parallel}$.  The dotted lines show the effect of varying $\alpha_{\perp}$ (triangular data points) and $\alpha_{\parallel}$ (diamond data points) individually, from 1.5 to 2.5, in increments of 0.1.  }
	\label{fig:alphatp1}
\end{figure}
\addtocounter{figure}{-1}
\renewcommand{\thefigure}{\arabic{figure}(b)}
\begin{figure}
	\plotone{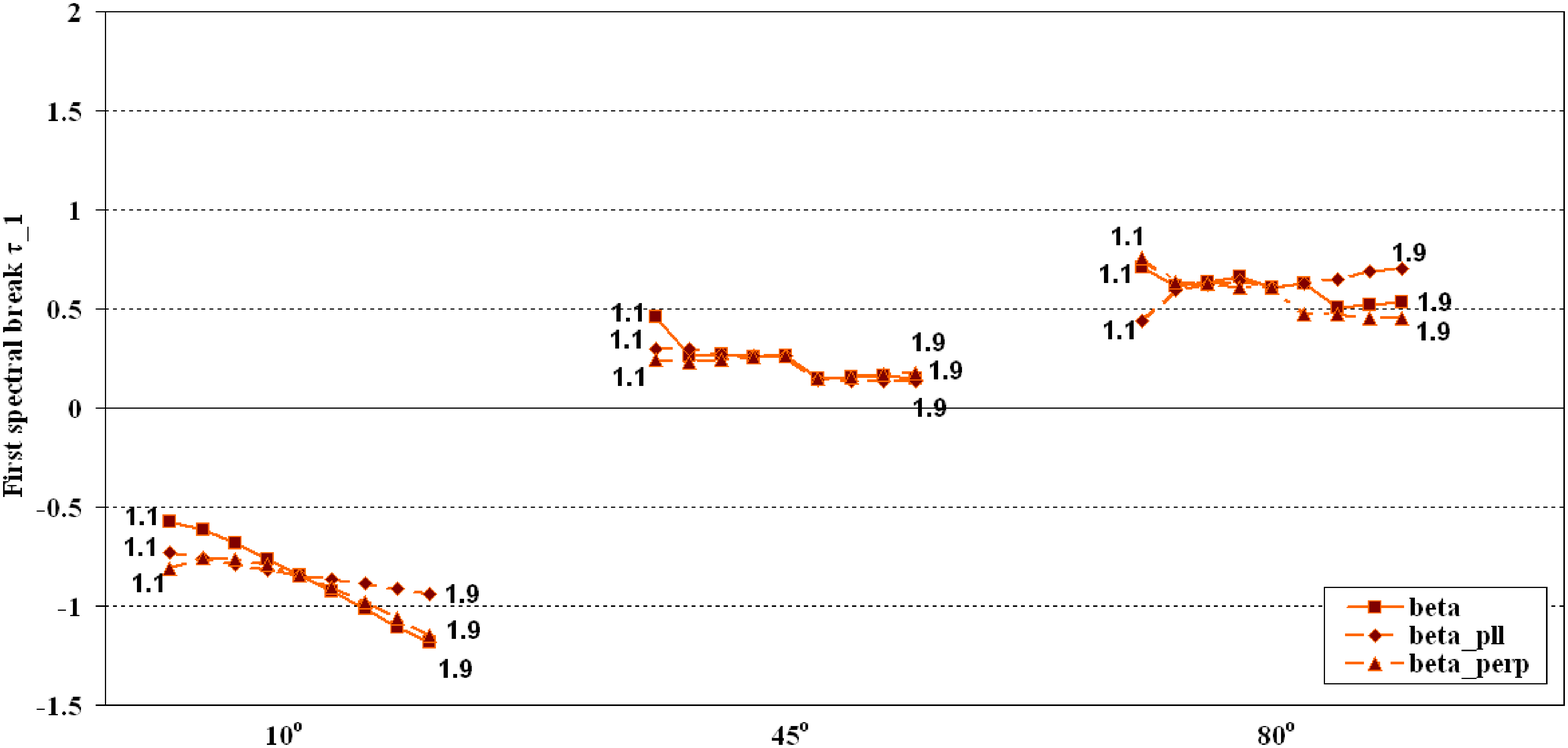}
	\caption{This figure shows the variation of the first spectral transition point with changes in the magnetic field parameters $\beta_i$, for $\theta = 10^o$, $45^o$, and $80^o$.  The solid line with square data points indicates the behavior when $\beta$ is jointly varied from 1.1 to 1.9, in increments of 0.1, in both the transverse and parallel magnetic field equation: $\beta = \beta_{\perp}=\beta_{\parallel}$.  The dotted lines show the effect of varying $\beta_{\perp}$ (triangular data points) and $\beta_{\parallel}$ (diamond data points) individually, also from 1.1 to 1.9, in increments of 0.1.}
	\label{fig:betatp1}
\end{figure}
\clearpage
\addtocounter{figure}{-1}
\renewcommand{\thefigure}{\arabic{figure}(c)}
\begin{figure}
	\plotone{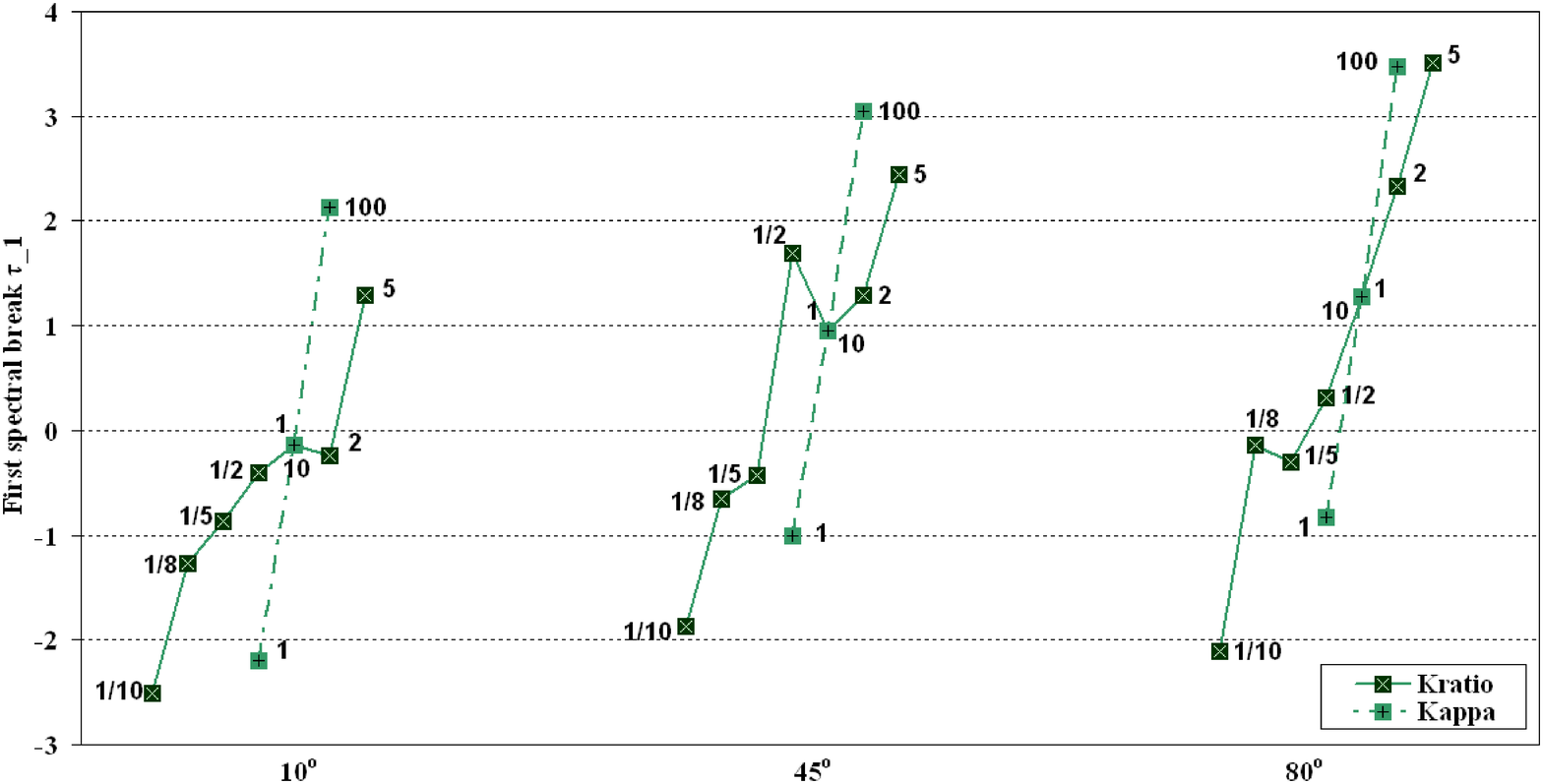}
	\caption{This figure shows the variation of the first spectral transition point with changes in the magnetic field parameters $\kappa_i$, for $\theta = 10^o$, $45^o$, and $80^o$.  We vary $\kappa$ jointly ($\kappa = \kappa_{\perp} = \kappa_{\parallel}$) by powers of 10, from 1 to 100, as indicated by the dotted lines.  We also vary $\kappa_{\perp}$ and $\kappa_{\parallel}$ relative to one another by changing the ratio $K = \kappa_{\perp}/\kappa_{\parallel}$ through a range of values as indicated by the solid lines.}
	\label{fig:kappatp1}
\end{figure}
\renewcommand{\thefigure}{\arabic{figure}}
\begin{figure}
	\plotone{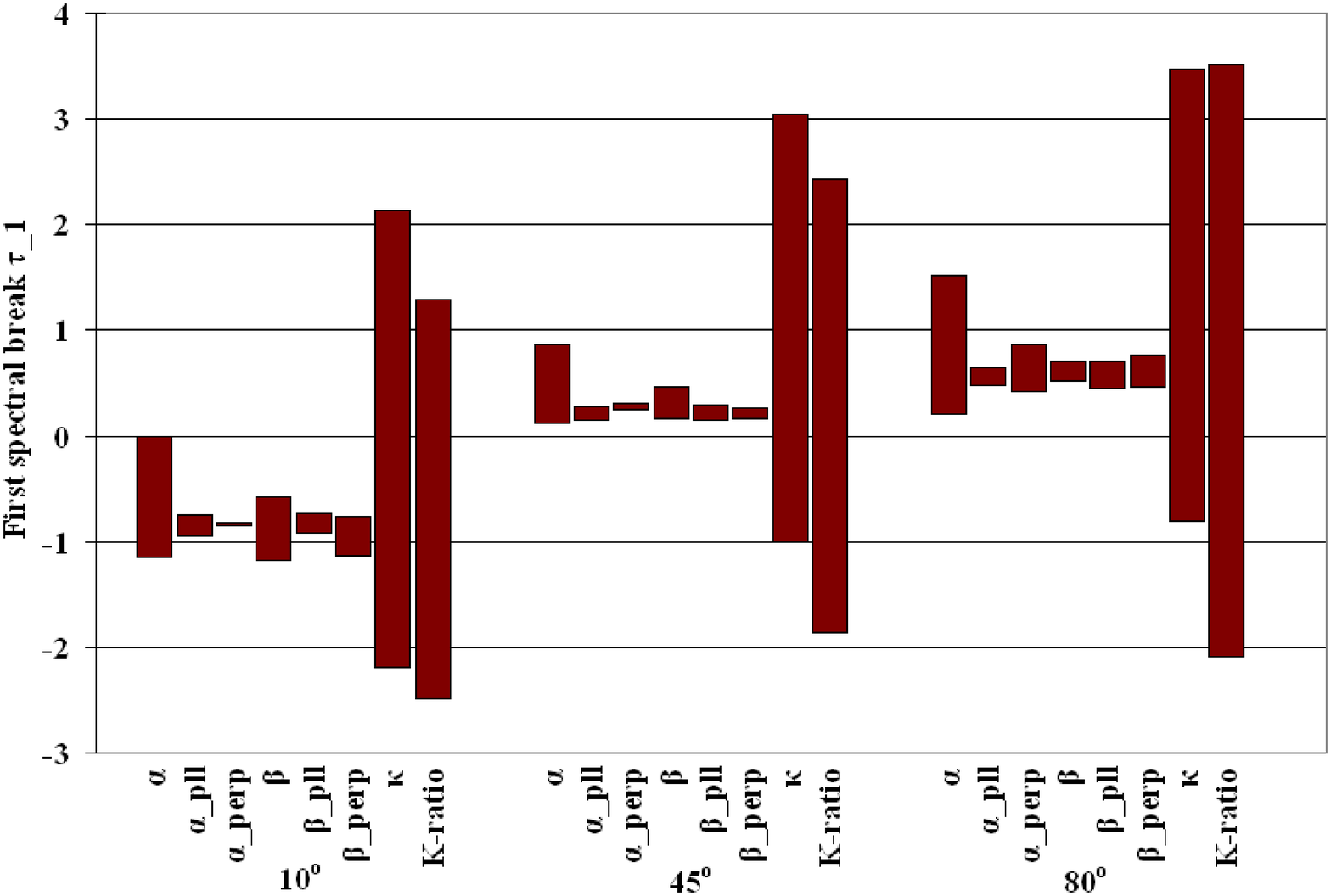}
	\caption{This figure compares the influences of different magnetic field spectral parameters on the first spectral break $\log_e(\omega/\omega_o\gamma^2)=\tau_1$ of the radiation spectrum obtained for representative viewing angles $\theta = 10^o$, $45^o$, and $80^o$.  For each spectral parameter (indicated on the bottom axis) the graph indicates the range between the maximum and minimum values of $\tau_1$ obtained by our variations of that parameter. (The parameter variations are as indicated in the previous figure and described in detail in section \ref{s:sfvarparam}.)}
	\label{fig:tp1comp}
\end{figure}
%
\renewcommand{\thefigure}{\arabic{figure}(a)}
\begin{figure}
	\plotone{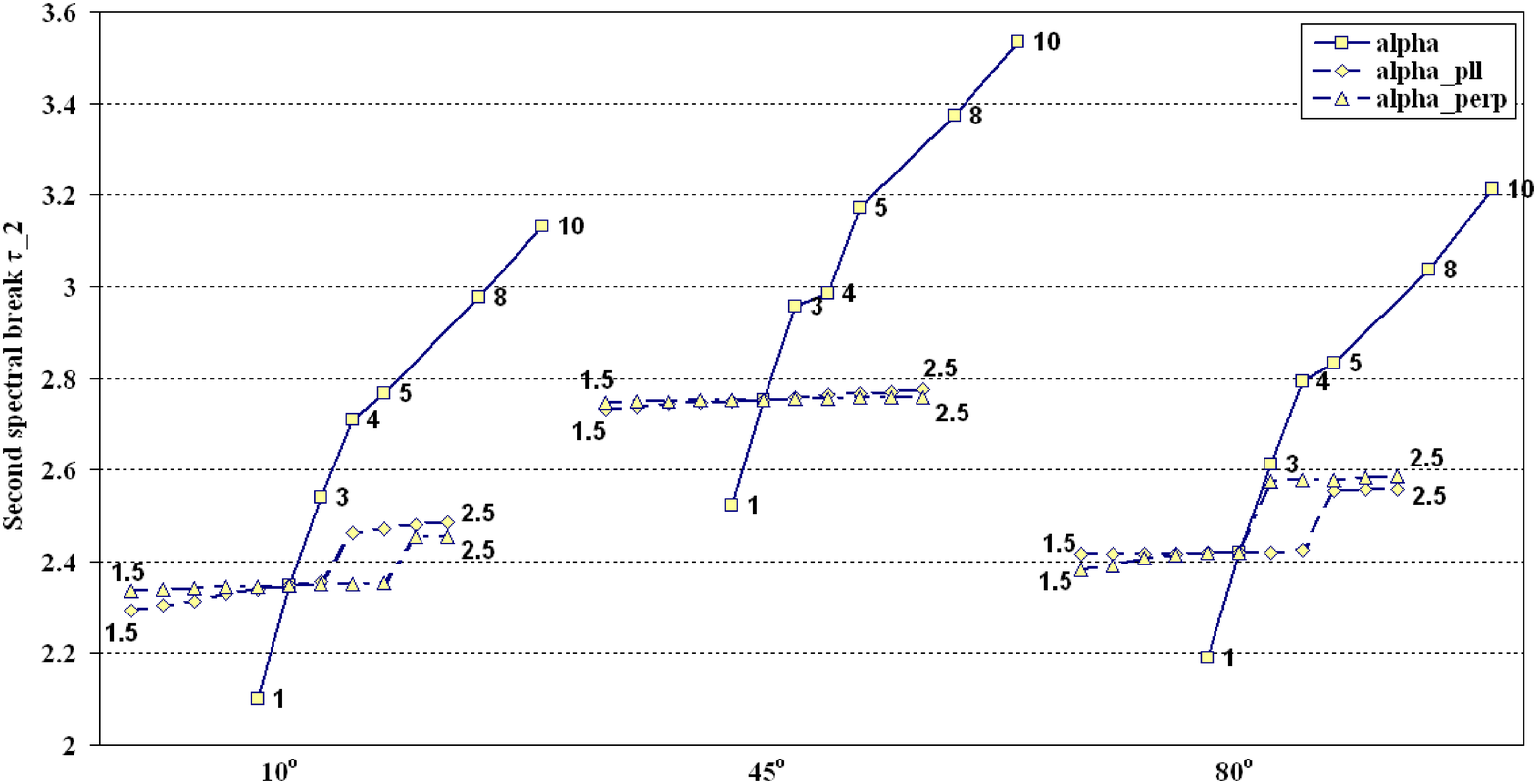}
	\caption{This figure shows the variation of the second spectral break (as determined by the fit described in section \ref{s:sfvarparam}) with changes in the magnetic field parameters $\alpha_i$.  The solid line with square data points indicates the behavior when $\alpha$ is jointly varied (over a range from 1 to 10, as indicated) in both the transverse and parallel magnetic field equation: $\alpha = \alpha_{\perp}=\alpha_{\parallel}$.  The dotted lines show the effect of varying $\alpha_{\perp}$ (triangular data points) and $\alpha_{\parallel}$ (diamond data points) individually, from 1.5 to 2.5, in increments of 0.1.  }
	\label{fig:alphatp2}
\end{figure}
\addtocounter{figure}{-1}
\renewcommand{\thefigure}{\arabic{figure}(b)}
\begin{figure}
	\plotone{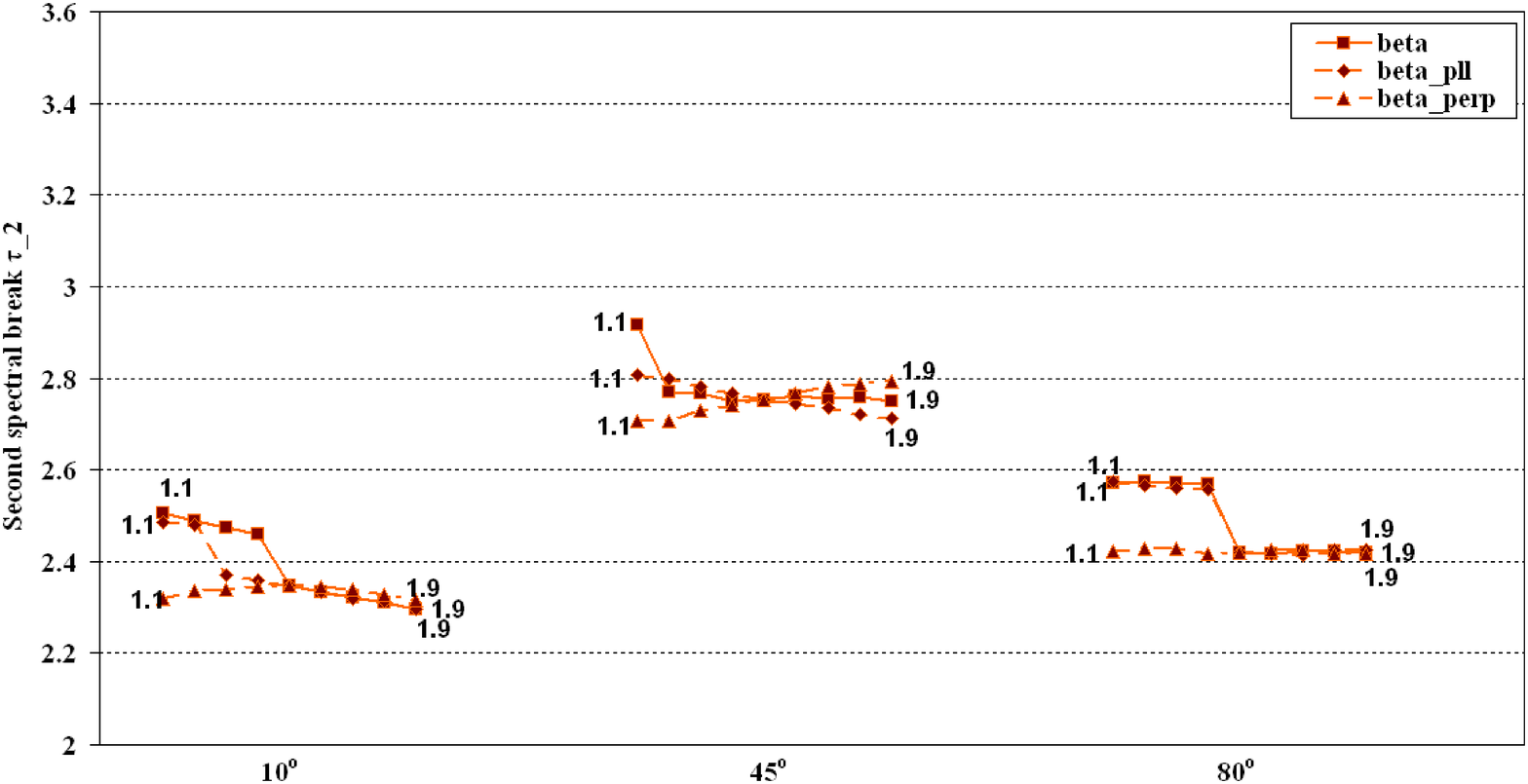}
	\caption{This figure shows the variation of the second spectral break with changes in the magnetic field parameters $\beta_i$.  The solid line with square data points indicates the behavior when $\beta$ is jointly varied from 1.1 to 1.9, in increments of 0.1, in both the transverse and parallel magnetic field equation: $\beta = \beta_{\perp}=\beta_{\parallel}$.  The dotted lines show the effect of varying $\beta_{\perp}$ (triangular data points) and $\beta_{\parallel}$ (diamond data points) individually, also from 1.1 to 1.9, in increments of 0.1.}
	\label{fig:betatp2}
\end{figure}
\addtocounter{figure}{-1}
\renewcommand{\thefigure}{\arabic{figure}(c)}
\begin{figure}
	\plotone{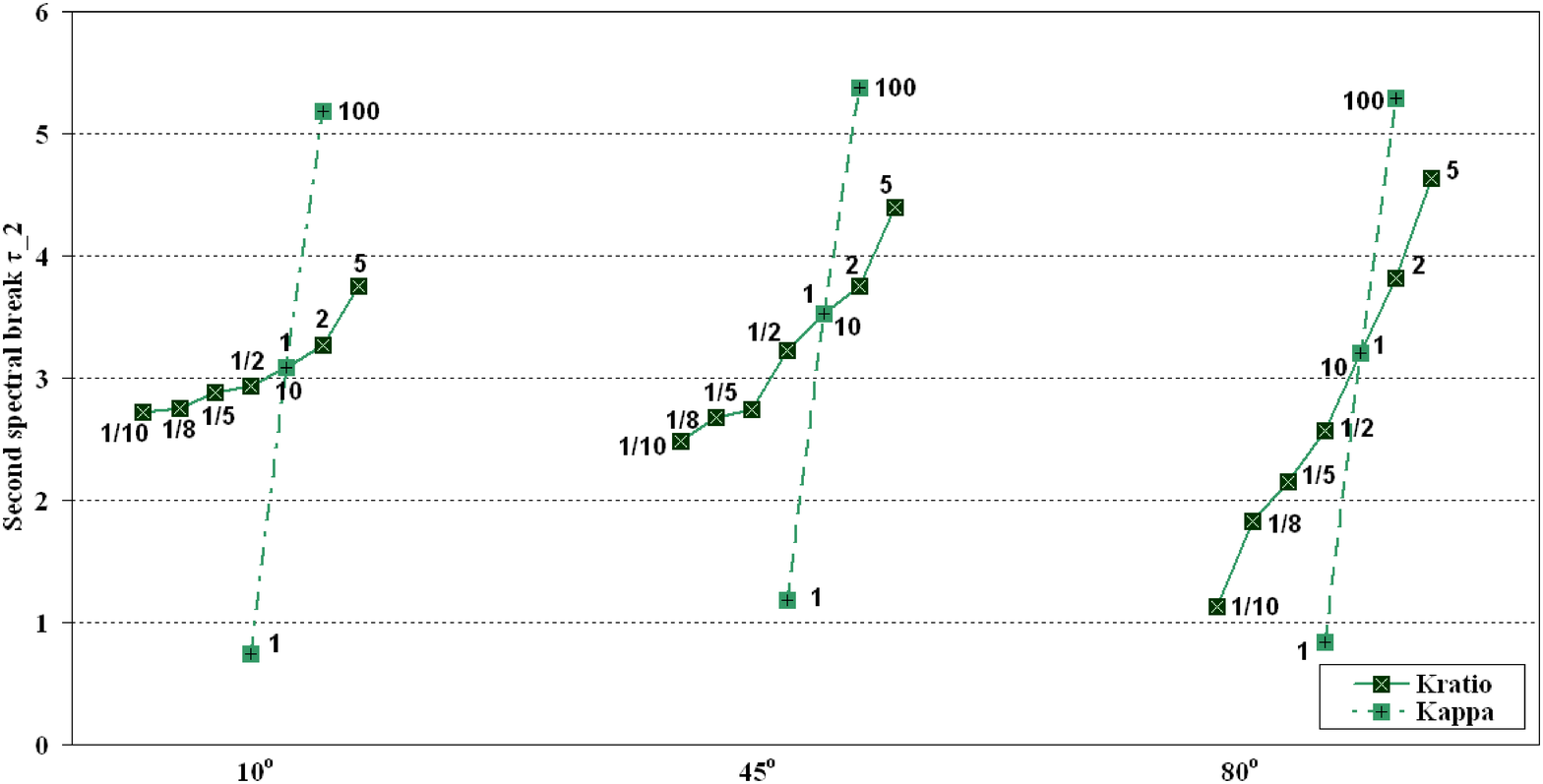}
	\caption{This figure shows the variation of the second spectral break with changes in the magnetic field parameters $\kappa$.  We vary $\kappa$ jointly ($\kappa = \kappa_{\perp} = \kappa_{\parallel}$) by powers of 10, from 1 to 100, as indicated by the dotted line.  We also vary $\kappa_{\perp}$ and $\kappa_{\parallel}$ relative to one another by changing the ratio $K = \kappa_{\perp}/\kappa_{\parallel}$ through a range of values as indicated by the solid line.}
	\label{fig:kappatp2}
\end{figure}
\begin{figure}
	\plotone{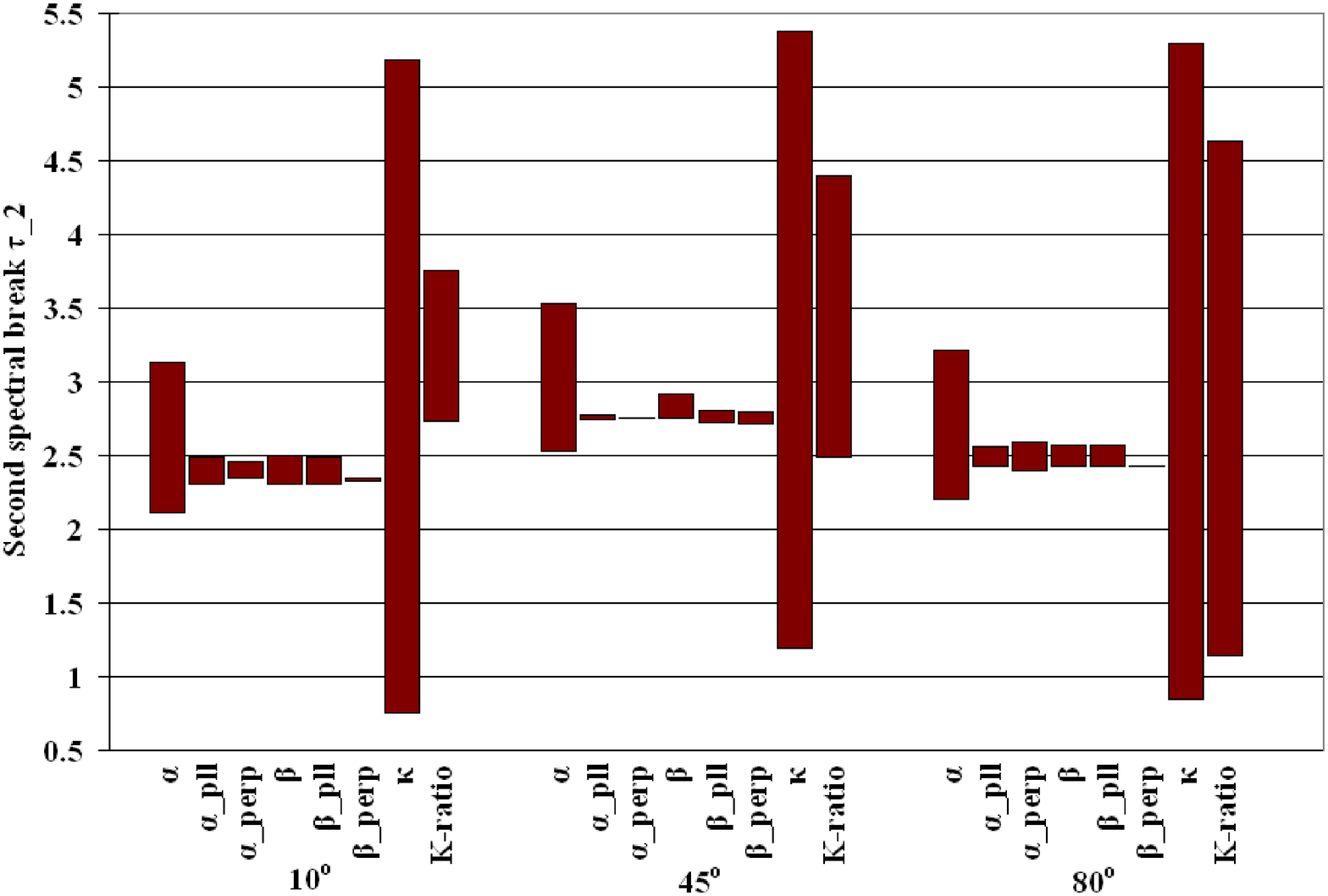}
	\caption{This figure compares the influences of different magnetic field spectral parameters on the second spectral break $\log_e(\omega/\omega_o\gamma^2)=\tau_2$ of the radiation spectrum.  For each spectral parameter (indicated on the bottom axis) the graph indicates the range between the maximum and minimum values of $\tau_2$ obtained by our variations of that parameter (The parameter variations are as indicated in the previous figure and described in detail in section \ref{s:sfvarparam}.)}
	\label{fig:tp2comp}
\end{figure}
%
\renewcommand{\thefigure}{\arabic{figure}(a)}
\begin{figure}
	\plotone{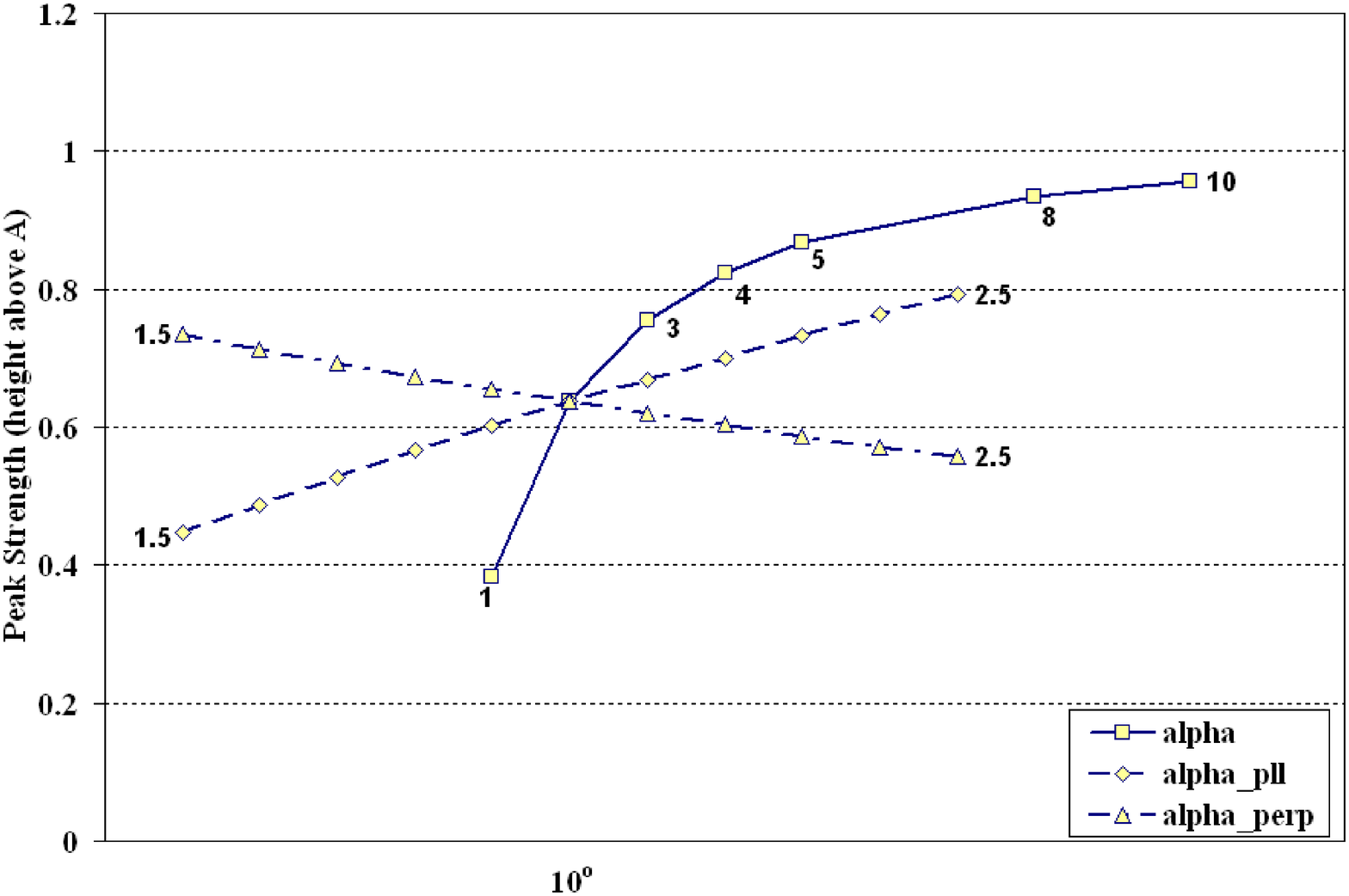}
	\caption{This figure shows the variation of peak strength, namely the peak height relative to the initial amplitude, for changes in the magnetic field parameters $\alpha_i$.  We present results only for $\theta = 10^{o}$ because at $\theta = 45^o$ and $\theta = 80^o$ the spectrum is unpeaked.  The solid line with square data points indicates the behavior when $\alpha$ is jointly varied (over a range from 1 to 10, as indicated) in both the transverse and parallel magnetic field equation: $\alpha = \alpha_{\perp}=\alpha_{\parallel}$.  The dotted lines show the effect of varying $\alpha_{\perp}$ (triangular data points) and $\alpha_{\parallel}$ (diamond data points) individually, from 1.5 to 2.5, in increments of 0.1.}
	\label{fig:alphapkheight}
\end{figure}
\addtocounter{figure}{-1}
\renewcommand{\thefigure}{\arabic{figure}(b)}
\begin{figure}
	\plotone{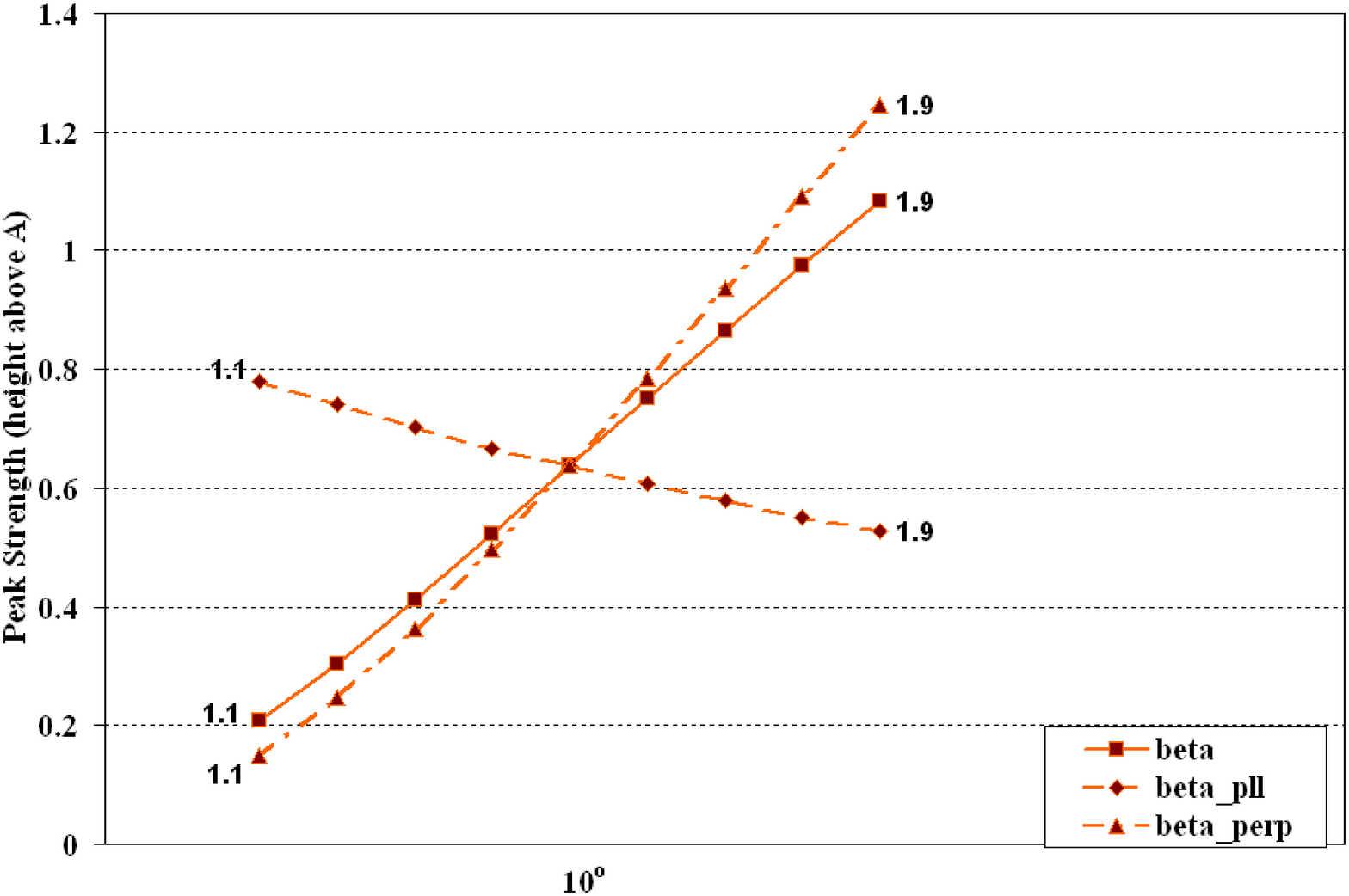}
	\caption{This figure shows the variation of peak strength (the peak height relative to the initial amplitude) with changes in the magnetic field parameters $\beta_i$.  The solid line with square data points indicates the behavior when $\beta$ is jointly varied from 1.1 to 1.9, in increments of 0.1, in both the transverse and parallel magnetic field equation: $\beta = \beta_{\perp}=\beta_{\parallel}$.  The dotted lines show the effect of varying $\beta_{\perp}$ (triangular data points) and $\beta_{\parallel}$ (diamond data points) individually, also from 1.1 to 1.9, in increments of 0.1.}
	\label{fig:betapkheight}
\end{figure}
\addtocounter{figure}{-1}
\renewcommand{\thefigure}{\arabic{figure}(c)}
\begin{figure}
	\plotone{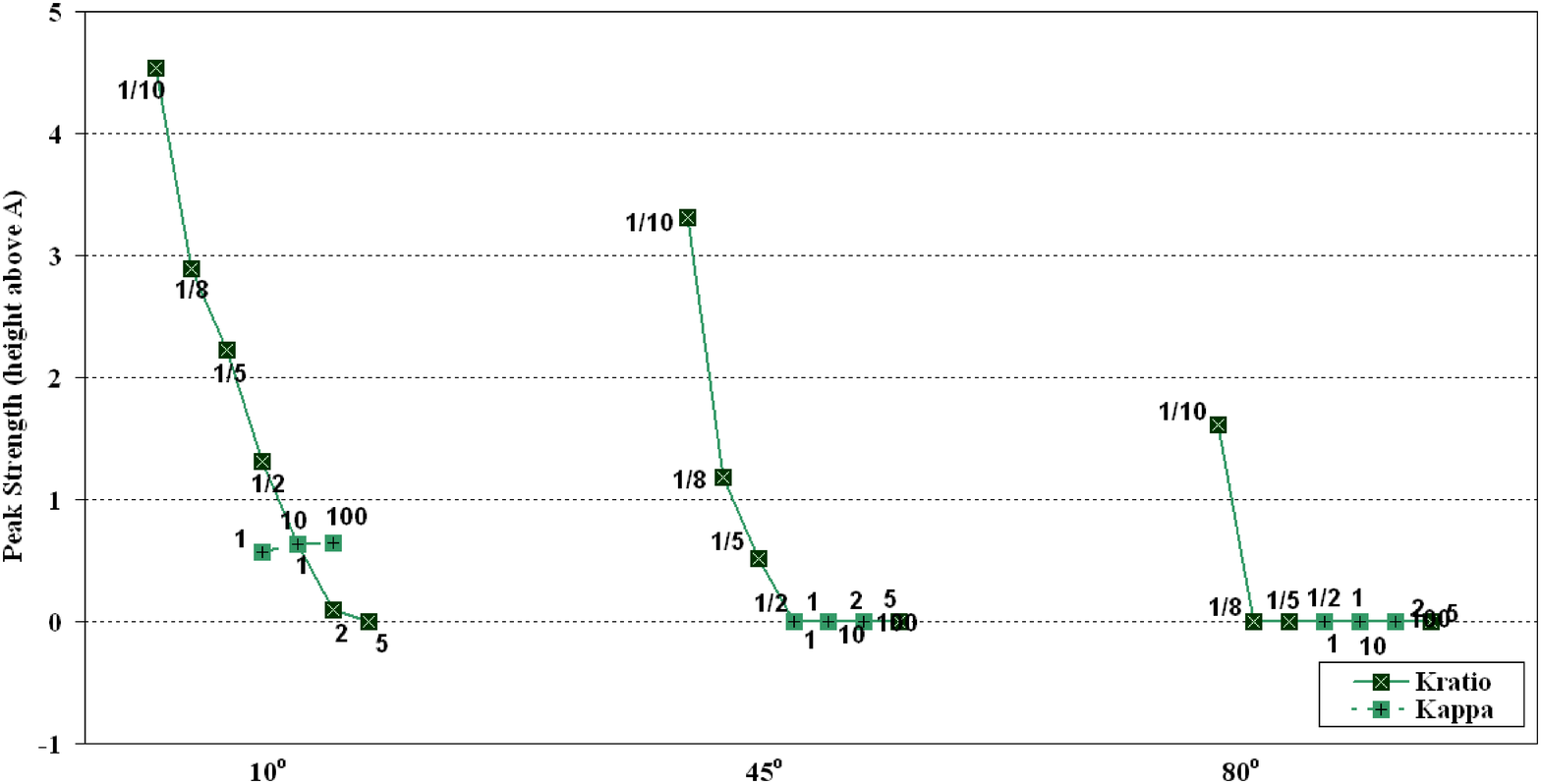}
	\caption{This figure shows the variation of peak strength (the peak height relative to the initial amplitude) with changes in the magnetic field parameters $\kappa$.  We vary $\kappa$ jointly ($\kappa = \kappa_{\perp} = \kappa_{\parallel}$) by powers of 10, from 1 to 100, as indicated.  We also vary $\kappa_{\perp}$ and $\kappa_{\parallel}$ relative to one another by changing the ratio $K = \kappa_{\perp}/\kappa_{\parallel}$ through a range of values as indicated.  Note that we obtain peaked forms of the spectra for larger $\theta$ when we have small values of the ratio $K$.}
	\label{fig:kappapkheight}
\end{figure}


\begin{thebibliography}{dummy}
%
\bibitem[Band et al.(1993)]{band}
Band, D.L., et al. 1993, \apj, 413, 281
%
\bibitem[Chang et al.(2008)]{CSA07}
Chang, P., Spitkovski, A., Arons, J. 2008, \apj, 674, 378
%
\bibitem[Frederiksen et al.(2004)]{fred04} Frederiksen, J. T.,
Hededal, C. B., Haugb\o lle, T., Nordlund, \AA. 2004 \apjl, 608, L13
%
\bibitem[Huntington et al.(2008)]{GRB+Hercules08} Huntington, C., et 
al. 2008, Bull. AAS, 40, 192 
%
\bibitem[Landau \& Lifshitz(1971)]{LL} Landau, L., \& Lifshitz, E. M. 1971, The classical theory of fields, (Oxford: Pergamon Press)
%
\bibitem[Kaneko et al.(2006)]{Kaneko} Kaneko, Y., et al. 2006, \apjs, 166, 298
%
\bibitem[Medvedev \& Loeb(1999)]{ML99}
Medvedev, M. V., \& Loeb, A. 1999, \apj, 526, 697
%
\bibitem[Medvedev(2000)]{M00} 
Medvedev, M. V. 2000, \apj, 540, 704
%
\bibitem[Medvedev(2006)]{M06} 			
Medvedev, M. V.  2006, \apj, 637, 869
%
\bibitem[Medvedev, et al.(2005)]{M+05} 
Medvedev, M. V., Fiore, M., Fonseca, R. A., Silva, L O., Mori, W. B. 2005, \apjl, 618, L75
%
\bibitem[Medvedev \& Spitkovsky(2009)]{MS09} 
Medvedev, M. V. \& Spitkovsky, A.  2009, \apj, 700, 956
%
\bibitem[Medvedev et al.(2009)]{MPR09}
Medvedev, M.V., Pothapragada, S., Reynolds, 2009, \apjl, 702, L91
%
\bibitem[Nishikawa, et al.(2003)]{Nish+03}
Nishikawa, K.-I., Hardee, P., Richardson, G., Preece, R., Sol, H., \&
Fishman, G. J. 2003, \apj, 595, 555 
%
\bibitem[Preece, et al.(1998)]{Preece+98}
Preece, R. D., Briggs, M. S., Malozzi, R. S., Pendleton, G. N., 
Paciesas, W. S., Band, D. L. 1998, \apjl, 506, 23
%
\bibitem[Reynolds, et al.(2007)]{HerculesKU07} Reynolds, S., 
Pothapragada, S., Graham, S., 
\& Medvedev, M.~V.\ 2007, APS Meeting Abstracts, 1021 
%
\bibitem[Silva, et al.(2003)]{Silva+03} Silva, L.~O., Fonseca, R.~A.,
Tonge, J.~W., Dawson, J.~M., Mori, W.~B., \& Medvedev, M.~V.\ 2003,
\apjl, 596, L121
%
\bibitem[Spitkovsky(2005)]{Spit05}
Spitkovsky, A. 2005, in AIP Conf. Proc. 801, Astrophysical Sources of High
Energy Particles and Radiation, ed. T. Bulik \& B. Rudak (Melville: AIP), 345
%
\bibitem[Spitkovsky(2007)]{Spit07}
Spitkovsky, A. 2007, \apj, 673, L39
%
\bibitem[Swisdak, et al.(2008)]{Swis+08}
Swisdak, M., Liu, Y.-H., \& Drake, J.F. 2008, \apj, 680, 999
%
\bibitem[Zenitani \& Hesse(2008)]{ZH08}
Zenitani, S., \& Hesse, M. 2008, Phys. Plasmas, 15, 022101
%
\end{thebibliography}
\end{document}